\documentclass[numberedappendix]{emulateapj}
\usepackage{apjfonts}
\usepackage{lscape}
\newcommand{\rotator}{}
\newcommand{\lscapeclose}{\end{landscape}}
\newcommand{\lscapeopen}{\begin{landscape}}
\newcommand{\sizer}{\tabletypesize{\tiny}}
\newcommand{\scaleup}{\epsscale{1.1}}
\newcommand{\plotter}{\plotone}
\newcommand{\plotterr}{\plotone}
\newcommand{\breaker}{}
\newcommand{\tableast}{$\ast$}
\newcommand{\longtabler}{\LongTables}
\newcommand{\appendixcolumns}{\twocolumngrid}

\newcommand{\tableset}{deluxetable}

\newcommand{\Mdot}{\dot{M}}

\newcommand{\etal}{et al.}
\newcommand{\mbh}{M_{\rm BH}}
\newcommand{\mstar}{M_{\ast}}
\newcommand{\msun}{M_{\sun}}
\newcommand{\lstar}{L_{\ast}}

\newcommand{\qeos}{q_{\rm eos}}
\newcommand{\fgas}{f_{\rm gas}}
\newcommand{\mdyn}{M_{\rm dyn}}
\newcommand{\re}{R_{\rm e}}
\newcommand{\fsb}{f_{\rm sb}}
\newcommand{\paperone}{Paper \textrm{I}}

\newcommand{\dmu}{\sigma_{\mu}}
\newcommand{\Sersic}{S\'ersic}

\shorttitle{Cusps in Elliptical Galaxies}
\shortauthors{Hopkins \etal}
\slugcomment{Submitted to ApJ, January 23, 2008}
\begin{document}

\title{Dissipation and Extra Light in Galactic Nuclei: \textrm{II}.\ ``Cusp'' Ellipticals}
\author{Philip F. Hopkins\altaffilmark{1}, 
Thomas J. Cox\altaffilmark{1,2}, 
Suvendra N. Dutta\altaffilmark{1}, 
Lars Hernquist\altaffilmark{1}, 
John Kormendy\altaffilmark{3}, 
\&\ Tod R.\ Lauer\altaffilmark{4}
}
\altaffiltext{1}{Harvard-Smithsonian Center for Astrophysics, 
60 Garden Street, Cambridge, MA 02138}
\altaffiltext{2}{W.~M.\ Keck Postdoctoral Fellow at the 
Harvard-Smithsonian Center for Astrophysics}
\altaffiltext{3}{Department of Astronomy, University of Texas, 1 
University Station, Austin, Texas 78712}
\altaffiltext{4}{National Optical Astronomy Observatory, Tucson, AZ 85726}

\begin{abstract}

We study the origin and properties of ``extra'' or ``excess'' central
light in the surface brightness profiles of cusp or power-law 
elliptical galaxies. Dissipational mergers give rise to two-component 
profiles: an outer profile established by violent relaxation
acting on stars already present in the progenitor galaxies prior to
the final stages of the merger, and an inner stellar
population comprising the extra light,
formed in a compact central starburst. 
By combining a large set of hydrodynamical simulations with
data that span a broad range of profiles at
various masses, we show that observed cusp ellipticals appear 
consistent with the predicted ``extra light'' structure, 
and we use our simulations to motivate a two-component 
description of the observations that allows us to 
examine how the properties and mass of this component 
scale with e.g.\ the mass, gas content, and other properties of the galaxies. 
We show how to robustly separate the physically meaningful extra 
light and outer, violently relaxed profile, and demonstrate that 
the observed cusps and ``extra light'' are reliable tracers of the degree of 
dissipation in the spheroid-forming merger. 
We show that the typical degree of dissipation is a strong function of 
stellar mass, roughly tracing the observed gas fractions of disks 
of the same mass over the redshift range $z\sim0-2$. 
We demonstrate a correlation between the strength of this 
component and effective radius at fixed mass, in the sense that 
systems with more dissipation are more compact, 
sufficient to explain the discrepancy in the maximum phase-space 
and mass densities of
ellipticals and their progenitor spirals. 
We show that the outer shape of the light profile in simulated and observed systems
(when fit to properly account for the central light) does not depend 
on mass, with a mean outer \Sersic\ index $\sim2.5$.
We also explore how this relates to e.g.\ the shapes, kinematic properties, 
and stellar population gradients of 
ellipticals. Extra light contributes to making remnants rounder and diskier, 
and imprints stellar population gradients. Simulations with the gas content
needed to match observed surface brightness profiles reproduce the 
observed age, metallicity, and color gradients of cusp ellipticals, and 
we make predictions for how these can be used as tracers of the degree of dissipation 
in spheroid formation. 

\end{abstract}

\keywords{galaxies: elliptical and lenticular, cD --- galaxies: evolution --- 
galaxies: formation --- galaxies: nuclei --- galaxies: structure --- 
cosmology: theory}

\section{Introduction}
\label{sec:intro}

Thirty years ago, \citet{toomre77} proposed the ``merger hypothesis,''
that major mergers between spirals could result in elliptical galaxies, 
and the combination of detailed observations of 
recent merger remnants \citep{schweizer82,LakeDressler86,Doyon94,ShierFischer98,James99,
Genzel01,tacconi:ulirgs.sb.profiles,dasyra:mass.ratio.conditions,dasyra:pg.qso.dynamics,
rj:profiles,rothberg.joseph:kinematics} and e.g.\ faint shells and tidal 
features around ellipticals \citep{malin80,malin83,schweizer80,
schweizerseitzer92,schweizer96} have lent considerable 
support to this picture \citep[e.g.][]{barneshernquist92}.
Furthermore, in the now established $\Lambda$CDM 
cosmology, structure grows hierarchically \citep[e.g.][]{whiterees78}, making 
mergers an inescapable element in galaxy formation.

However, it has long been recognized 
that purely dissipationless (collisionless) mergers of stellar 
disks cannot explain the high mass and 
phase-space densities of nearby ellipticals (especially 
those of relatively low mass $\lesssim\lstar$), which 
are far more dense than local 
stellar disks of the same mass 
\citep{ostriker80,carlberg:phase.space,gunn87,kormendy:dissipation}. Of course, 
spiral galaxies are not purely collisionless systems, but contain
interstellar gas in addition to stars and dark matter. 
Furthermore, most ellipticals have old stellar populations, implying 
that their last gas-rich mergers occurred at $z\gtrsim1$, and therefore 
their progenitors may be high-redshift spirals. These 
were likely more dense than nearby spirals, and observational 
evidence \citep{erb:lbg.gasmasses} indicates they 
had even larger gas fractions than their present-day 
counterparts ($f_{\rm gas}\sim0.5$, with some
approaching $f_{\rm gas}\sim0.8-0.9$). Because gas
can radiate, it is not subject to Liouville's Theorem, and 
processes related to gas dynamics and
star formation can reconcile the high phase space densities of
ellipticals relative to spirals
\citep{gunn87,lake:merger.remnant.phase.space,schweizer98}. 
In detail,
\citet{hernquist:phasespace} estimated that $\sim10\%$ 
of the stellar mass must be added in a compact dissipational 
component to account for the central densities of typical ellipticals.

The possible importance of gas dynamics and triggered star formation in
mergers is reinforced by observations of ultraluminous infrared
galaxies (ULIRGs)
\citep[e.g.][]{soifer84a,soifer84b}, which are always associated 
with mergers in the local Universe \citep{joseph85,sanders96:ulirgs.mergers}. 
The infrared emission from
ULIRGs is thought to be powered by intense starbursts in the nuclei of
these objects, originating in compact, central concentrations of gas
\citep[e.g.][]{scoville86, sargent87,sargent89}, which will leave 
dense stellar remnants \citep{kormendysanders92,hibbard.yun:excess.light,rj:profiles}. 
Moreover, observations of merging systems and gas-rich merger remnants
\citep[e.g.,][]{LakeDressler86,Doyon94,ShierFischer98,James99}, as
well as post-starburst (E+A/K+A) galaxies
\citep{goto:e+a.merger.connection}, have shown that their kinematic
and photometric properties are consistent with them eventually
evolving into typical $\sim L_{\ast}$ elliptical galaxies. The
correlations obeyed by these mergers and remnants
\citep[e.g.,][and references 
above]{Genzel01,rothberg.joseph:kinematics,rothberg.joseph:rotation}
are similar to e.g.\ the observed fundamental plane and 
\citet{kormendy77:correlations} 
relations for relaxed ellipticals, and consistent with evolution onto
these relations as their stellar populations age, as well as the
clustering and mass density of ellipticals \citep{hopkins:clustering}.

The link between these processes and the formation of ellipticals 
may be manifest in their surface brightness profiles. 
Early work by e.g.\ 
\citet{kormendy77:photometry,king78,young78,lauer:85} and 
\citet{kormendy85:profiles} \citep[see][for a review]{kormendy:cores.review} 
showed 
that typical elliptical surface brightness 
profiles were not as simple as uniform $r^{1/4}$ laws. 
Typically, central profiles interior to $\sim1\,$kpc 
deviate from $r^{1/4}$ laws fitted to the envelopes of ellipticals, 
falling both above and below the 
inward extrapolation of the $r^{1/4}$ law. 
Contemporaneously with the discovery of black
hole-host galaxy correlations \citep[e.g.][]{KormendyRichstone95,fm00,Gebhardt00}, 
{\em Hubble Space Telescope}
observations of the centers of elliptical galaxies established that
typical $\lesssim \lstar $ ellipticals exhibit 
central ``cusps'' -- i.e.\ a continued rise in 
power-law like fashion towards small radii \citep{lauer91,lauer92:m32,crane93,
ferrarese:type12,kormendy94:review,lauer:95,kormendy99}, 
whereas the most massive ellipticals appear to exhibit central 
flattening or ``cores.''
\citet{kormendy99} demonstrated in a number of cases that these 
``cusps'' appeared in some sense to be ``extra'' light, i.e.\ a distinct 
component above the inward extrapolation of the outer profile measured 
at large radii.
With the combination of 
HST and ground-based photometry \citep{jk:profiles}, it now appears that this 
excess is ubiquitous in ``cuspy'' or ``power-law'' ellipticals, 
with mass ranges and spatial extents comparable to those
expected from observations of ongoing merger-induced 
starbursts \citep{hibbard.yun:excess.light,rj:profiles} 
and numerical simulations \citep{mihos:cusps,hopkins:cusps.mergers}.

\citet{faber:ell.centers} showed that the presence or 
absence of a cusp or power-law nuclear profile is strongly correlated 
with other, global properties of ellipticals -- cusp ellipticals 
tend to be more rotationally supported, diskier, and have 
slightly higher ellipticities. 
They argued that these differences reinforce the idea that at least the 
cusp\footnote{There are some differences in the literature in 
the use of the term ``cusp ellipticals.'' Unless otherwise stated, we will use 
it to refer to ellipticals without a central resolved core/flattening -- 
i.e.\ ``power-law'' ellipticals.}
ellipticals are the direct product of gas-rich 
mergers, with dissipation forming the central ``cusp'' and 
giving rise to correlated kinematic and photometric properties.
The central excess or ``extra light'' in these cases 
may therefore represent a distinct imprint of the degree of dissipation in 
the spheroid-forming merger. 

Numerical modeling over the past twenty years has also indicated that
gas physics and star formation play key roles in shaping elliptical
galaxies and has elucidated the relationship of mergers to the various
phenomena described above. 
The possible relevance of these additional processes was anticipated
already by \citet{toomre72}, who asked whether mergers would not
``... tend to bring {\it deep} into a galaxy a fairly {\it sudden}
supply of fresh fuel in the form of interstellar material ...''
\citet{barnes.hernquist.91,barneshernquist96} showed that tidal
torques excited during {\it major} mergers excite rapid inflows of gas
into the centers of galaxies, providing the fuel to power intense
starbursts \citep{mihos:starbursts.94, mihos:starbursts.96} and to
feed rapid black hole growth \citep{dimatteo:msigma,hopkins:lifetimes.letter,
hopkins:lifetimes.methods}. 
Gas consumption
by the starburst and dispersal of residual gas by supernova-driven
winds and feedback from black hole growth
\citep{springel:red.galaxies}, culminating in a pressure-driven
blast-wave \citep[e.g.][]{hopkins:faint.slope,hopkins:seyferts},
terminate star formation so that the
remnant quickly evolves from a blue to a red galaxy. 
Provided that the interaction involved a
``major'' merger,\footnote{In a major merger, tidal forces are
sufficiently strong to drive nuclear inflows of gas and build
realistic spheroids.  The precise meaning of major merger in this
context is blurred by a degeneracy between the progenitor mass ratio
and the orbit
\citep{hernquist.89,hernquist.mihos:minor.mergers,bournaud:minor.mergers},
but both numerical \citep{younger:minor.mergers} and observational
\citep{dasyra:mass.ratio.conditions,woods:tidal.triggering} studies
indicate that massive inflows of gas and morphological transformation
are typical for mass ratios only below $\sim 3:1$.  Unless otherwise
noted, we generally take the term ``mergers'' to refer to major
mergers.} the remnant will resemble an elliptical galaxy, with the
bulk of its mass on large scales made from progenitor stars which
experienced violent relaxation 
\citep[e.g.][]{barnes:disk.halo.mergers,barnes:disk.disk.mergers,
hernquist:bulgeless.mergers,hernquist:bulge.mergers},
and the dissipation-induced starburst
appropriately boosting the concentration and central phase space
density \citep{hernquist:phasespace,robertson:fp,naab:gas,cox:kinematics}.
Moreover, \citet{mihos:cusps}
predicted that this process should leave an observable signature
in the surface brightness profiles of remnants, in the form of an
upwards departure from the outer \citet{devaucouleurs} $r^{1/4}$-law
distribution in the inner regions: i.e. a central ``extra
light'' above the inwards extrapolation of the outer profile.

Understanding the processes responsible for establishing the
structural properties of ellipticals and their correlations has the
potential of revealing the formation histories of these objects.
However, notwithstanding major observational and numerical advances,
little effort has been made to use the extra light content of
ellipticals in this manner, and most studies have restricted 
their focus to
determining whether or not some extra light component is evident.
This owes largely to the absence of a detailed theoretical framework:
while the original work of \citet{mihos:cusps}
predicted that such cusps should
exist, a more refined treatment of star formation and feedback, along
with better resolution, is required for more detailed interpretation
and modeling. For example, owing to limited spatial and
temporal resolution and 
a simplified treatment of star formation, the ``extra light'' profiles 
predicted by 
\citet{mihos:cusps} generally exhibited much more severe breaks
than predicted by state-of-the-art simulations and seen in 
recent observations (see Appendix~\ref{sec:appendix:resolution}). 
There have been considerable improvements in these areas in recent
years \citep[e.g.][]{springel:multiphase,springel:models,cox:feedback}, and we take
advantage of these refinements here, and in companion papers,
\citet{hopkins:cusps.mergers} (hereafter \paperone) and
\citet{hopkins:cores,hopkins:cusps.fp}, to study galaxy
cusps or extra light in both simulations and observed systems.  Our
objective in this effort is to identify the existence and understand
the origin of different components that contribute to the surface
density profiles of ellipticals, their cosmological scalings and
relevance for the formation history of such galaxies, and their
implications for global galaxy properties.

In this paper, we focus on the extra light in our simulations and in
known cuspy elliptical galaxies.  In \S~\ref{sec:sims} and
\S~\ref{sec:data} we describe our set of gas-rich merger simulations
and the observational data sets we consider, respectively.  In
\S~\ref{sec:fits}, we compare different approaches for fitting the
surface density profile, and attempt to calibrate various methods in
order to recover the physically distinct (dissipational versus
dissipationless) components in merger remnants.  In \S~\ref{sec:obs}
we compare our simulations with and apply our fitted galaxy
decomposition to a wide range of observed systems. In
\S~\ref{sec:properties} we use these comparisons to study how
structural parameters of the outer stellar light and inner extra light
component scale with galaxy properties. In \S~\ref{sec:structural.fx}
we examine how the existence and strength of the extra light component
is related to galaxy structure, global shape and rotation, and 
show that it
drives galaxies along the fundamental plane.  We investigate how this
extra light component influences and is related to stellar population
gradients in ellipticals in \S~\ref{sec:ssp.fx}. Finally,
in \S~\ref{sec:discuss} we discuss our results and outline future
explorations of these correlations.

Throughout, we adopt a $\Omega_{\rm M}=0.3$, $\Omega_{\Lambda}=0.7$,
$H_{0}=70\,{\rm km\,s^{-1}\,Mpc^{-1}}$ cosmology, and appropriately
normalize all observations and models shown, but this has
little effect on our conclusions.  We also adopt a
\citet{chabrier:imf} initial mass function (IMF), and convert all
stellar masses and mass-to-light ratios to this choice. The exact IMF
systematically shifts the normalization of stellar masses herein, but
does not substantially change our comparisons. All magnitudes are in
the Vega system, unless otherwise specified.

\section{The Simulations}
\label{sec:sims}

Our merger simulations were performed with the parallel TreeSPH code
{\small GADGET-2} \citep{springel:gadget}, based on a fully conservative
formulation \citep{springel:entropy} of smoothed particle hydrodynamics (SPH),
which conserves energy and entropy simultaneously even when smoothing
lengths evolve adaptively \citep[see e.g.,][]{hernquist:sph.cautions,oshea:sph.tests}. 
Our simulations account for radiative cooling, optional heating by
a UV background \citep[as in][although it 
is not important for the masses of interest here]{katz:treesph,dave:lyalpha}, and
incorporate a sub-resolution model of a multiphase interstellar medium
(ISM) to describe star formation and supernova feedback \citep{springel:multiphase}.
Feedback from supernovae is captured in this sub-resolution model
through an effective equation of state for star-forming gas, enabling
us to stably evolve disks with arbitrary gas fractions \citep[see, e.g.][]{springel:models,
springel:spiral.in.merger,robertson:disk.formation,robertson:msigma.evolution}. 
This is described by the parameter $\qeos$,
which ranges from $\qeos=0$ for an isothermal gas with effective
temperature of $10^4$ K, to $\qeos=1$ for our full multiphase model
with an effective temperature $\sim10^5$ K. We also compare with a subset of 
simulations which adopt the star formation and feedback prescriptions 
from \citet{mihos:cusps,mihos:starbursts.94,mihos:starbursts.96}, in which the ISM is treated as a
single-phase isothermal medium and feedback energy is deposited in a 
purely kinetic radial impulse (for details, 
see, e.g.\ \cite{mihos:method}).

Although we find that they make little difference to 
the extra light component, most of our simulations include 
supermassive black holes at the centers of both progenitor galaxies.  
The black holes are represented by ``sink'' particles
that accrete gas at a rate $\Mdot$ estimated from the local gas
density and sound speed using an Eddington-limited prescription based
on Bondi-Hoyle-Lyttleton accretion theory.  The bolometric luminosity
of the black hole is taken to be $L_{\rm bol}=\epsilon_{r}\dot{M}\,c^{2}$,
where $\epsilon_r=0.1$ is the radiative efficiency.  We assume that a
small fraction (typically $\approx 5\%$) of $L_{\rm bol}$ couples dynamically
to the surrounding gas, and that this feedback is injected into the
gas as thermal energy, weighted by the SPH smoothing kernel.  This
fraction is a free parameter, which we determine as in \citet{dimatteo:msigma}
by matching the observed $M_{\rm BH}-\sigma$ relation.  For now, we do
not resolve the small-scale dynamics of the gas in the immediate
vicinity of the black hole, but assume that the time-averaged
accretion rate can be estimated from the gas properties on the scale
of our spatial resolution (roughly $\approx 20$\,pc, in the best
cases). In any case, repeating our analysis for simulations with no black 
holes yields identical conclusions. 

The progenitor galaxy models are described in
\citet{springel:models}, and we review their properties here.  For each
simulation, we generate two stable, isolated disk galaxies, each with
an extended dark matter halo with a \citet{hernquist:profile} profile,
motivated by cosmological simulations \citep{nfw:profile,busha:halomass}, 
an exponential disk of gas and stars, and (optionally) a
bulge.  The galaxies have total masses $M_{\rm vir}=V_{\rm
vir}^{3}/(10GH[z])$ for an initial redshift $z$, with the baryonic disk having a mass
fraction $m_{\rm d}=0.041$, the bulge (when present) having $m_{\rm
b}=0.0136$, and the rest of the mass in dark matter.  The dark matter
halos are assigned a
concentration parameter scaled as in \citet{robertson:msigma.evolution} appropriately for the 
galaxy mass and redshift following \citet{bullock:concentrations}. We have also 
varied the concentration in a subset of simulations, and find it has little 
effect on our conclusions because the central regions of the 
galaxy are baryon-dominated. 
The disk scale-length is computed
based on an assumed spin parameter $\lambda=0.033$, chosen to be near
the mode in the $\lambda$ distribution measured in simulations \citep{vitvitska:spin},
and the scale-length of the bulge is set to $0.2$ times this. Modulo explicit 
variation in these parameters, these choices ensure that the initial disks 
are consistent with e.g.\ the observed baryonic 
Tully-Fisher relation and estimated halo-galaxy mass 
scaling laws \citep[][and references therein]{belldejong:tf,kormendyfreeman:scaling,
mandelbaum:mhalo}.

Typically, each galaxy initially consists of 168000 dark matter halo
particles, 8000 bulge particles (when present), 40000 gas and 40000
stellar disk particles, and one black hole
(BH) particle.  We vary the numerical
resolution, with many simulations using twice, and a subset up to 128
times, as many particles. We choose the initial seed
mass of the black hole either in accord with the observed $M_{\rm
BH}$-$\sigma$ relation or to be sufficiently small that its presence
will not have an immediate dynamical effect, but we have varied the seed
mass to identify any systematic dependencies.  Given the particle
numbers employed, the dark matter, gas, and star particles are all of
roughly equal mass, and central cusps in the dark matter and bulge
are reasonably well resolved. 
The typical gravitational 
softening in our simulations is $\sim20-50\,$pc in the 
$\lesssim L_{\ast}$ systems of particular interest here, 
with a somewhat higher $\sim50-100\,$pc in the most massive 
systems (yielding an effectively constant resolution $\sim 0.01\,R_{e}$
in terms of the effective radius). In \paperone\ and Appendix~\ref{sec:appendix:resolution} 
we demonstrate that this is sufficient to properly resolve not only the mass 
fractions but also the spatial extent of the extra light components of 
interest here (although resolution may become an issue when attempting to 
model the very smallest galaxies, with $R_{e}\lesssim100$\,pc and 
$L<0.01\,L_{\ast}$, as discussed in \S~\ref{sec:obs}). The hydrodynamic 
gas smoothing length in the peak starburst phases of interest is 
always smaller than this gravitational softening. 

We consider a series of several hundred simulations of colliding
galaxies, described in \citet{robertson:fp,robertson:msigma.evolution} and
\citet{cox:xray.gas,cox:kinematics}.  We vary the numerical resolution, the orbit of the
encounter (disk inclinations, pericenter separation), the masses and
structural properties of the merging galaxies, initial gas fractions,
halo concentrations, the parameters describing star formation and
feedback from supernovae and black hole growth, and initial black hole
masses. 

The progenitor galaxies have virial velocities $V_{\rm vir}=55, 80, 113, 160,
226, 320,$ and $500\,{\rm km\,s^{-1}}$, and redshifts $z=0, 2, 3, {\rm
and}\ 6$, and our simulations span a range in final spheroid stellar mass
$M_{\ast}\sim10^{8}-10^{13}\,M_{\sun}$, covering essentially the
entire range of the observations we consider at all redshifts, and
allowing us to identify any systematic dependencies in our models.  We
consider initial disk gas fractions by mass of $\fgas = 0.05,\ 0.1,\ 0.2,\ 0.4,\ 0.6,\ 
0.8,\ {\rm and}\ 1.0$ (defined as the fraction of disk baryonic mass which is gas) 
for several choices of virial velocities,
redshifts, and ISM equations of state. The results described in this
paper are based primarily on simulations of equal-mass mergers;
however, by examining a small set of simulations of unequal mass
mergers, we find that the behavior does not change dramatically for
mass ratios to about 3:1 or 4:1. The mass ratios we study are appropriate for the
observations of ellipticals used in this paper, which are only formed 
in our simulations in major merger events. At higher mass ratios, 
the result is a small bulge in a still disk-dominated galaxy 
\citep[see e.g.][]{younger:minor.mergers,hopkins:disk.survival,
hopkins:disk.heating},
which we do not study here. 

Each simulation is evolved until the merger is complete and the remnants are 
fully relaxed, typically $\sim1-2$\,Gyr after the final merger 
and coalescence of the BHs. We then analyze the 
remnants following \citet{cox:kinematics}, in a manner designed to mirror 
the methods typically used by observers. For each remnant, we project the 
stars onto a plane as if observed from a particular direction, and consider 
100 viewing angles to each remnant, which uniformly sample the unit sphere. 
Given the projected stellar mass distribution, we calculate the iso-density contours 
and fit ellipses 
to each (fitting major and minor 
axis radii and hence ellipticity at each iso-density contour), 
moving concentrically from $r=0$ until the entire stellar mass 
has been enclosed. This is designed to mimic observational isophotal fitting 
algorithms \citep[e.g.][]{bender:87.a4,bender:88.shapes}. The radial deviations 
of the iso-density contours from the fitted ellipses are 
expanded in a Fourier series in the standard fashion to determine 
the boxyness or diskyness of each contour (the $a_{4}$ parameter). 
Throughout, we show profiles and quote our results in 
terms of the major axis radius. For further details, we refer to \citet{cox:kinematics}.

We directly extract the effective radius $\re$ as the projected half-mass stellar 
effective radius, and the velocity dispersion $\sigma$ as the average 
one-dimensional velocity dispersion within a circular 
aperture of radius $\re$. This differs from what is sometimes adopted 
in the literature, where $\re$ is determined from the best-fitting
\Sersic\ profile, but because 
we are fitting \Sersic\ profiles to the observed systems we usually quote both the 
true effective radius of the galaxy and effective radii of the fitted \Sersic\ components. 
Throughout, the stellar mass $M_{\ast}$ refers to the total stellar mass of the galaxy, and 
the dynamical mass $\mdyn$ refers to the 
traditional dynamical mass estimator 
\begin{equation}
\mdyn\equiv k\,\frac{\sigma^{2}\,\re}{G},
\end{equation}
where we adopt $k=3.8$ (roughly what is 
expected for a \citet{hernquist:profile} profile, and the choice that most accurately 
matches the true enclosed stellar plus dark matter mass within $\re$ in our 
simulations; 
although this choice is irrelevant as long as we apply it 
uniformly to both observations and simulations). 
When we plot quantities such as $\re$, $\sigma$, and $\mdyn$, we 
typically show just 
the median value for each simulation across all $\sim100$ sightlines. The sightline-to-sightline 
variation in these quantities is typically smaller than the 
simulation-to-simulation scatter, but we explicitly note where it is large.

\section{The Data}
\label{sec:data}

We compare our simulations to and test our predictions on an ensemble
of observed surface brightness profiles of ellipticals.  Specifically,
we consider three samples of cusp or extra light ellipticals
and a compilation of remnants of recent gas-rich mergers.  The first
is the $V$-band Virgo elliptical survey of \citet{jk:profiles}, based
on the complete sample of Virgo galaxies down to extremely faint
systems $M_{B}\sim-15$ in \citet{binggeli:vcc} 
\citep[the same sample studied in][]{cote:virgo,ferrarese:profiles}. 
\citet{jk:profiles} combine observations from a
large number of sources
\citep[including][]{bender:data,bender:06,caon90,caon:profiles,davis:85,jedrzejewski:87,
jedrzejewski:87b,kormendy:05,
lauer:85,lauer:95,lauer:centers,liu:05,peletier:profiles} 
and new photometry from McDonald Observatory, the HST archive, and 
the SDSS 
for each of their objects which (after careful conversion to a single
photometric standard) enables accurate surface brightness measurements
over a wide dynamic range (with an estimated 
zero-point accuracy of $\pm0.04\,V\,{\rm mag\, arcsec^{-2}}$). 
Typically, the galaxies in this sample have
profiles spanning $\sim12-15$ magnitudes in surface brightness,
corresponding to a range of nearly four orders of magnitude in
physical radii from $\sim10\,$pc to $\sim100\,$kpc, permitting the
best simultaneous constraints on the shapes of both the outer and
inner profiles of any of the objects we study.  The profiles include
e.g.\ ellipticity, $a_{4}/a$, and $g-z$ colors as a function of
radius.  
Unfortunately, since this is restricted to Virgo ellipticals,
the number of galaxies is limited, especially at the intermediate and high end of the
mass function.

We therefore add surface brightness profiles from \citet{lauer:bimodal.profiles}, 
further supplemented by \citet{bender:data}. 
\citet{lauer:bimodal.profiles} compile $V$-band measurements of a
large number of nearby systems for which HST imaging of the galactic
nuclei is available.  These include the 
\citet{lauer:centers} WFPC2 data-set, the \citet{laine:03} WFPC2 BCG
sample (in which the objects are specifically selected as brightest
cluster galaxies from \citet{postmanlauer:95}), and the \citet{lauer:95}
and \citet{faber:ell.centers} WFPC1 compilations 
\citep[see also][]{quillen:00,rest:01,ravindranath:01}. 
Details
of the treatment of the profiles and conversion to a single standard
are given in \citet{lauer:bimodal.profiles}. 
The sample includes ellipticals over
a wide range of luminosities, down to $M_{B}\sim-15$, but is dominated
by intermediate and giant ellipticals, with typical magnitudes $M_{B}
\lesssim -18$. This therefore greatly extends our sampling of the
intermediate and 
high-mass end of the mass function, but at the cost of some dynamic
range in the data. The HST images alone,
while providing information on the central regions, typically extend
to only $\sim1$\,kpc outer radii, which is insufficient to fit the
outer profile. \citet{lauer:bimodal.profiles} 
therefore combine these data with ground-based
measurements from a number of sources (see the references for the
\citet{jk:profiles} sample) to construct profiles that typically span
physical radii from $\sim10\,$pc to $\sim10-20$\,kpc. Although the 
composite profiles 
were used in \citet{lauer:bimodal.profiles} to estimate effective radii, they were not 
actually shown in the paper. 
It should also be noted that there is
no single criterion that characterizes galaxies included in this
sample, but they generally
comprise luminous nearby ellipticals and S0 galaxies for which
detailed imaging is available.  We emphasize that issues of completeness
and e.g.\ environment are not important for any of our conclusions.

We occasionally 
supplement the profiles from \citet{lauer:bimodal.profiles} with additional 
profiles used in \citet{bender:data,bender:ell.kinematics,bender:ell.kinematics.a4,
bender:velocity.structure}, and in some cases subsequently updated. 
These are more limited: typically the profiles cover $\sim7$ magnitudes in
surface brightness, extending from $\sim30-50\,$pc out to $\sim$ a few
kpc (typically $\sim3$\,kpc in low-luminosity systems, and $\sim
15$\,kpc in the brightest systems, sufficient for acceptable, but not
strong constraints on the outer profile shapes).
However, the measurements are usually
in each of the $V$, $R$, and $I$ bands, and hence allow us to
construct multicolor surface brightness, ellipticity, and $a_{4}/a$
profiles. We use this to estimate e.g.\ the 
sensitivity of the fitted parameters 
and galaxy profiles on the observed waveband and on the 
quality and dynamic range of the photometry.

In various places, we compare our results from the study of these ellipticals 
to our results in \paperone\ from a study of local remnants of gas-rich 
merger remnants \citep{rj:profiles}. For these objects, \citet{rj:profiles} compile
$K$-band imaging, surface brightness, ellipticity, and $a_{4}/a$
profiles, where the profiles typically range from $\sim100\,$pc
to $\sim10-20$\,kpc. These span a moderate range in luminosity
(including objects from $M_{K}\sim-20$ to $M_{K}\sim-27$, but with
most from $M_{K}\sim-24$ to $M_{K}\sim-26$) and a wide range in merger
stage, from ULIRGs and (a few) unrelaxed systems to shell
ellipticals. As demonstrated in \citet{rj:profiles} and 
argued in \paperone, these systems will almost all
become (or already are, depending on the classification scheme used)
typical $\sim \lstar$ ellipticals, with appropriate phase space
densities, surface brightness profiles, fundamental plane relations,
kinematics, and other properties. For a detailed discussion of the modeling 
of these systems and the profiles themselves, we refer to \paperone\ 
(all of the results shown for these systems are derived therein). We 
show the results from \paperone\ here in order to test the continuity of 
merger remnant and (cusp) elliptical populations. 

Because we are here specifically interested in extra light or cusps in
observed ellipticals, and because the generally accepted belief is
that core ellipticals are not directly formed in 
gas-rich major mergers but are
subsequently modified by dry re-mergers
\citep[see e.g.][]{faber:ell.centers,
vandokkum:dry.mergers,bell:dry.mergers}, we restrict our attention
only to those ellipticals which are confirmed via HST observations as
being cusp ellipticals. 
We include all the
confirmed gas-rich merger remnants, but note there are a small number
of extreme unrelaxed cases for which sharp features in the surface
brightness profiles prevented derivation of meaningful quantities
(note, however, as shown in \paperone, that almost all of the objects
in this sample are sufficiently well-relaxed at the radii of interest
for our fitting). We exclude dwarf spheroidals, as they are not
believed to form in major mergers as are ellipticals \citep[e.g.][]{kormendy:spheroidal1,
kormendy:spheroidal2,jk:profiles}, 
and in any case they dominate at
extremely low masses where our simulations do not sample the
population (they also predominate as satellite galaxies, whose effects
we do not model).

We also exclude S0 galaxies (adopting the morphological 
classifications from \citet{jk:profiles} and \citet{lauer:bimodal.profiles}, 
although it makes little difference exactly which classifications we 
consider). This is not because of a physical
distinction: observations suggest that these likely form a continuous
family with the low-luminosity cusp ellipticals, and in fact a number
of our simulated gas-rich merger remnants would, from certain viewing
angles, be classified as S0s. However, in order to derive e.g.\ the
parameters of the outer, violently relaxed profile and central extra
light, it would be necessary to remove the contribution of the
large-scale disk from the surface brightness profiles of these
objects.  Our two-component (outer dissipationless and inner
dissipational) \Sersic\ models (described in \S~\ref{sec:fits}) then
become three-component fits, and the degeneracies involved with three
independent components, even with our best data and simulations, are
so large as to render the results meaningless. We have, however,
re-visited all of the S0s in these samples in light of our results,
and find that they are, in all cases, consistent with our predicted
and observed trends.  However, it is too difficult to infer these
trends directly from the S0s themselves without ideal disk
subtraction.

This yields a final sample of $\approx 80$ unique elliptical
galaxies, and $\approx 50$ confirmed remnants of gas-rich mergers. Most of the
sample spans a range of three orders of magnitude in stellar mass,
from $\lesssim0.1\,\mstar$ to $\sim10\,\mstar$, and a wide range in
extra light properties.  There is, of course, some overlap in the 
samples that define our compilation; we have
$\sim300$ surface brightness profiles for our collection of $\approx80$ unique
ellipticals, including (for many objects) repeated measurements in 
multiple bands and with various instruments. 
This turns out to be quite useful, as
it provides a means to quantify error estimates in fits to these
profiles. The variations between fit parameters 
derived from observations 
in different bands or made using different
instruments are usually much larger than the formal statistical errors in the
fits to a single profile. There are no obvious systematic effects
(i.e.\ systematic changes in profile fits from $V$ to $I$ band), but as
demonstrated in \paperone\ the effects of
using different bands or changing
dynamic range (from different instruments) can be complex, depending
on the structure and degree of relaxation of the outer regions of a
system. On the other hand, there are well-relaxed objects for which
almost no significant change in the fits occurs from band to band.  
It is therefore useful to have multiple observations of the same system, 
as it allows us to get some idea of how sensitive our fits are to 
differences in e.g.\ the choice of observed wavelength or dynamic 
range from instrument to instrument. 

In Table~\ref{tbl:cusp.fits}, we list the names and properties of our 
sample ellipticals, including the relevant sources of 
photometry, stellar masses, absolute magnitudes, stellar 
velocity dispersions and effective radii, ellipticities, isophotal shapes, 
and rotation properties. 
We have converted all the observations to 
physical units given our adopted cosmology, 
and compile global parameters (where not available in the original papers) 
including e.g.\ kinematic properties, luminosities, and black hole masses
from the literature. 
We determine stellar masses ourselves in a uniform manner for 
all the objects, based on their total $K$-band luminosities and 
$(B-V)$ color-dependent mass-to-light ratios from \citet{bell:mfs}, 
corrected for our adopted IMF. We have 
also repeated our analysis using stellar masses derived from a 
mean $M/L$ as a function of luminosity or from fitting 
the integrated $UBVRIJHK$ photometry of each object to a 
single stellar population with the models of \citet{BC03}, and 
find this makes no difference to our conclusions. 

Throughout, we will usually refer interchangeably to the observed surface
brightness profiles in the given bands and the surface stellar mass
density profile. Of course, stellar light is not exactly the same as
stellar mass, but in \paperone\ and \S~\ref{sec:ssp.fx} herein, 
we consider the differences between
the stellar light and the stellar mass density profiles as a function
of time, wavelength, and properties of the merger remnant, and show
that the $V$ and $K$-band results introduce little bias (i.e.\ are good tracers of
the stellar mass); the \Sersic\ indices and extra light fractions fitted
to the $K$-band profiles of the simulations are good proxies for the
\Sersic\ index of the stellar mass profile and extra mass/starburst mass
fraction, even close in time to the peak episode of star formation.

Although we are not concerned about the
absolute normalization of the profile (i.e.\ mean $M/L$), since we
derive total stellar masses separately from the integrated photometry,
we must account for systematics that might be induced by a change in
$M/L$ as a function of radius. 
The results from our merger remnant 
sample (observed in $K$-band) are, on average, robust in 
this sense, but they 
should be treated with care, especially in the most 
extreme cases (namely the few LIRGs and ULIRGs in the sample), 
where younger stellar populations may decrease $M/L$ towards their center 
(see \S~\ref{sec:ssp.fx}). 
We emphasize though that many of these systems are much 
older and more relaxed (e.g.\ ellipticals 
with faint shells or tidal debris). 
The profiles in optical bands such as
$V$ require more care -- when the system is very young ($\lesssim
1-2$\,Gyr after the major merger-induced peak of star formation),
there can be considerable bias or uncertainty owing to stellar
population gradients and dust. However, once the system is relaxed,
the optical bands also become good proxies for the stellar mass
distribution.

In fact, in \paperone\ and \S~\ref{sec:ssp.fx} we demonstrate that once the system reaches
intermediate age, the bias in e.g.\ $B$ or $V$ band is often less than
that in $K$ band, because systems tend to be both younger and more
metal rich in their centers.  In $K$-band,
these both increase $L/M$, leading to a (small) systematic bias. In
optical bands, however, the two have opposite effects (younger age
increases $L/M$, but higher metallicity decreases $L/M$), and they
tend to mostly cancel.  
Since essentially all of our ellipticals are older than this stellar population
age (even in their centers), and they have been carefully vetted and
either corrected for
the effects of e.g.\ dust lanes in the sources 
or (where correction was too difficult) excluded from our samples 
\citep[see][]{jk:profiles,lauer:bimodal.profiles}, we are not concerned that
significant bias might persist. Furthermore, comparison of systems
observed in different bands demonstrates that our conclusions are
unchanged (modulo small systematic offsets) regardless of the observed
bands in which we analyze these systems. As has been noted 
in other works,
most of these objects have weak color gradients, indicating little
variation in $M/L$ with radius.

\section{Recovering the Physically Appropriate ``Extra Light''}
\label{sec:fits}

\begin{figure*}
    \centering
    \plotone{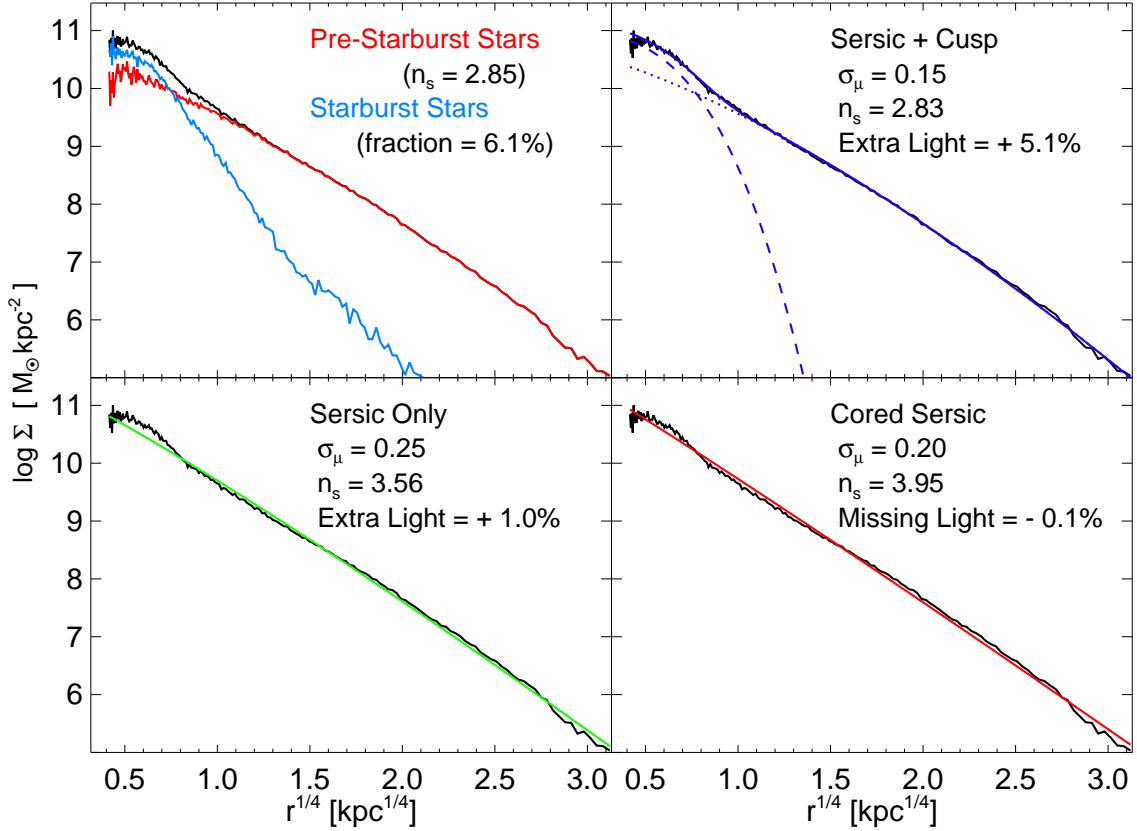}
    \caption{{\em Upper Left:} Surface mass density of a typical merger 
    remnant from our simulation library (black), 
    decomposed into stars formed prior to the final merger (which 
    are then violently relaxed; red) and stars formed in the 
    dissipational starburst (blue). The \Sersic\ index fitted to the pre-starburst component 
    alone is shown, with the stellar mass fraction of the 
    starburst component. {\em Upper Right:} Two-component (\Sersic\ plus 
    cusp or extra light) fit (inner exponential 
    and outer \Sersic) to the total light profile, with the \Sersic\ index of the outer component 
    and mass fraction of the inner component, and rms scatter ($\dmu$, in ${\rm mag\,arcsec^{-2}}$) 
    about the 
    fit. {\em Lower Left:} Single \Sersic\ function fit to the profile. {\em Lower Right:} 
    Core-\Sersic\ function fit. Our two-component, 
    cusp plus \Sersic\ function fit ({\em top right}) accurately 
    recovers the profile of the violently relaxed component and mass fraction 
    of the starburst component. The other fits give 
    less intuitive results in this case.
    \label{fig:demo.fit.danger}}
\end{figure*}

We would like to use the surface brightness profiles of merger
remnants to estimate the contribution from extra light and, in
particular, to infer the fraction of stellar mass that was formed in a
compact central starburst. However, as noted in \paperone, the light
profiles in our simulation remnants are quite smooth, even where the
extra light fraction is large. This makes recovering the extra light
component a non-trivial procedure, which can be sensitive to the
assumptions made in fitting \citep[see also][]{naab:profiles}. 
We discuss various procedures and their
consequences in \paperone, but briefly review them to highlight the
most important decomposition we will adopt.

Figure~\ref{fig:demo.fit.danger} shows the surface density profile of
a typical merger remnant from our simulation library, with a gas fraction 
of $\sim10\%$, which 
happens to provide a good match to
several observed ellipticals. We begin by reducing the profile to the
two most physically relevant components: the ``pre-starburst'' or
``disk'' stars, i.e.\ those formed in the rotationally supported disks
before the final coalescence of the galaxies, and the ``starburst''
stars, produced in the final, compact starburst. We operationally 
define this as stars formed within $\pm125$\,Myr of the peak in the 
starburst star formation, but since the starburst is usually very distinct, 
changing this definition within reason makes almost no difference. 
We combine both the stars present in the stellar disks at the beginning of our
calculations and the stars formed in the disks over the course of the
simulations (the disk stars and pre-starburst stars in
Figure~\ref{fig:demo.fit.danger} of \paperone) into a single
pre-starburst stellar population, because there is no robust physical
distinction between the two. Not only do both populations experience
violent relaxation in the final merger/coalescence, but it is also
arbitrary where (or at what time) we initialize our simulations,
relative to the final merger, and therefore what fraction of the
stellar mass forms before our simulation begins. 

We also neglect
embedded stellar disks which can be formed by gas remaining after the
merger; as demonstrated in \paperone, these contribute negligibly to
the surface mass density profile in even the most gas-rich merger
remnants (and we are primarily concerned with the surface mass density
profiles, not the early-time optical/UV light profiles which more strongly
reflect the light from new stars). This does not mean such disks 
are unimportant -- indeed they are ubiquitous in gas-rich merger 
remnant simulations and contribute critically to the 
kinematics \citep[especially the rotation and $a_{4}$, see][]{cox:kinematics}, 
and observations suggest that low-level disks may be present in 
nearly all cusp galaxies \citep[perhaps all, given 
projection effects; see][]{ferrarese:type12,lauer:centers}. However, 
especially given our exclusion of S0 galaxies, these are not a significant 
component of the surface brightness profiles. 

We showed in \paperone\ that the total profile of the system can be 
robustly represented as a two-component sum of two \Sersic\ 
distributions. 
If we fit each known physical component separately, 
we find that the pre-starburst component follows 
a nearly exact \Sersic\ law (here, $n_{s}=2.85$), while the inner component 
(given its shape and origin in a gas-rich, dissipational event) 
can be well fit by a lower-$n_{s}\sim1$ law.\footnote{Formally, we consider our simulation profiles outside of 
some multiple $\sim3-5$ times the resolution limit, or with a seeing correction 
appropriate for the comparison observed samples. We equally sample 
the profile in $\log{r}$ over a dynamic range extending to the largest radii in the 
observed samples (\S~\ref{sec:data}), 
and weight each point equally assuming an intrinsic $\sim0.1\,$mag point-to-point variance 
in the SB profile (the typical magnitude of residuals fitting arbitrary splines to the data). 
We have varied these choices and find that our fits and conclusions are 
not sensitive to them.} 
Based on this behavior, we therefore adopt a two-component decomposition of the 
observed quantity, the total surface brightness profile. 
This is defined as a \Sersic\ plus cusp or extra light model, with an outer
component for the pre-starburst stars with a free \Sersic\ index, 
and an inner component reflecting the starburst stars 
(with fixed $n_{s}=1$ when the shape of the starburst is not 
well-resolved, or free -- albeit still generally low -- $n_{s}$ when 
resolution permits).
We have studied this decomposition in \paperone, and
show that it provides a good description of both simulated and
observed merger remnant profiles, accurately separating the central
light of younger stellar populations (as observationally
determined) from the light of older stars, which form a more extended
distribution.  The total surface brightness profile is then
\begin{eqnarray} 
\nonumber I_{\rm tot} &=&
I^{\prime}\,\exp{{\Bigl\{}-b_{n}^{\prime}\,{\Bigl(}\frac{r}{R_{\rm extra}}{\Bigr)}^{1/n_{s}^{\prime}}{\Bigr\}}}\\
& &+I_{\rm o}\,\exp{{\Bigl\{}-b_{n}\,{\Bigl(}\frac{r}{R_{\rm outer}}{\Bigr)}^{1/n_{s}}{\Bigr\}}},
\label{eqn:fitfun}
\end{eqnarray} 
where $R_{\rm extra}$ and $R_{\rm outer}$ are the effective radii of the 
inner ($n_{s}^{\prime}\sim1$) and outer (free $n_{s}$) components 
(which we identify with the starburst and
old bulge or pre-starburst components, respectively), $I^{\prime}$ and
$I_{\rm o}$ are the corresponding normalizations, 
$n_{s}^{\prime}$ is the \Sersic\ index of the inner (extra light) component 
(fixed $n_{s}^{\prime}=1$ where resolution limits apply)
and $n_{s}$ is the \Sersic\
index of the outer bulge or pre-starburst component.  The constant
$b_{n}$ is the appropriate function of $n_{s}$ such that $R_{\rm extra}$ and
$R_{\rm outer}$ correspond to to the projected half-mass radii. 

The upper right panel of 
Figure~\ref{fig:demo.fit.danger} shows an example of the outcome of
such a fit.  The resulting model of the surface density profile fits
the simulation well, with a rms deviation ($\dmu\equiv\langle \Delta\mu^{2} \rangle^{1/2}$)
of only $\dmu\sim0.15$\,mag
(assuming $\mu\propto-2.5\,\log{I_{\rm tot}}$).  This is comparable to
the point-to-point variance in the profile of this simulation if we
fit an arbitrary spline to the data, and thus reflects a genuinely
good fit.  More important, this fit, despite having no direct
information about the physical components into which we decompose the
brightness profile, recovers almost exactly the appropriate parameters
for both components. The best-fit \Sersic\ index ($n_{s}=2.83$ compared
to $2.85$) and effective radius of the outer or bulge component reproduce
well those from fitting directly to the pre-starburst stellar
population. Likewise, the inner or extra light component is a good
match to the starburst component, and the fit recovers the extra light
fraction accurately ($5.1\%$ compared to $6.1\%$; a smaller difference
than reasonable uncertainties in our physical definition of the
starburst component). Again, we emphasize that we have simply fit a function 
of the form in Equation~(\ref{eqn:fitfun}) to the total surface brightness profile, 
ignoring our knowledge of the genuine physical breakdown; but we find 
that we recover an accurate reflection of that true decomposition.

\begin{figure}
    \centering
    \scaleup
    \plotter{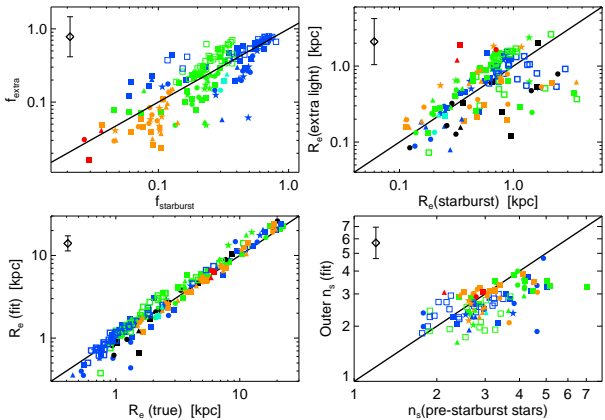}
    \caption{Success of our proposed two-component empirical decomposition 
    at recovering the known physical parameters of the galaxy starburst and pre-starburst 
    (violently relaxed) components. 
    {\em Top Left:} Mass fraction in the fitted ``extra light'' component $f_{\rm extra}$ 
    versus the known mass fraction of the physical starburst $f_{\rm sb}$. 
    Each point is the average across $\sim100$ sightlines to a given simulation, 
    although the sightline-to-sightline variance is moderate ($\approx0.15\,$dex). 
    Different colors and symbols denote different initial disk gas fractions, orbital 
    parameters, and merger redshifts (see key in Figure~\ref{fig:ns.mass}). 
    The fitted $f_{\rm extra}$ recovers the physical $f_{\rm sb}$ on average, with a 
    factor $\sim2$ scatter (plotted error bar), and without any 
    significant bias from any varied simulation parameters. 
    {\em Top Right:} Fitted effective radius of the 
    extra light component versus the projected half-mass radius of the known starburst stars. 
    Again, the true values are recovered with a factor $\sim2$ scatter, independent 
    of simulation choices. 
    {\em Bottom Left:} Fitted effective radius of the entire galaxy 
    versus $R_{e}$ from direct profile integration. The fits recover $R_{e}$ to 
    better than $\sim0.1$\,dex. 
    {\em Bottom Right:} Fitted \Sersic\ index of the outer (dissipationless) component 
    versus that fitted directly to the known pre-starburst component profile. 
    The outer profile shape $n_{s}$ is recovered to within $\sim0.1$\,dex 
    (the only significant bias is when there is a large disk in the remnant; but 
    those cases are not relevant for this paper). 
    The empirical decompositions do well at recovering the known 
    parameters in the simulations, across the entire range of simulation 
    parameter space we have surveyed. 
    \label{fig:check.fe.fsb}}
\end{figure}

Figure~\ref{fig:check.fe.fsb} shows the results of repeating 
this procedure for
several hundred simulations; we directly compare 
the fitted extra light mass fraction and size ($R_{\rm extra}$) 
to the mass fraction and size of the known
physical starburst component, and 
find that the fitted components recover the physical values in the mean with a 
factor $\sim2$ scatter. This result is robust with respect to e.g.\ the mass, 
orbital parameters, mass ratios, initial gas content, treatment of feedback 
and model for the ISM equation of state, and redshift of our simulations. 
We similarly compare the effective radius of the entire galaxy determined 
from these fits to that known from the direct integration of the profile. 
The fits recover $R_{e}$ to within $\sim20-25\%$, given a dynamic range 
comparable to the observations with which we compare. We also 
compare the \Sersic\ index of the fitted outer component 
to that fitted directly to the known physical pre-starburst component 
and find the two agree within $\sim30-40\%$, without any systematic dependences 
except where the remnants have large embedded disks (biasing the 
fitted decompositions to lower $n_{s}$). This gives us confidence that such an 
empirical approach can be used, in a statistical sense, to recover 
physically meaningful parameters describing the galaxy components. 

We note that, so long as we are fitting radii $\gtrsim50\,$pc, we obtain 
similar results for a free inner component shape parameter $n_{s}^{\prime}$ 
or fixed inner $n_{s}^{\prime}=1$. In \paperone\ we demonstrate that
the choice of $n_{s}^{\prime}=1$ for the inner component recovers, on average,
the correct physical mass fraction which participated in the
starburst, and mitigates against degeneracies in fitting to the outer
\Sersic\ profile (so that we recover the same value if we fit the two
stellar populations independently). We therefore adopt this choice for
our decompositions in situations where we cannot reliably resolve the 
innermost extra light shape/structure (namely our simulations 
and the observed samples of \citet{rj:profiles} and \citet{bender:data}).
We emphasize, however, that this choice for the inner
component does {\em not} imply that this reflects the true shape of
the central extra light, which can be complex
\citep[see e.g.\ the range of central 
profile shapes in][]{lauer:centers,cote:virgo,jk:profiles}. 
In fact, caution should be taken when
considering the central $\sim30-50\,$pc, which our simulations do not
generally resolve (for most of the galaxy observations 
considered here, this corresponds to $\sim0.5''$, a factor 
$\sim10$ larger than the HST diffraction limit). 
We include a detailed discussion of profile shapes as $r\rightarrow0$ 
in Appendix~\ref{sec:appendix:resolution}. 

For now, we simply note 
that the quantities of interest here 
are well-converged in resolution studies, and our 
numerical tests (smoothing extremely high-resolution simulations 
over various seeing) find that $n_{s}^{\prime}=1$ choice 
is robust (in the sense that the mean properties are recovered 
similarly in either case). 
However, where the information is 
available in the central regions of the galaxy (namely the 
samples of \citet{jk:profiles} and \citet{lauer:bimodal.profiles}), we find 
the best results using all of the information and freeing the \Sersic\ 
index of the inner light component, to accommodate the real, observed 
shapes and structure in the inner light component. 
Despite these caveats, our treatment describes the starburst mass
profile well where it is important to the overall surface 
density (from $\sim100\,$pc to $\sim1\,$kpc) and 
accurately recovers the total mass in the starburst component and its
effective radius.

We apply this formalism to the observed systems 
in \S~\ref{sec:obs} below\footnote{We fit the observed 
points in the same manner as the simulations, weighting 
each point with the (quadratic) sum of the same intrinsic 
$\sim0.1\,$mag point-to-point variance and the observational 
errors. These errors are however generally small (much smaller 
than the plotted points shown in the observed profiles), so 
weighting the data points equally gives almost identical results.}, 
and we present the results of our 
fits to each elliptical in our sample in Table~\ref{tbl:cusp.fits}. 
For sources with 
multiple independent observations, we define error bars for each 
fit parameter representing 
the $\sim1\,\sigma$ range in 
parameters derived from various observations, typically from three
different surface brightness profiles but in some cases from as many
as $\approx 5-6$ sources (where there are just $2$ sources, the ``error'' 
is simply the range between the two fits). 
In many cases the different observations are comparable; 
in some there are clearly measurements 
with larger dynamic range and better resolution: the errors derived 
in this manner should in such cases be thought of as the 
typical uncertainties introduced by lower dynamic range or less 
accurate photometry. 

In terms of direct comparison with our
simulations, the data often cover a dynamic range and have resolution
comparable to our simulations, provided we do not heavily weight the
very central ($\lesssim30\,$pc) regions of HST nuclear profiles. 
Experimenting with different smoothings and imposed dynamic range
limits, we find it is unlikely that resolution or seeing differences
will substantially bias our comparisons. They do introduce
larger scatter; the robustness of our results increases considerably 
as the dynamic range of the observed profiles is increased.

As we demonstrate in \paperone, care should be taken to adopt and test 
physically motivated interpretations of different functional 
forms that could be considered when fitting these profiles. 
For example,
the lower left panel of
Figure~\ref{fig:demo.fit.danger} shows the results of fitting a
pure \Sersic\ function to the entire surface density profile (including 
the central starburst component).  There is
a reasonable fit to the entire profile with a single \Sersic\ index
$n_{s}=3.56$, quite different from the \Sersic\ index which describes
either the pre-starburst or starburst light components.  Likewise, if
we consider the excess light to be that light in the real profile
above the prediction of the best-fit \Sersic\ model, we would infer only
a tiny extra light fraction $\sim1.0\%$.  Although the fit is
technically worse, with variance $\dmu=0.25$, the difference is
not dramatic, and by many observational standards would be considered
a good fit.  Clearly, however, the results do not have the same physical
meaning in this case -- the ``extra light'' determined in this manner is 
no longer a direct tracer of the physical starburst component.  
The differences grow if we add a
degree of freedom and fit a ``core-\Sersic'' profile, of the form
$I \propto [1 + (r_{b}/r)^{\alpha}]^{\gamma/\alpha}\,
\exp{\{ -b_{n}\, [(r^{\alpha}+r_{b}^{\alpha})/r_{e}^{\alpha}]^{1/\alpha n} \}}$
\citep[e.g.][]{graham:core.sersic}, which behaves as a single \Sersic\ profile outside of 
$r_{b}$ and breaks to a power-law of slope $\gamma$ within $r_{b}$. 
The lower right panel of
Figure~\ref{fig:demo.fit.danger} shows the
outcome of this fit, which is again good in a pure statistical sense,
albeit worse than our best fit \Sersic+extra light fit
($\dmu=0.20$\,mag). However, here the derived parameters are even
less intuitively related to the known physical 
decomposition -- the best fit \Sersic\ index is a much steeper
$n_{s}=3.95$ and one actually infers that the system is a {\em core}
galaxy, with {\em missing} light relative to the best-fit \Sersic\
profile. Furthermore, by comparison with our results in \paperone, we
find that the discrepancy between the physical parameters which
accurately describe the outer and starburst components and those
recovered by the pure \Sersic\ or core-\Sersic\ profile fits becomes even
worse when the mass fraction of starburst component is larger.

The cause of these differences is that the extra light
component blends smoothly with the outer pre-starburst light profile.
By increasing the central surface brightness, the extra light
component makes the overall profile appear steeper (concave up in the
$\mu-r^{1/4}$ projection), owing to the rise at small $r$.  However,
the cusp itself does not continue to rise steeply inwards (in most
cases), so after steepening the best-fit \Sersic\ index to fit the outer
part of the extra light component, one is often forced to infer the
existence of a core in the central regions.  Again, these fits do not directly 
reflect the physical two-component nature of the profiles (rather reflecting 
some combination of the components, with the extra light no longer 
apparent in an excess with respect to the fit but in the higher fitted \Sersic\ indices), 
but they are not terrible matches to the
light profile.  {\em This emphasizes that a physically motivated profile
must be adopted when fitting a parameterized model to the data, if one wishes 
to translate these parameters into robust physical properties.}
Fortunately, there are some indications from observations that the
\Sersic\ only and core-\Sersic\ fits are not physically motivated as a means 
to decompose the two-component nature of the observed systems. 

First, they are technically worse fits, although the difference is not
large (and in some rare cases, core-\Sersic\ profile is
a better match to our simulations than the \Sersic+extra light
profile). With photometry accurate to 
$\sim0.01\,{\rm mag\,arcsec^{-2}}$ it is possible to 
robustly distinguish the quality of the fits shown in Figure~\ref{fig:demo.fit.danger}, 
especially to note the fact that the errors in the \Sersic\ and core-\Sersic\ 
models are a strong function of radius (an indication of 
the less appropriate choice of fitting function), but with the exception of \citet{jk:profiles}, 
our data sets do not attain such high accuracy. 
Second, they begin to fail at large radii -- however, this
is where the true nature of the \Sersic\ profile of the outer light
component is most prominent, so any failure at large $r$ should be
especially worrisome.  Furthermore, when we examine the kinematics
(e.g.\ ellipticity, boxy/diskyness, rotation properties) along the
major axis, one can often see a transition in these properties where
the extra light begins to dominate (see \S~\ref{sec:obs}), whereas
we would expect no such change if the ability to fit a continuous \Sersic\ profile were 
taken to imply that there is only a single physical component 
constituting the galaxy. 
Finally, when fitting a core-\Sersic\ profile to a system with significant 
extra light, a large {\it missing} light fraction (defined as the difference 
between the core-\Sersic\ fit and the inwards extrapolation of the 
\Sersic\ portion of the fit) is sometimes
seen, even relative to what is typically observed in genuine core
galaxies \citep[e.g.\ massive, boxy, slow-rotating ellipticals;
see][]{jk:profiles}.  This is true when the dissipational component is
large -- but Figure~\ref{fig:demo.fit.danger} demonstrates that it is
not always the case, so again, care must be taken to employ a
physically well-motivated decomposition and interpretation.

We do not claim that a pure \Sersic\ or core-\Sersic\ profile
is never a physically motivated parameterization of the galaxy light profile.
However, for gas-rich merger remnants, we know from our simulations
and have good reason to believe observationally that there is some
excess light component. In these cases, which we investigate here, the
results of these fits are demonstrably less physically intuitive and can be
misleading.

We also emphasize that although there are some superficial similarities 
between our adopted parametric profile decomposition and 
that in e.g.\ \citet{cote:virgo} and \citet{ferrarese:profiles}, 
the two are in detail significantly different and address 
very different spatial scales and physical properties of the galaxies. 
Typically, the ``outer profile'' we refer to extends to and beyond (in 
our simulations) the limits of our ground-based photometry, corresponding 
to physical radii of $\sim20-100\,$kpc, and our ``inner profile'' refers to 
the residual from a central starburst at scales where a significant 
fraction of the galaxy mass becomes self-gravitating (see \S~\ref{sec:properties}), 
at $\sim0.5-1\,$kpc. We stress again that we are not resolving 
inwards of the central $\sim30-50\,$pc, and our modeling should 
not be extrapolated to within these radii without considerable care. 
In contrast, the ``outer profile'' in \citet{ferrarese:profiles} is 
based on the HST ACS profiles, which extent to outer 
radii $\sim1$\,kpc, and their ``inner profiles'' typically dominate 
the light profile at very small radii $\sim0.01-0.02\,R_{e}$ ($\sim10-40\,$pc 
for most of their sample). 
This is more akin to separating our 
``inner'' component itself into multiple sub-components -- i.e.\ a starburst 
stellar component that blends (as we have shown) relatively smoothly 
onto the outer, violently relaxed stars and an innermost nuclear 
component. The authors themselves address this, and denote these 
nuclear excesses as central stellar clusters. In Appendix~\ref{sec:appendix:nuclei} 
we demonstrate that such clusters are clearly distinct, orders-of-magnitude smaller 
objects than the starburst/extra light components we study in this paper. 
Such systems may indeed be present (and could be formed 
in the same dissipational starburst which we model): but if so they are distinct subsystems 
sitting on top of the starburst light component, 
which we do not have the ability to model or resolve in our simulations.
Therefore, while the two approaches may yield complementary constraints, 
we caution that our 
results are not directly comparable and are specifically designed 
to trace distinct physical structures.

\section{Comparison with Observations: Extra Light in Ellipticals}
\label{sec:obs}

\begin{figure*}
    \centering
    \plotone{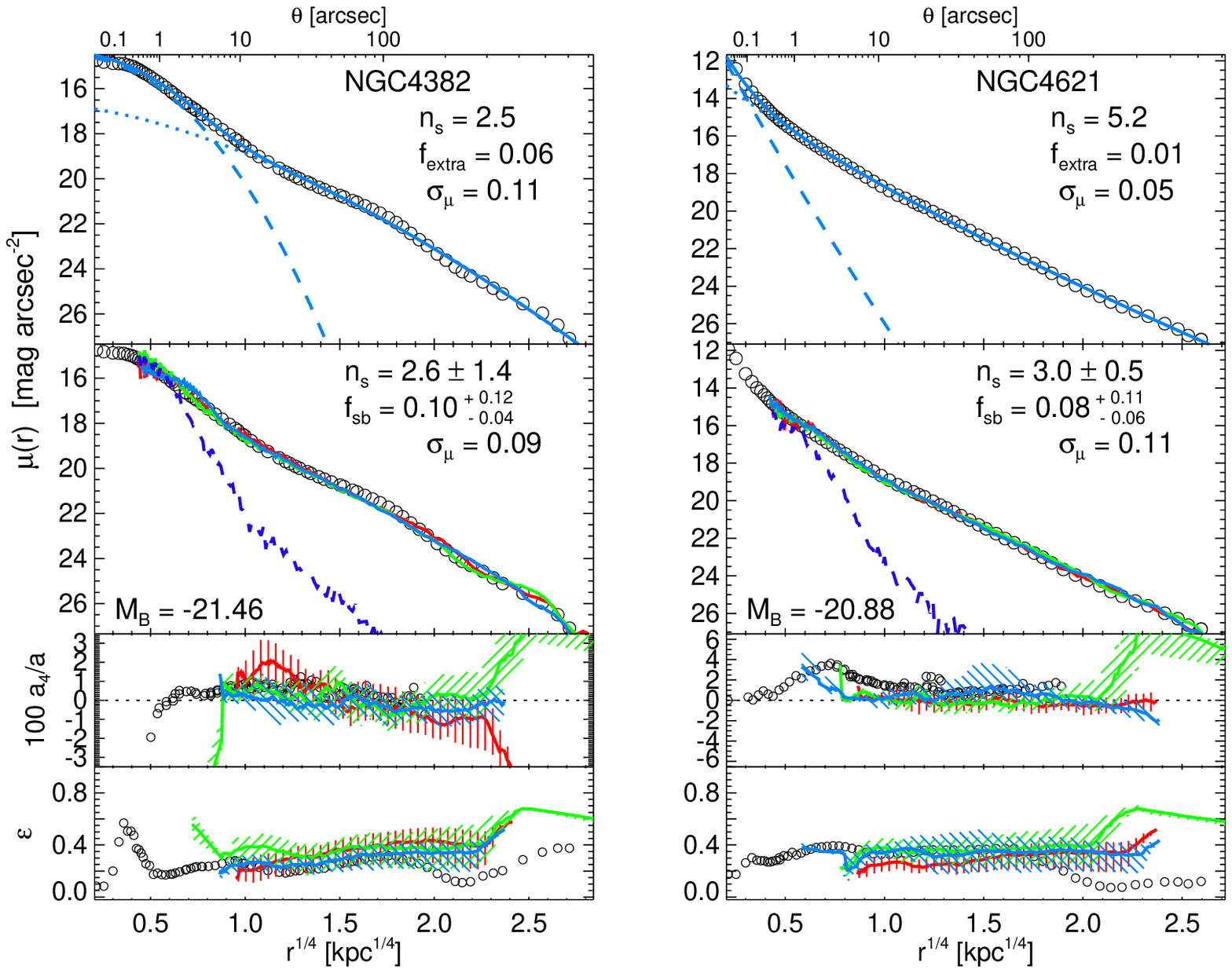}
    \caption{Surface brightness profiles are shown for cuspy ellipticals in the 
    Virgo cluster.
    Open circles show the observations, from \citet{jk:profiles}. 
    These are the highest-mass cusp or extra light ellipticals in Virgo$^{\ref{foot:4382}}$
    ($\sim2\,\mstar$).
    {\em Top:} Observed V-band surface brightness profile with our 
    two component best-fit model (solid, dashed, and dotted lines show the 
    total, inner/extra light component, and outer/pre-starburst component). 
    The best-fit outer \Sersic\ index, extra light fraction, and variance about the 
    fit are shown.
    {\em  Middle:} Colored lines show the corresponding surface brightness 
    profiles from the three simulations in our library which correspond 
    most closely to the observed system (shown outside to the gravitational 
    softening length, $\sim30$\,pc). Dashed line shows the 
    profile of the starburst light in the best-matching simulation. 
    The range of outer \Sersic\ indices in the simulations (i.e.\ across sightlines for 
    these objects) and range of starburst mass fractions ($f_{\rm sb}$) 
    which match the 
    observed profile are shown$^{\ref{foot:explainfits}}$, 
    with the variance of the observations about the 
    best-fit simulation ($\dmu$, in ${\rm mag\,arcsec^{-2}}$). 
    {\em Bottom:} Observed disky/boxy-ness ($a_{4}$) and ellipticity profiles, 
    with the median (solid) and $25-75\%$ range (shaded) corresponding profile 
    from the best-fitting simulations above. Note that these are not fitted for in any sense. 
    Figures~\ref{fig:jk2}-\ref{fig:jk10}
    show the other cusp ellipticals in the sample, ranked from most to least massive.
    \label{fig:jk1}}
\end{figure*}
\begin{figure*}
    \centering
    \plotone{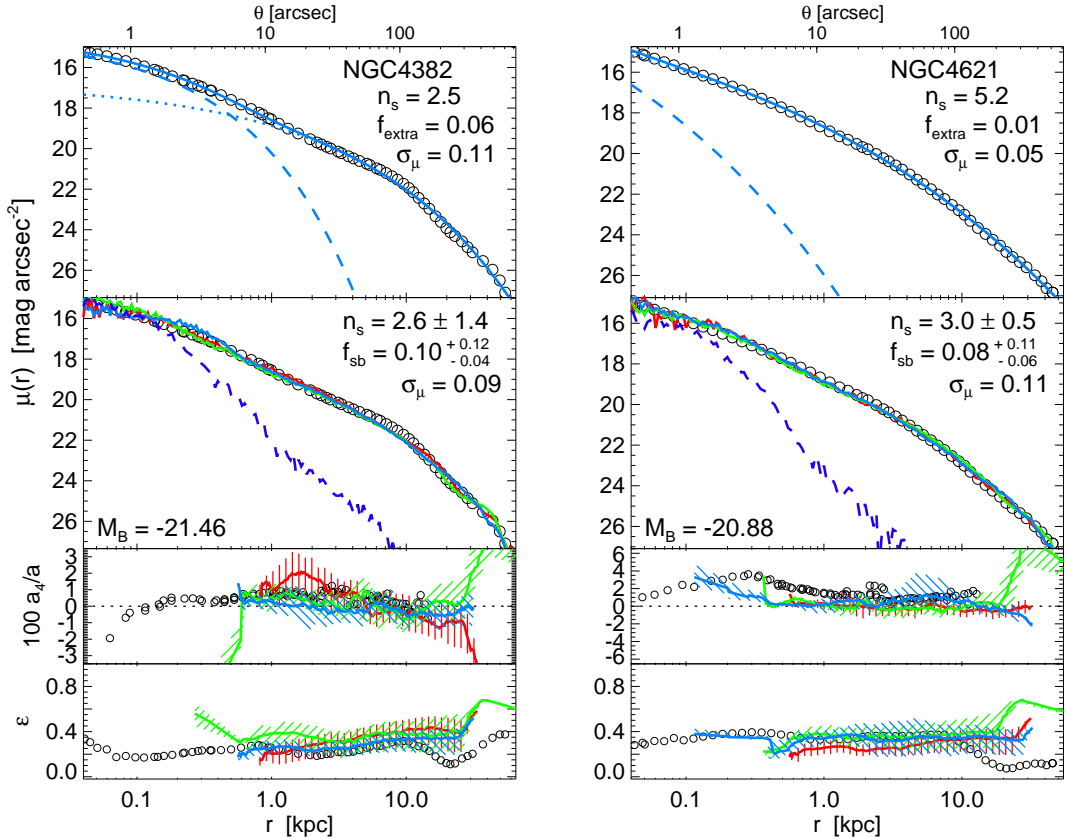}
    \caption{As Figure~\ref{fig:jk1}, but in log-log space. The two-component nature of the 
    profiles is somewhat less obvious in this projection, but the dynamic range is 
    more clear. We show the systems from the outermost observed radii down to 
    our best simulation resolution limits ($\sim30\,$pc). Over nearly four orders of 
    magnitude in radius (and $\sim14$\,mag in surface brightness), 
    simulations and observed systems agree. We show all of 
    Figures~\ref{fig:jk1}-\ref{fig:jk10} in this projection in Appendix~\ref{sec:appendix:jk}. 
    \label{fig:jk1.log}}
\end{figure*}

\begin{figure*}
    \centering
    \plotone{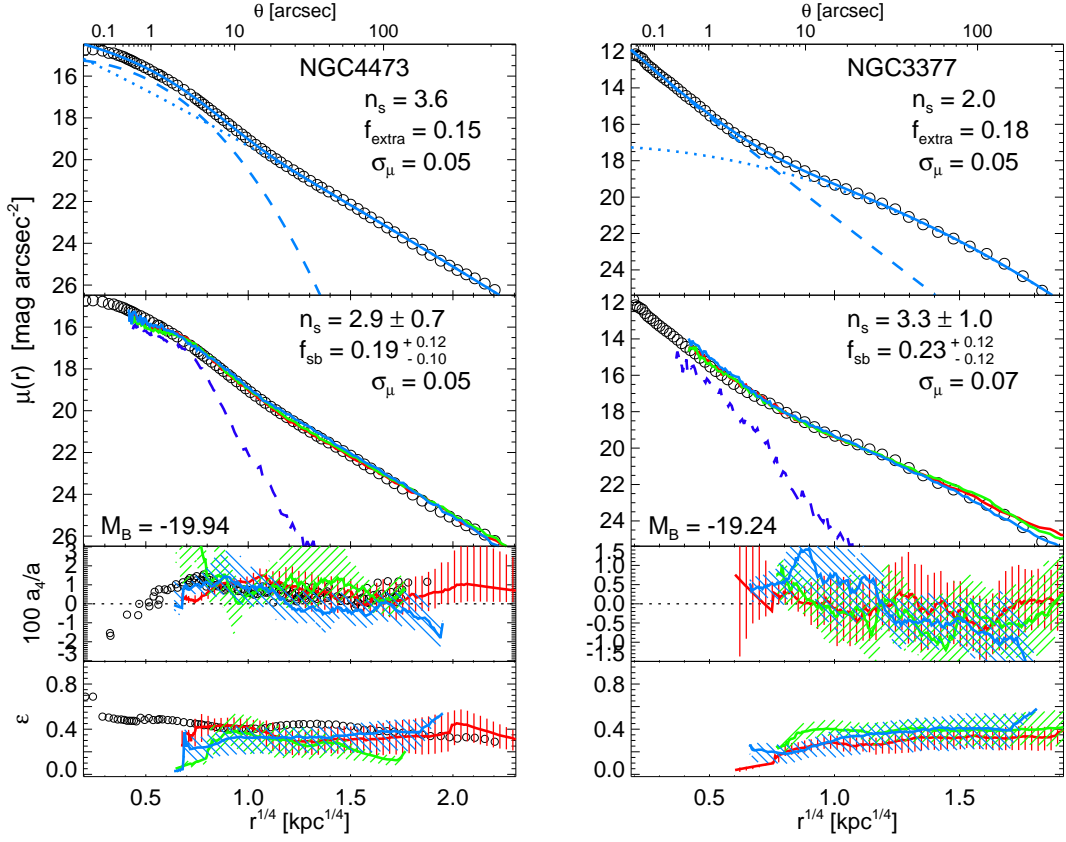}
    \caption{The next most massive cusp ellipticals ($\sim1\,\mstar$). Note that NGC 3377 is not 
    a Virgo member.
    \label{fig:jk2}}
\end{figure*}
\begin{figure*}
    \centering
    \plotone{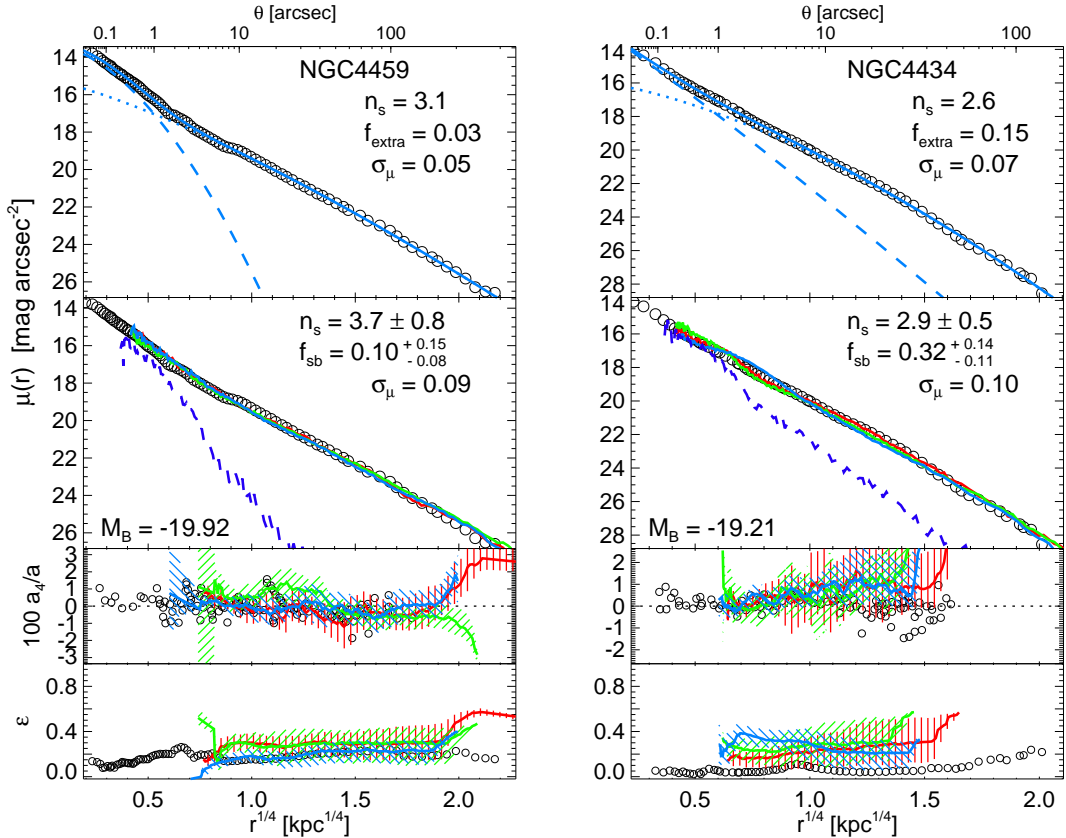}
    \caption{The next most massive cusp ellipticals ($\sim0.5\,\mstar$).
    \label{fig:jk3}}
\end{figure*}
\begin{figure*}
    \centering
    \plotone{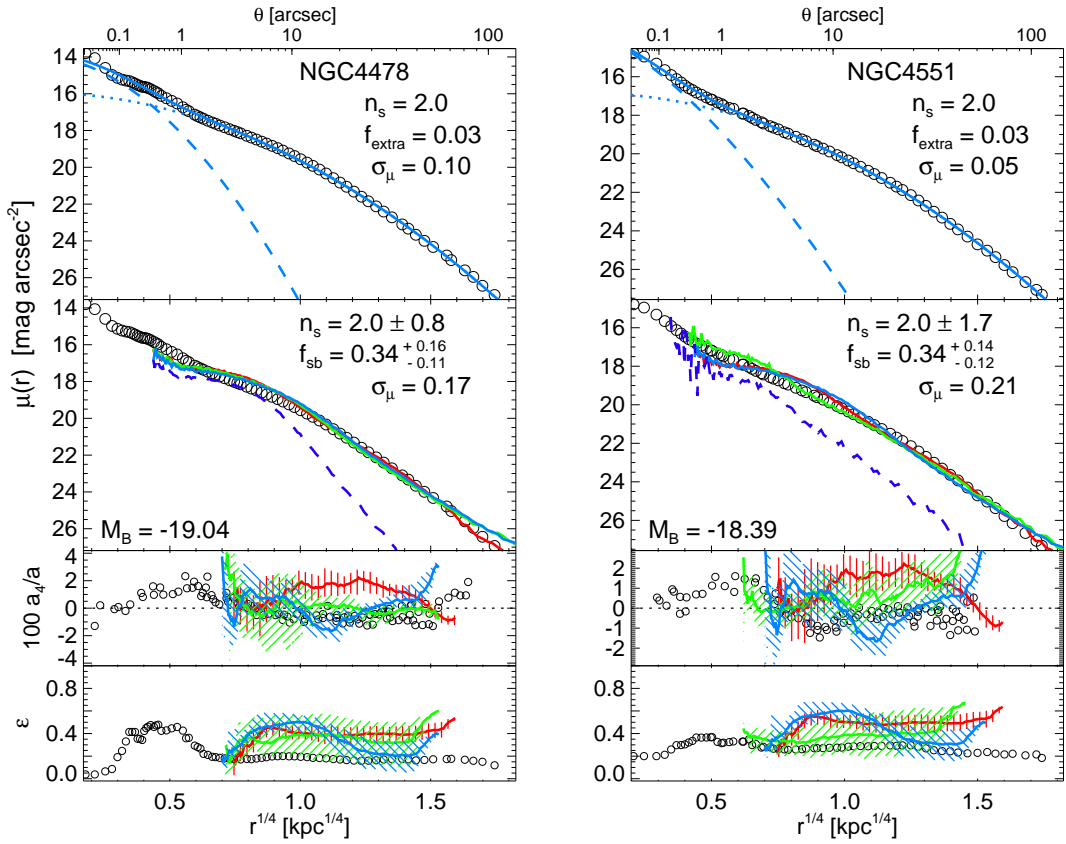}
    \caption{Lower-mass cusp ellipticals ($\sim0.2-0.3\,\mstar$). Our simulations
    reproduce the observed outer profiles and kinematic properties of such galaxies, but
    do not resolve the stellar cluster nuclei at small radii. The extra light recovered by our 
    two-component fits therefore may be misleading at such low mass. 
    \label{fig:jk4}}
\end{figure*}
\begin{figure*}
    \centering
    \plotone{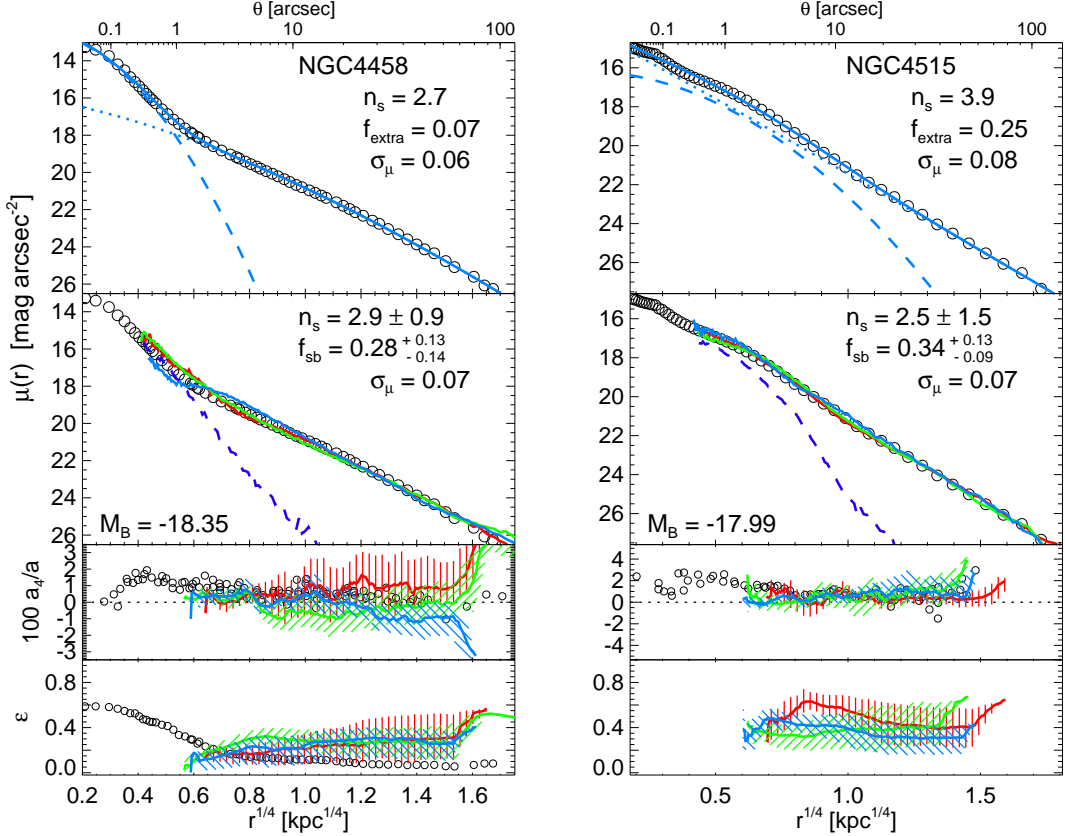}
    \caption{Additional low-mass ($\sim0.2\,\mstar$) cusp ellipticals. Our 
    fits perform better in this case.
    \label{fig:jk5}}
\end{figure*}
\begin{figure*}
    \centering
    \plotone{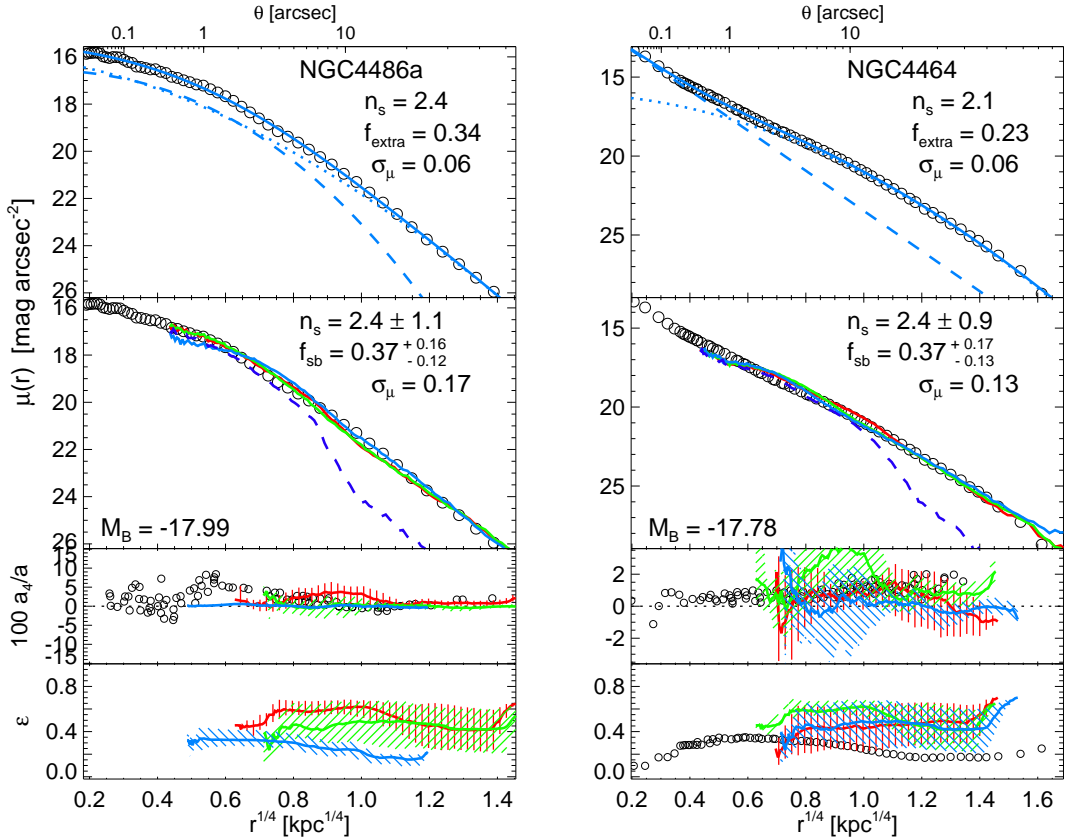}
    \caption{Additional low-mass ($\sim0.1-0.2\,\mstar$) cusp ellipticals, but 
    in this case without prominent stellar clusters in their nuclei. In this case 
    our parameterized fitting is not misled and we recover similar starburst 
    fractions to our simulations.
    \label{fig:jk6}}
\end{figure*}
\begin{figure*}
    \centering
    \plotone{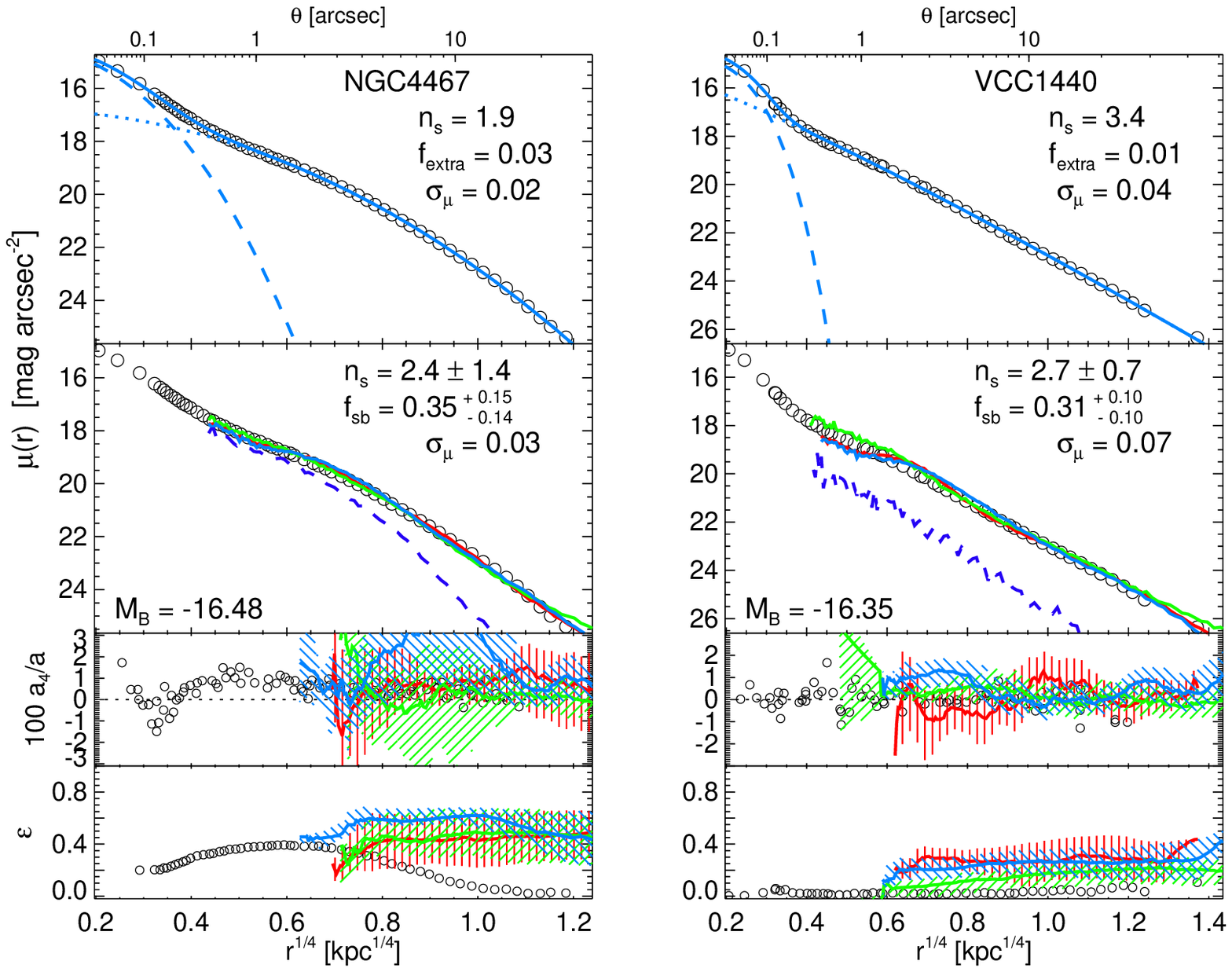}
    \caption{Very low-mass cusp ellipticals ($\sim0.03-0.1\,\mstar$). 
    Our simulations provide less good matches at 
    these luminosities, where dwarf galaxies dominate the spheroid 
    population (ellipticals at these masses are very rare). 
    Robustly resolving the extra light in these 
    very small systems probably requires $\lesssim10\,$pc 
    spatial resolution.
    \label{fig:jk7}}
\end{figure*}
\begin{figure*}
    \centering
    \plotone{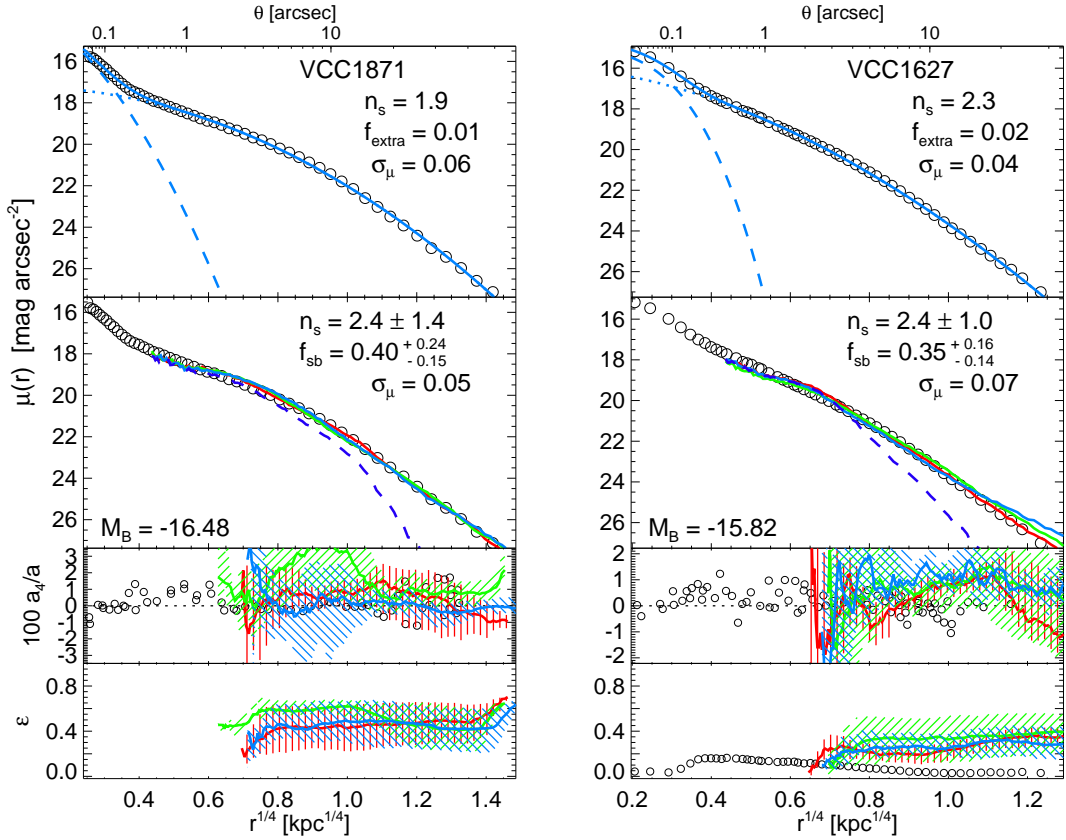}
    \caption{The lowest-luminosity cusp ellipticals in 
    Virgo ($\sim0.01\,\mstar$). The comparison with our 
    simulations is similar to Figure~\ref{fig:jk7}. 
    \label{fig:jk8}}
\end{figure*}
\begin{figure*}
    \centering
    \plotone{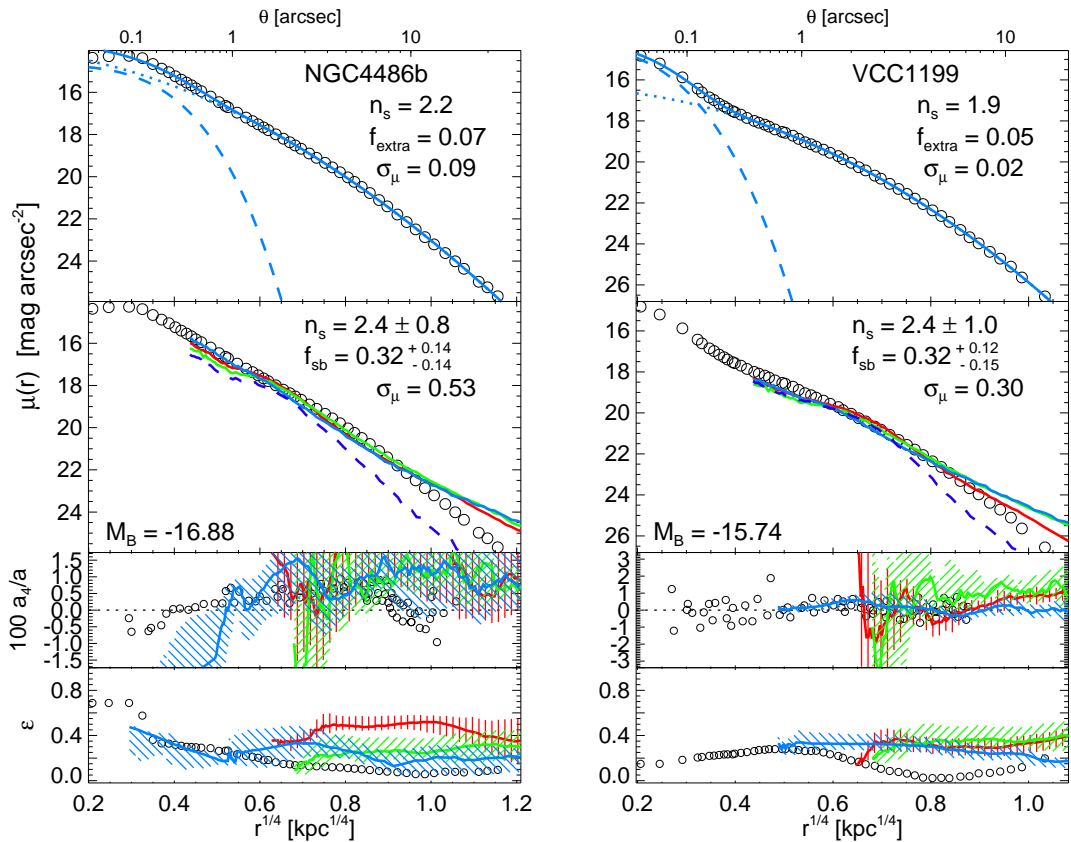}
    \caption{``Compact ellipticals.'' None of our simulations are 
    as compact as these objects (effective radii $\sim200\,$pc). 
    \label{fig:jk10}}
\end{figure*}
\breaker

We now extend our analysis to the sample of observed ellipticals and
merger remnants described in \S~\ref{sec:data}. 
Figures~\ref{fig:jk1}-\ref{fig:jk10} show surface brightness profiles
of objects in the Virgo elliptical sample of \citet{jk:profiles}, in
order of most massive to least massive.\footnote{\label{foot:4382}
NGC 4382 is typically classified as a 
core or intermediate galaxy in the literature. We show it because 
it is also sometimes considered a gas-rich merger remnant, based on 
the presence of ripples and shells formed by cold tidal material
\citep[e.g.][]{hernquist.spergel.92}, a 
high fine structure index \citep{schweizerseitzer92}, 
young ($\sim2\,$Gyr old) stellar populations 
\citep{trager:ages,mcdermid:sauron.profiles}, disky isophotes 
and high ellipticity \citep{jk:profiles}, and rapid rotation \citep{emsellem:sauron.rotation}. 
Excluding it from our sample, however, makes no difference to our conclusions.}
For each object, we plot the
surface brightness profile with the best-fit two component model, and
the corresponding fitted outer \Sersic\ index and extra light fraction.
It is a reassuring consistency check that our fitted outer 
\Sersic\ indices agree well with those estimated 
independently (and with a slightly different methodology) 
in \citet{jk:profiles} -- i.e.\ accounting for the fact that the extra light is 
an independent component (and with sufficient dynamic range to 
well-resolve the outer profile), the results in most cases are 
not highly sensitive to the exact fitting procedure. 
We refer to \paperone\ for the same comparisons
with local observed gas-rich merger remnants from \citet{rj:profiles}.
A complete list of fit parameters and compiled galaxy properties 
is included in Table~\ref{tbl:cusp.fits} for the ellipticals and 
Table~\ref{tbl:mgr.fits} for the merger remnants.

We also compare each observed system with our library of simulations,
in a non-parametric fashion. We do this by allowing the normalization
of the simulated light profile to vary (within $\pm0.5$\,dex), and
quantifying the $\chi^{2}$ (variance of the observed points with
respect to the simulated light curve at
$>1$ gravitational softening length) of each simulation. We allow the
normalization to vary because we have a finite number of simulations
and therefore do not sample a continuum in e.g.\ total brightness, but
instead discretely sample at factor $\sim2$ intervals (we do not allow
the simulated profiles to vary by more than this amount, to avoid an
unphysical match to a simulation with very different total mass). We
do not allow any other parameters to vary -- i.e.\ we allow limited 
rescaling in the surface brightness of the simulated galaxies, but 
{\em not} their radii or other properties. 
Despite the allowed surface brightness rescaling, 
the best-fit simulations almost
always have similar total luminosities to the observed system, because
they must have a similar effective radius in order to be a good
match. Considering $\sim100$ sightlines to each of our simulations
(although, as noted in \paperone, the observed surface brightness
profile varies by a small amount sightline-to-sightline), we find the
best fit to each observed system.

In the middle panels of 
Figures~\ref{fig:jk1}-\ref{fig:jk10} we show the three simulations
which most closely match the observed light profile. For the best-fit
simulation, we also show the profile of the stars formed in the final,
central, merger-driven starburst, as described in
\S~\ref{sec:fits}. We show in the figures the outer \Sersic\ indices
fitted to these simulations, along with the typical range both across
sightlines and across the best-fitting simulations (which together
give some rough approximation to the range of $n_{s}$ which might be
observed for these galaxies along different sightlines). We also show the
best-fit starburst mass fraction, along with the range across the
best-fitting simulations (described below), and the variance of the
observed points with respect to the best fit\footnote{\label{foot:explainfits}
The values shown 
in Figures~\ref{fig:jk1}-\ref{fig:jk10} are based 
on comparison only to the profiles shown, from \citet{jk:profiles}. In 
Table~\ref{tbl:cusp.fits}, the values represent the results from all available 
data sets, including multiple different observations of the systems shown here, 
and so can be slightly different (however the differences are generally small).}.

In addition, for these simulations we show the isophotal shape and
ellipticity as a function of major axis radius, compared to that
observed.  Note that we do {\em not} fit these quantities, only the
surface brightness profile.  We show, for each simulation, the range
across sightlines in these quantities -- it is clear that these depend
much more strongly on sightline than the surface brightness profile
(this is primarily why we do not fit these quantities).  In every
case, there is a significant fraction of sightlines with shape and
ellipticity profiles roughly consistent with those observed, but the
simulations highlight the range of profile shapes for similar
spheroids to those observed.

For the intermediate and higher mass Virgo ellipticals, $\gtrsim0.1\,\lstar$, we easily
find simulations which provide an excellent match to the observed
profiles, with variance $\dmu$ often less than even a
multi-component parameterized fit. The fits are good over the entire 
dynamic range from the largest observed radii ($\sim100\,$kpc) 
down to our resolution limits ($\sim30$\,pc)\footnote{\label{foot:loglog}
The dynamic range of the fits is somewhat difficult to discern in 
Figures~\ref{fig:jk1}-\ref{fig:jk10} owing to the plotting versus 
$r^{1/4}$; we therefore reproduce these figures plotting $\mu$ versus $r$ in 
Appendix~\ref{sec:appendix:jk}.}. At radii below our softening limits, 
the simulation profiles artificially flatten; but we show in Appendix~\ref{sec:appendix:resolution} 
that the agreement continues down to smaller and smaller radii 
as we increase our resolution. 
At moderate and large masses, the starburst
fraction recovered by our two-component fit is usually a good match to
the physical starburst mass fraction in the best-fitting
simulations (see \S~\ref{sec:properties}).  

At the lowest masses $L\lesssim0.01\,L_{\ast}$, 
the fitted extra light components tend to be smaller than 
our simulated starbursts. This is at least in part a 
resolution issue, both in our simulations and in the observations 
(the extra light in the lowest-luminosity ellipticals is poorly resolved 
even with HST data). 
Below $M_{B}\sim-18$ (i.e.\ roughly an order of magnitude below
$\sim\lstar$), it is also no longer clear that the fitted 
extra light components are the same physical entities -- some of the 
sharp central features in the profiles may in fact be 
nuclear stellar clusters, modeling of which would require
resolving individual star-forming complexes in our simulations.  
(See Appendix~\ref{sec:appendix:nuclei} for a discussion of the 
differences between extra light components studied herein 
and the stellar nuclei studied in e.g.\ \citet{cote:virgo} and 
\citet{ferrarese:profiles}. Those nuclear components 
are on much smaller scales and bear little resemblance to the 
starburst component we are interested in here.) The two lowest-mass 
``compact ellipticals'' in our sample, NGC4486b and VCC1199 
\citep{binggeli:virgo.center}, with $R_{e}\sim150-200\,$pc, 
are somewhat smaller than any of our simulated merger remnants. 
Given our resolution limits and 
limited sampling of e.g.\ initial disk sizes at these very low 
masses, it is premature to say whether some different physics 
\citep[e.g.\ tidal stripping;][]{faber:compact.ellipticals.origin} is needed, 
but the systems lie on the fundamental 
plane and their outer profile shapes appear normal.

\begin{figure*}
    \centering
    \scaleup
    \plotone{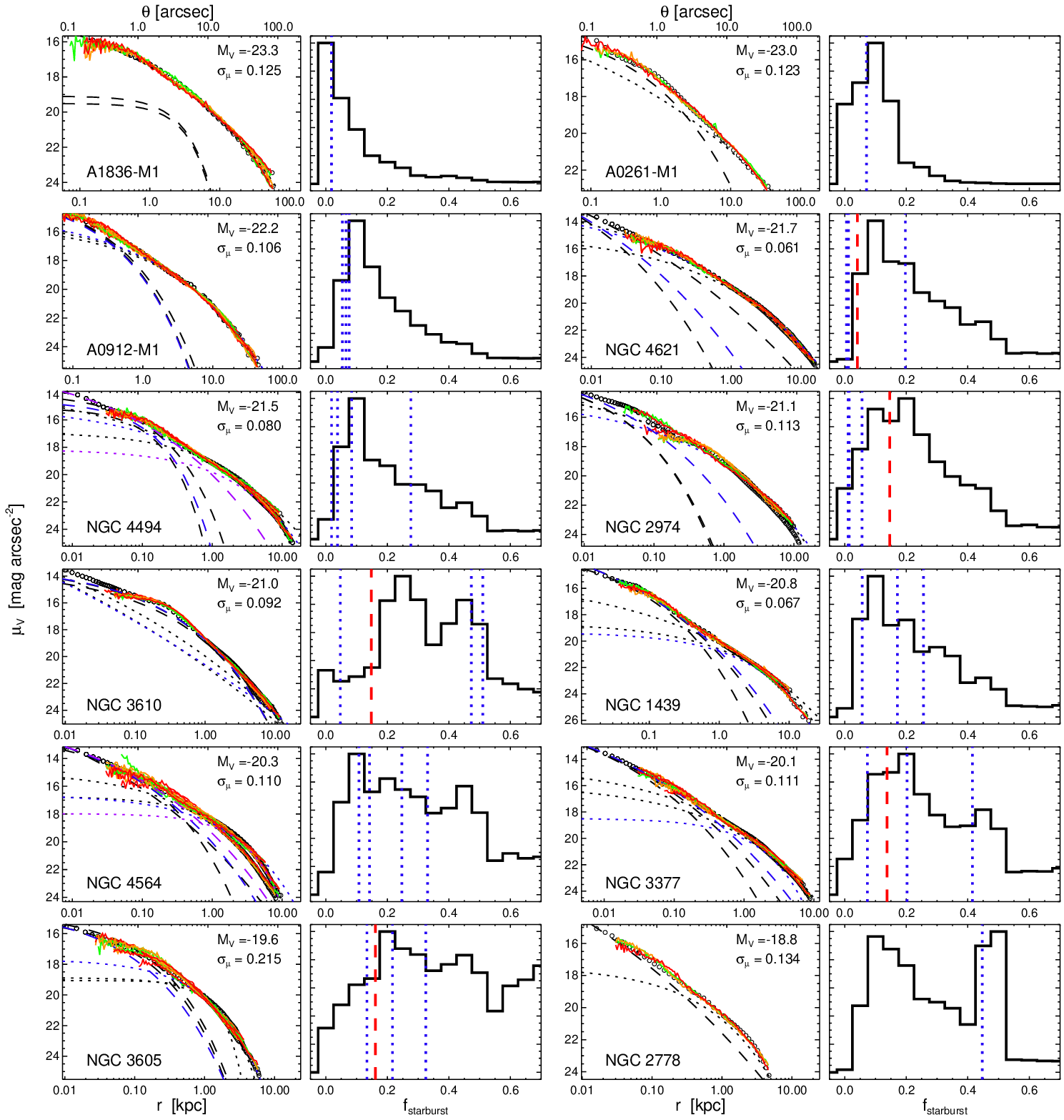}
    \caption{Observed surface brightness profiles of 
    a subset of the confirmed cuspy 
    ellipticals from the sample of \citet{lauer:bimodal.profiles}, 
    with the best-matching two component 
    parameterized fit (dashed and dotted lines)
    and best-fitting simulations (red, orange, and green lines), as in Figure~\ref{fig:jk1}. 
    Where multiple sources of photometry are available, independent fits to each are 
    shown.
    The objects are ranked from brightest to faintest in $V$-band (as shown). 
    Profiles are shown over a constant angular scale (top axis; bottom axis shows 
    physical radius in kpc).
    The corresponding ({\em right}) panel for each shows the distribution of physical 
    starburst fractions for the simulations which provide a good fit to the 
    observed profile (as described in the text), with the fitted (parameterized) 
    extra light fraction (blue dotted line; one for each source of 
    photometry). Where available, red dashed lines show the mass fraction 
    of a secondary (recent) starburst population independently 
    estimated from stellar population studies in other works. 
    Our simulation resolution 
    limits do not extend within the central $\sim30-50\,$pc, and our fits 
    are not intended to describe these radii. 
    \label{fig:lauerpp1}}
\end{figure*}

Figure~\ref{fig:lauerpp1} again shows the observed and best-fit simulated 
surface brightness profiles, for a subset of confirmed cusp ellipticals in 
the sample of \citet{lauer:bimodal.profiles}, in order of $V$-band magnitude from 
brightest to faintest. For each simulation in our library, we 
have a $\chi^{2}$ corresponding to its goodness of fit to the observed 
profile, and the genuine physical starburst mass fraction $\fsb$. We can 
therefore construct a $\chi^{2}$-weighted distribution of $\fsb$ 
for each observed system -- essentially, the probability, across a 
uniform sample of initial conditions, that the observed profile 
was drawn from a simulation with the given starburst mass fraction. 
These are shown, and compared to the fitted extra light fraction for our 
two-component models. 
In general, the fitted extra light fraction corresponds well to 
the characteristic starburst mass fractions in simulations which produce 
similar light profiles. 

There are a small number of observed objects for which detailed 
spectral energy distributions (SEDs)
have enabled two-component stellar population models to be fit, in
which there are generally an older, smoother distribution and a
younger, metal-enriched single burst population
\citep{titus:ssp.decomp,schweizer:7252,schweizer96,schweizer:ngc34.disk,
reichardt:ssp.decomp,michard:ssp.decomp}. For these objects, we
plot the observationally estimated mass fraction (from these 
studies) in the secondary
burst population (the vertical red dashed lines), 
which we expect should correspond (roughly) to the
starburst population in the spheroid-forming merger, if this is a good
description of the formation history. Although there are only a few
systems for which this comparison is possible, the agreement
between this estimate and our inferred extra light or starburst
fractions is surprisingly good. Of course, there are a number of
uncertainties and degeneracies in an attempt to observationally
decompose stellar populations, but this gives us confidence that there
is physical meaning to our decompositions.

\breaker
\section{Properties of ``Extra Light'' Profiles}
\label{sec:properties}

Having fit both our simulations and observed cuspy ellipticals to an 
outer violently relaxed component and an inner starburst profile, we 
now compare these fits as a function of galaxy properties.

\subsection{Outer Profiles: \Sersic\ Indices}
\label{sec:properties.outer}

\begin{figure*}
    \centering
    \epsscale{0.9}
    \plotone{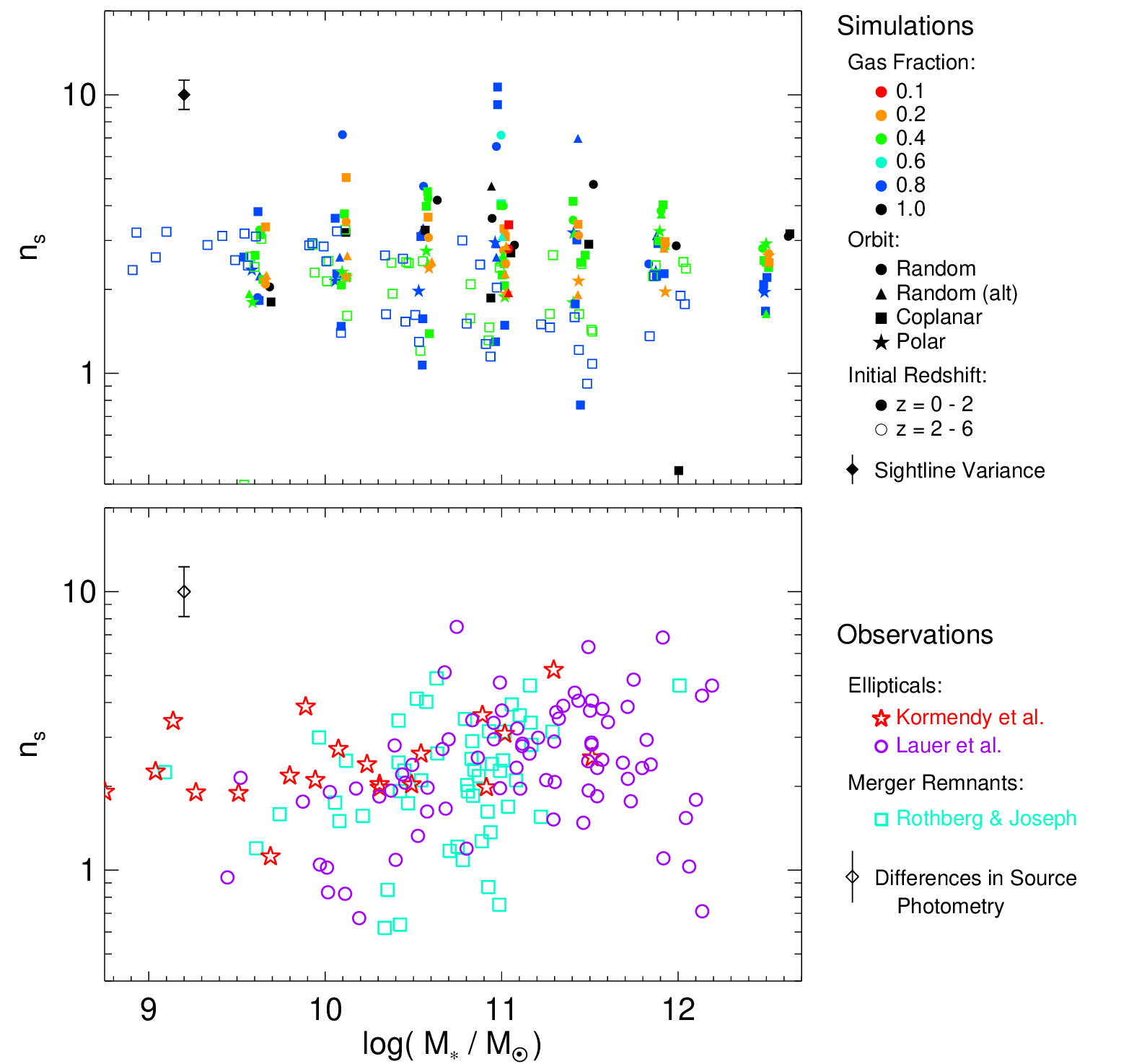}
    \caption{Outer \Sersic\ indices in cuspy ellipticals and simulated gas-rich merger 
    remnants, using our 
    two-component decomposition. Gas-rich merger remnants have characteristic 
    $n\sim2-3$, without a strong systematic dependence on mass or other properties. 
    {\em Top:} Simulations: color encodes gas fraction, symbol encodes orbital parameters 
    (the two ``random'' orbits are two different common random orbits, one somewhat 
    closer to prograde, the other -- ``alt'' -- somewhat closer to polar), 
    and filled/unfilled encodes the initial redshift of the simulations (as in plotted key). 
    The plotted $n_{s}$ is the median across $\sim100$ sightlines, 
    typical sightline-to-sightline differences are shown as the filled plotted error bar.
    {\em Bottom:}
    Open colored points are observed systems: 
    red stars are the cusps in the Virgo elliptical sample of \citet{jk:profiles}, purple circles 
    the confirmed cusps in the local elliptical sample of \citet{lauer:bimodal.profiles} 
    (supplemented by the sample of \citet{bender:data}), and  
    cyan squares are the gas-rich merger remnants from \citet{rj:profiles}. 
    Open plotted error shows the typical differences in $n_{s}$ derived from different sources of 
    photometry and/or different observed wavelengths.
    We use this point notation throughout. 
    \label{fig:ns.mass}}
\end{figure*}

Figure~\ref{fig:ns.mass} plots the outer \Sersic\ indices of our sample
of simulations as a function of galaxy stellar mass, compared with
those observed. The lack of any trend is striking -- we predict 
that there is {\em no dependence
of outer \Sersic\ index on galaxy mass for cusp ellipticals}. In fact, we have
searched our entire sample of mergers attempting to find a dependence
of outer \Sersic\ index on some galaxy property, including merger
redshift, gas fraction, halo concentration, baryon fraction, and the
presence or absence of initial bulges, and find no dependence. There
is a weak trend with gas content, but only in the sense that systems
with extremely high gas content even at late merger stages (e.g.\
$\gtrsim40\%$ gas at the time of final coalescence) can form or retain
massive disks, bringing them closer to $n_{s}=1$. There is also a weak
dependence on orbital parameters, but only in the sense of different,
extreme orbits changing the best fit outer \Sersic\ index by $\Delta
n_{s}\lesssim1$. The apparent difference between our low and 
high-redshift simulation $n_{s}$ distributions is in fact entirely 
attributable to these effects. 
Similarly, \citet{naab:profiles} find that in simulated 
collisionless (gas-free) disk merger 
remnants -- i.e.\ systems for which the entire profile is by definition part of 
the ``outer,'' violently relaxed component -- there is also no significant 
dependence of the \Sersic\ index on mass, effective radius, or merger mass 
ratio. 

The observations appear to confirm this prediction. \citet{jk:profiles} 
see no dependence of outer \Sersic\ index on galaxy luminosity
(within the extra light/cusp population), 
and our other data sets support this over a large 
baseline in luminosity and stellar mass (albeit with larger 
scatter, owing primarily to the lower quality of the data). 
Over more than three orders of magnitude in stellar mass, 
and two orders of magnitude in effective radius, the observations 
and simulations both show 
a typical $n_{s}\sim2-3$ with 
no dependence of outer \Sersic\ 
index on mass, luminosity, or radius in cusp ellipticals. 

This prediction appears to contradict some previous results 
that argue for a strong
dependence of \Sersic\ index on luminosity 
\citep{graham:bulges,trujillo:sersic.fits,ferrarese:profiles}
or effective radius 
\citep{caon:sersic.fits,prugniel:fp.non-homology}. 
However, we emphasize that these fits are
{\em not} directly comparable to ours.  First, 
these correlations were found considering samples of a broad
range of spheroids -- from dwarf spheroidals through cuspy, rapidly
rotating ellipticals through massive, cored, slowly rotating
ellipticals.  Here, we are only arguing that the specific subclass of
cuspy, true ellipticals formed in gas-rich mergers should have a mass-independent
\Sersic\ index distribution.  Different formation mechanisms \citep[for 
example, subsequent dry mergers;][]{hopkins:cores} can
systematically change the \Sersic\ index, giving rise (via cosmological
trends towards more mergers in higher-mass systems) 
to mean correlations between \Sersic\ index and galaxy mass or
size. To the extent that the observed \Sersic\ indices of 
the cusp population are relatively low and do not depend on 
mass or luminosity, it implies that they are generally formed in a small number 
of major mergers, without substantial subsequent re-merging. 

Second, these authors were often
fitting the entire galaxy light profile to a single \Sersic\ or
core-\Sersic\ law, whereas we have attempted to decompose the inner and
outer galaxy light profiles. This two component approach
will systematically yield different \Sersic\ indices, and in some cases
as demonstrated in \S~\ref{sec:fits} the difference can be dramatic.

\begin{figure}
    \centering
    \epsscale{1.15}
    \plotter{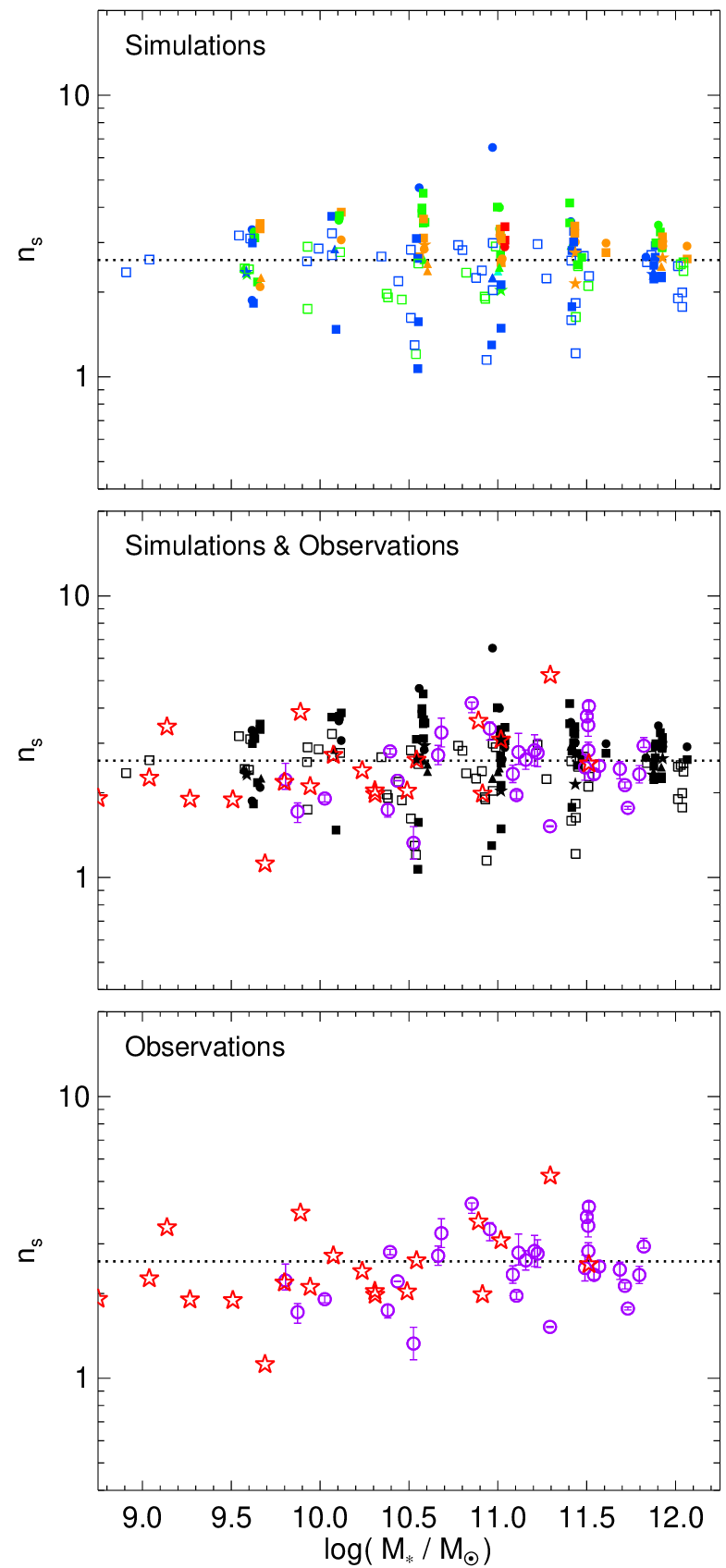}
    \caption{As Figure~\ref{fig:ns.mass}, but using a cleaned sample: we include 
    only those galaxies for which $\ge3$ sources of photometry (or photometry in 
    $\ge3$ bands) yield $n_{s}$ values different by $<20\%$. For the simulations, we 
    plot only those with sightline-to-sightline variance less than this amount (usually 
    eliminating those with significant tidal or unrelaxed features). Dotted line 
    shows the median $n_{s}\sim2.6$. 
    The distribution and 
    lack of dependence on mass is the same as Figure~\ref{fig:ns.mass}, but the 
    results here are more robust.
    \label{fig:ns.mass.cleaned}}
\end{figure}

To check if the effect illustrated in Figure~\ref{fig:ns.mass} is
caused by large scatter in our estimates (possibly obscuring an
underlying trend), we show in Figure~\ref{fig:ns.mass.cleaned} the
same \Sersic\ index as a function of mass, but using a strict, cleaned
sample. We include in this subsample only those galaxies for which
$\ge3$ sources of photometry (or photometry in $\ge3$ bands) yield
$n_{s}$ values different by $<20\%$. We exclude the recent merger
remnants, for which unrelaxed features may introduce additional
scatter or uncertainties.  For the simulations, we plot only those
with sightline-to-sightline variance less than this amount (usually
eliminating those with significant tidal or unrelaxed features). The
distribution and lack of dependence on mass is the same as
Figure~\ref{fig:ns.mass}, but the conclusion here is more robust.

\begin{figure}
    \centering
    \scaleup
    \plotter{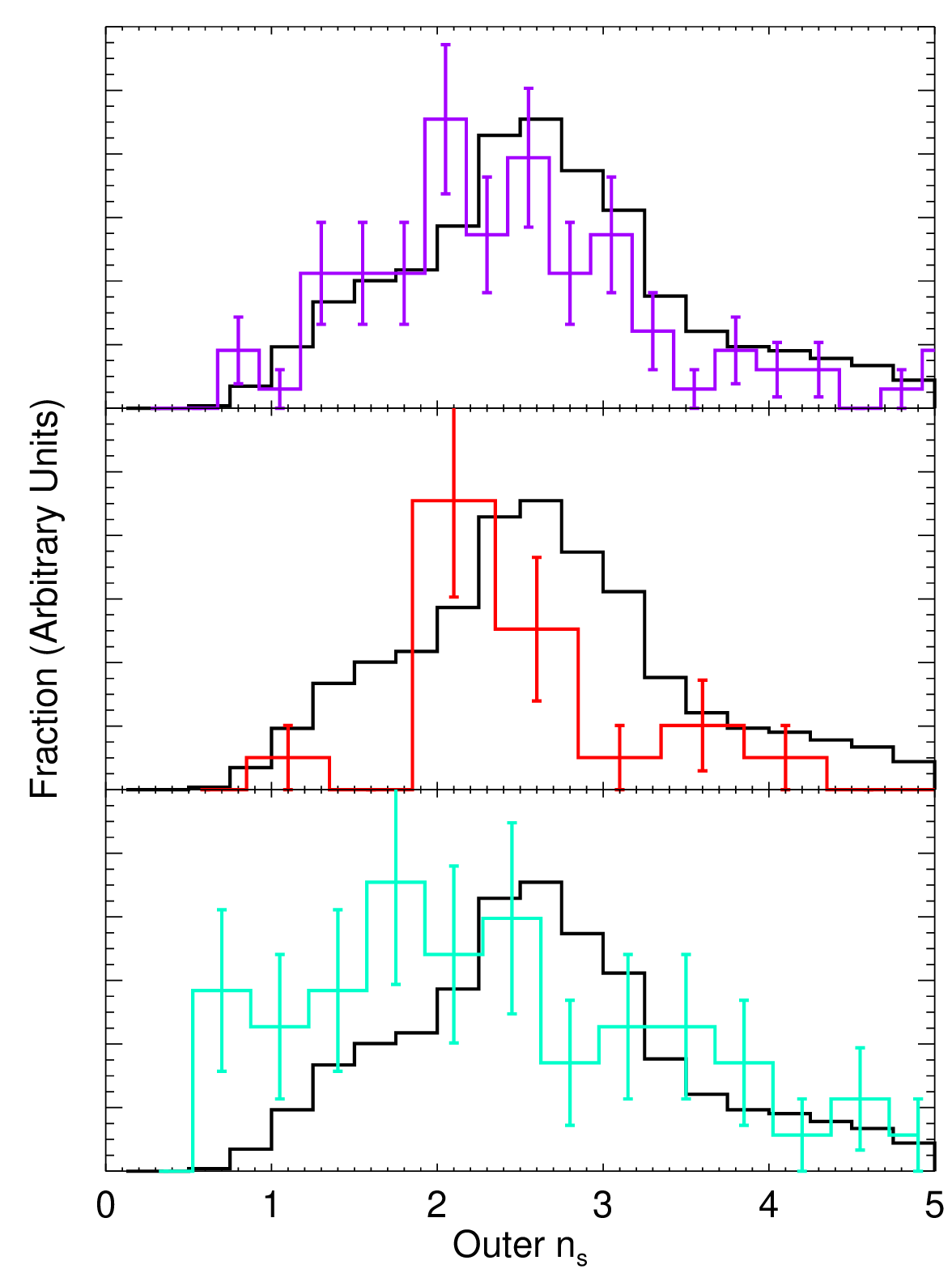}
    \caption{Distribution of outer \Sersic\ indices in cuspy ellipticals, using our 
    two-component decomposition. Solid black line shows the result 
    for our entire sample of simulations (each across $\sim100$ sightlines). Colored lines show the 
    results for the observed samples of \citet{lauer:bimodal.profiles}
    (top), \citet{jk:profiles} (middle), 
    and \citet{rj:profiles} (bottom), with Poisson error bars. \citet{rj:profiles} include 
    some likely S0s, yielding a larger fraction of $n=1$ systems. Cuspy ellipticals 
    have a fairly narrow range of $n\sim2.50-2.75\pm0.75$, in good agreement 
    with gas-rich merger simulations.
    \label{fig:ns.distrib}}
\end{figure}
\breaker

\begin{figure}
    \centering
    \scaleup
    \plotter{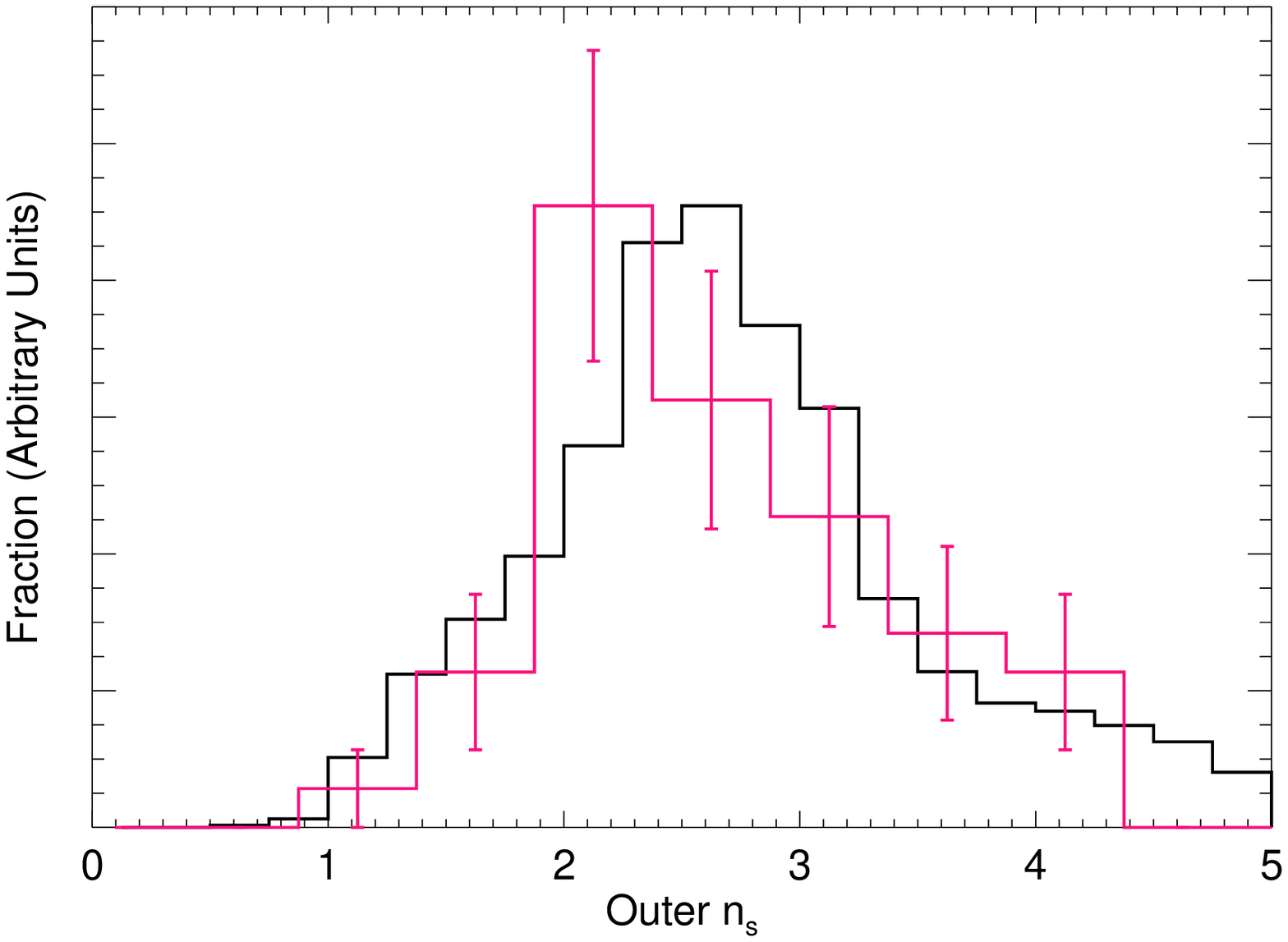}
    \caption{As Figure~\ref{fig:ns.distrib}, but for the robust 
    cleaned sample of Figure~\ref{fig:ns.mass.cleaned}.
    \label{fig:ns.distrib.cleaned}}
\end{figure}

Given that, at least for cuspy ellipticals formed in gas-rich mergers, there is 
no strong systematic dependence of $n_{s}$ on other properties or 
initial conditions, it is acceptable to place all such systems on the same footing 
and consider the overall distribution of $n_{s}$ values. Figure~\ref{fig:ns.distrib} 
shows this, for the simulations and our observed samples. 
Figure~\ref{fig:ns.distrib.cleaned} shows the same, but restricted to the 
cleaned subsample of Figure~\ref{fig:ns.mass.cleaned}. 
In each case, there is 
reasonable agreement, within the errors.

The sample of \citet{rj:profiles} shows somewhat more
$n_{s}\sim1$ systems than our simulations or observed cusp
ellipticals. This effect is only marginally significant, but probably
arises because a few of the systems in their sample will most likely
(once they are relaxed) be better classified as S0s than ellipticals.
There is a tentative suggestion that our predictions are shifted to
systematically higher $n_{s}$ than the observed systems, by $\Delta
n\sim0.25$ or so.  At this level, however, observational issues in the
measurements become important, as do the exact orbital parameters used
in the simulations and the dynamic range over which the fit is
performed \citep[see e.g.][]{boylankolchin:mergers.fp}.  For example,
\citet{blanton:env} find that for SDSS light profiles, the
observations may be biased to underestimate $n_{s}$ by $\sim0.2-0.5$,
and are sensitive to the sky subtraction \citep[see also][]{lauer:massive.bhs}, 
consistent with the offset we see.  It is therefore not surprising
that the agreement is not exact.  The important thing is that, with
few rare exceptions ($\lesssim5\%$ of cases), cuspy ellipticals and
gas-rich merger remnants have $n_{s}<4$ \citep[as in][]{jk:profiles}, 
i.e.\ are concave-down in $\mu-r^{1/4}$ space, without a significant 
dependence of $n_{s}$ on other galaxy properties.

\subsection{Dissipational (``Extra Light'') Mass Fractions}
\label{sec:properties.fextra}

\begin{figure}
    \centering
    \scaleup
    \plotter{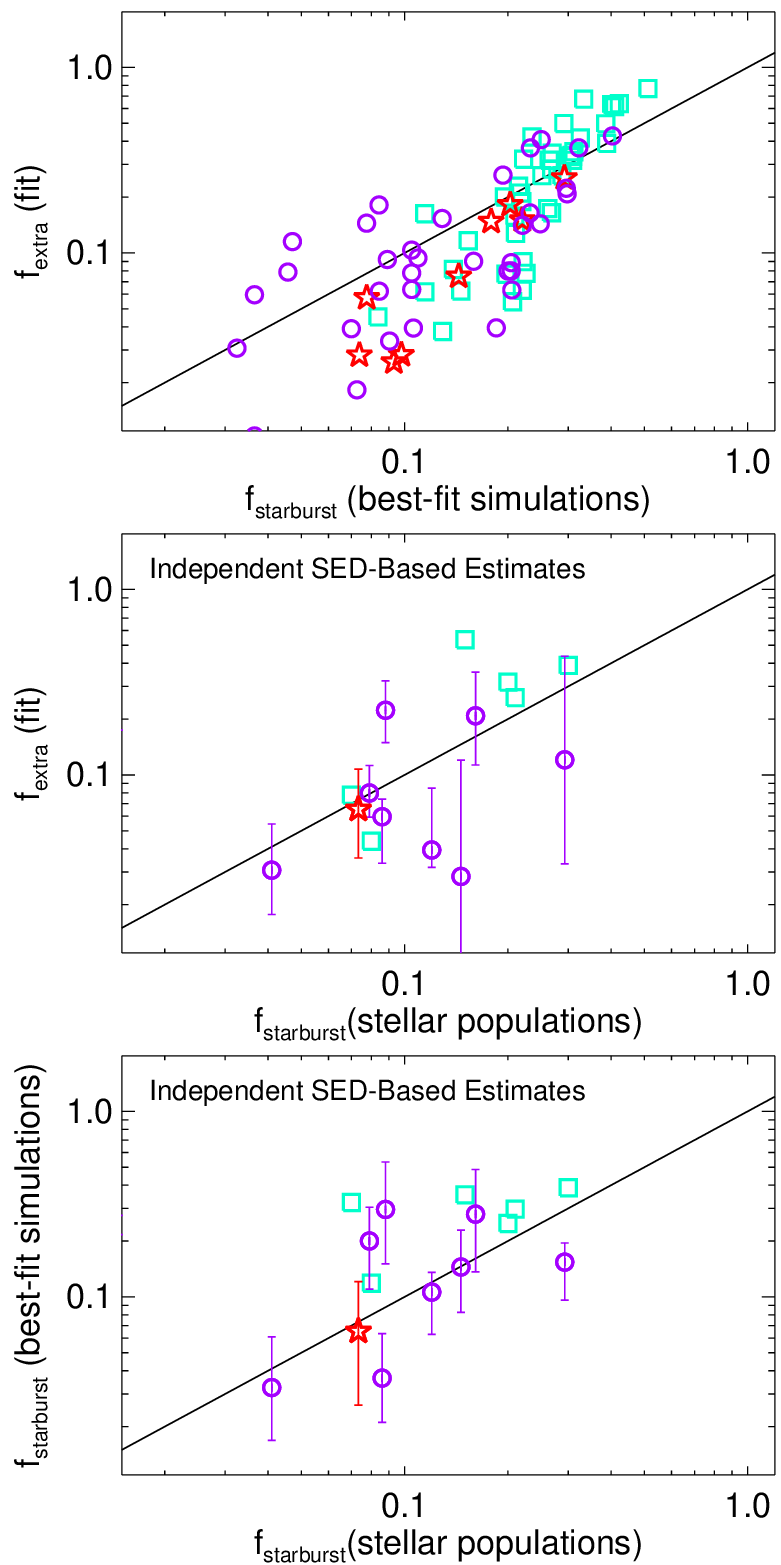}
    \caption{{\em Top:} Comparison of our estimated 
    mass in the fitted extra light component ($f_{\rm extra}$) versus 
    the starburst mass fraction in the best-fitting simulations ($f_{\rm sb}$; 
    as in Figures~\ref{fig:jk1}-\ref{fig:lauerpp1}). 
    For clarity, we show only observations from our ``cleaned'' sample and where 
    it would be possible for our simulations to resolve the extra light. 
    In these cases, the two estimates agree well, with a factor $\sim2-3$ scatter 
    in $f_{\rm extra}(f_{\rm sb})$ (similar to what we expect from our 
    simulations; see Figure~\ref{fig:check.fe.fsb}). 
    {\em Middle:} $f_{\rm extra}$ versus 
    independent observational estimates of 
    the mass fraction formed in a more recent starburst/star formation event, 
    from two-component stellar population model fits to the observed SEDs 
     \citep{titus:ssp.decomp,
    schweizer96,schweizer:7252,schweizer:ngc34.disk,
    reichardt:ssp.decomp,michard:ssp.decomp} 
    {\em Bottom:} Same, but comparing the stellar population estimates 
    to $f_{\rm sb}$ from the best-fitting simulations. 
    More observations are needed to independently test our 
    estimates, but the stellar population data independently 
    suggests our decompositions 
    are reasonable. 
    \label{fig:mass.vs.fsb}}
\end{figure}

Figure~\ref{fig:mass.vs.fsb} compares our estimates of the dissipational 
mass fraction in the observed ellipticals: the 
directly fitted extra light fraction $f_{\rm extra}$ 
and inferred starburst mass fraction $f_{\rm sb}$ from the best-fitting simulations. 
For clarity we restrict to the cleaned sample 
from Figure~\ref{fig:ns.mass.cleaned}. Our fitted decomposed extra 
light fraction reliably traces the inferred starburst 
mass fraction, with a factor $\sim2$ scatter similar to that predicted 
from our simulations (Figure~\ref{fig:check.fe.fsb}). The starburst 
fraction $f_{\rm sb}$ itself must, in some sense, reflect the cold gas 
mass available in the disks just before the final merger 
(and we show in \paperone\ that this is the case) -- in this 
physical sense, our fitted $f_{\rm extra}$ and inferred $f_{\rm sb}$ 
are a robust reflection of the gas content of the progenitors. 
Of course, changing simulation properties such as 
the presence or absence of an initial
bulge, the concentration of the progenitor halos and disks, the
presence or absence of a supermassive black hole, and the treatment of
star formation and the ISM equation of state 
can indirectly influence $f_{\rm sb}$ by altering how
efficiently gas is consumed and/or expelled before the final merger,
and therefore how much is available to participate in the
starburst. For a fixed gas mass at the time of the final starburst, however, 
the starburst component mass (and therefore also extra light mass, 
which traces the starburst) is independent of these effects. 

Figure~\ref{fig:mass.vs.fsb} also compares the results of our
fitting to independent stellar population-based 
estimates of the starburst fraction in observed ellipticals. 
It is in principle possible, by studying the stellar populations in
sufficient detail, to estimate the mass fraction which formed in a
recent, central starburst (as opposed to the more extended quiescent
star formation history), and this should provide an independent check 
of our decompositions. 
Unfortunately, there are still a number of
degeneracies, and this requires detailed observations, but it has been
attempted for several of the observed systems
\citep{titus:ssp.decomp,schweizer:7252,schweizer96,schweizer:ngc34.disk,
reichardt:ssp.decomp,michard:ssp.decomp}. Comparing our 
estimated $f_{\rm extra}$ or $f_{\rm sb}$ with these 
estimates for the mass fraction in the secondary (newly
formed/starburst) stellar populations, we find a reasonable correlation. 
Although there are only a
few objects for which sufficiently accurate stellar populations are
available to allow this comparison, they all suggest that our fitted
extra light component is indeed a good proxy for the mass fraction
which was involved in the central, merger-driven starburst.

\begin{figure*}
    \centering
    \scaleup
    \plotter{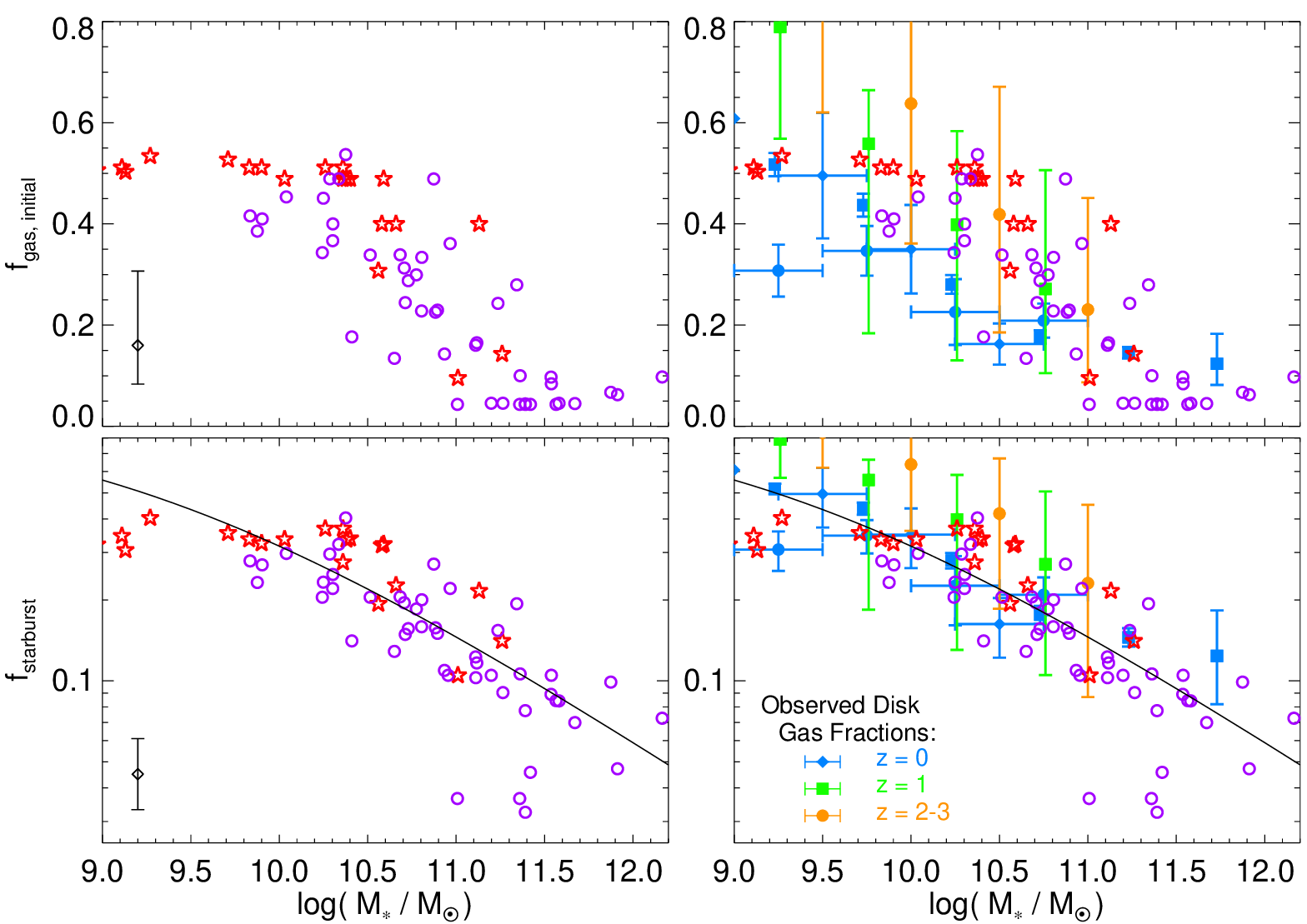}
    \caption{Inferred gas content (dissipational/starburst fraction) of 
    elliptical-producing mergers as a function of stellar mass. 
    Initial gas fraction ({\em top}) and physical final starburst mass 
    fraction ({\em bottom}) corresponding to the best-fit simulations to 
    each observed system in the samples of 
    \citet{lauer:bimodal.profiles} (circles) 
    and \citet{jk:profiles} (stars) are shown, 
    with the typical $25-75\%$ allowed range (error bar).
    Solid line shows the fit to the data (Equation~\ref{eqn:fgas.m}). 
    Colored points with error bars indicate the mean (and $\pm1\,\sigma$ 
    range in) disk gas fractions at the same stellar mass, at 
    $z=0$ \citep[][blue diamonds, squares, and circles, respectively]
    {belldejong:tf,kannappan:gfs,mcgaugh:tf}, 
    $z=1$ \citep[][green squares]{shapley:z1.abundances}, and 
    $z=2$ \citep[][orange circles]{erb:lbg.gasmasses}. There is a clear trend of increasing 
    dissipation 
    required to explain elliptical profiles at lower masses 
    (significant at $>8\,\sigma$), 
    in good agreement with the observed trend in progenitor disk 
    gas fractions over the redshift range where 
    cusp ellipticals are formed, and with what is invoked to explain  
    the observed densities and fundamental plane correlations of ellipticals 
    \citep[e.g.][]{kormendy:dissipation,hernquist:phasespace}.
    \label{fig:fgas.needed}}
\end{figure*}

Having some confidence that our estimates of $f_{\rm sb}$ are reasonable, 
Figure~\ref{fig:fgas.needed} plots the inferred starburst mass fraction 
$f_{\rm sb}$ for the observed systems as a function of stellar mass. In 
the same manner that we have defined a best-fit $f_{\rm sb}$ 
from the best-fit simulations, we can also define a best-fit 
``initial'' gas fraction (roughly the gas fraction $\sim1\,$Gyr before the 
final merger), and show this as well. We emphasize though (for the reasons above 
regarding the efficiency of pre-merger gas consumption and expulsion) that 
this is a much less robust quantity. In either case there is 
a clear trend of increasing dissipation 
(increasing fractional mass required in a dissipational starburst component) 
at lower masses. The significance of the correlation is unambiguous ($>8\,\sigma$). 
We can conveniently approximate the trend in dissipational mass fraction 
as a function of stellar mass with the fitted function 
\begin{equation}
\langle f_{\rm starburst} \rangle \approx 
{\Bigl[}1+{\Bigl(}\frac{M_{\ast}}{10^{9.15}\,\msun}{\Bigr)}^{0.4}{\Bigr]}^{-1}, 
\label{eqn:fgas.m}
\end{equation}
with roughly a constant factor $\sim2$ intrinsic scatter at each mass.  

\begin{figure}
    \centering
    \scaleup
    \plotter{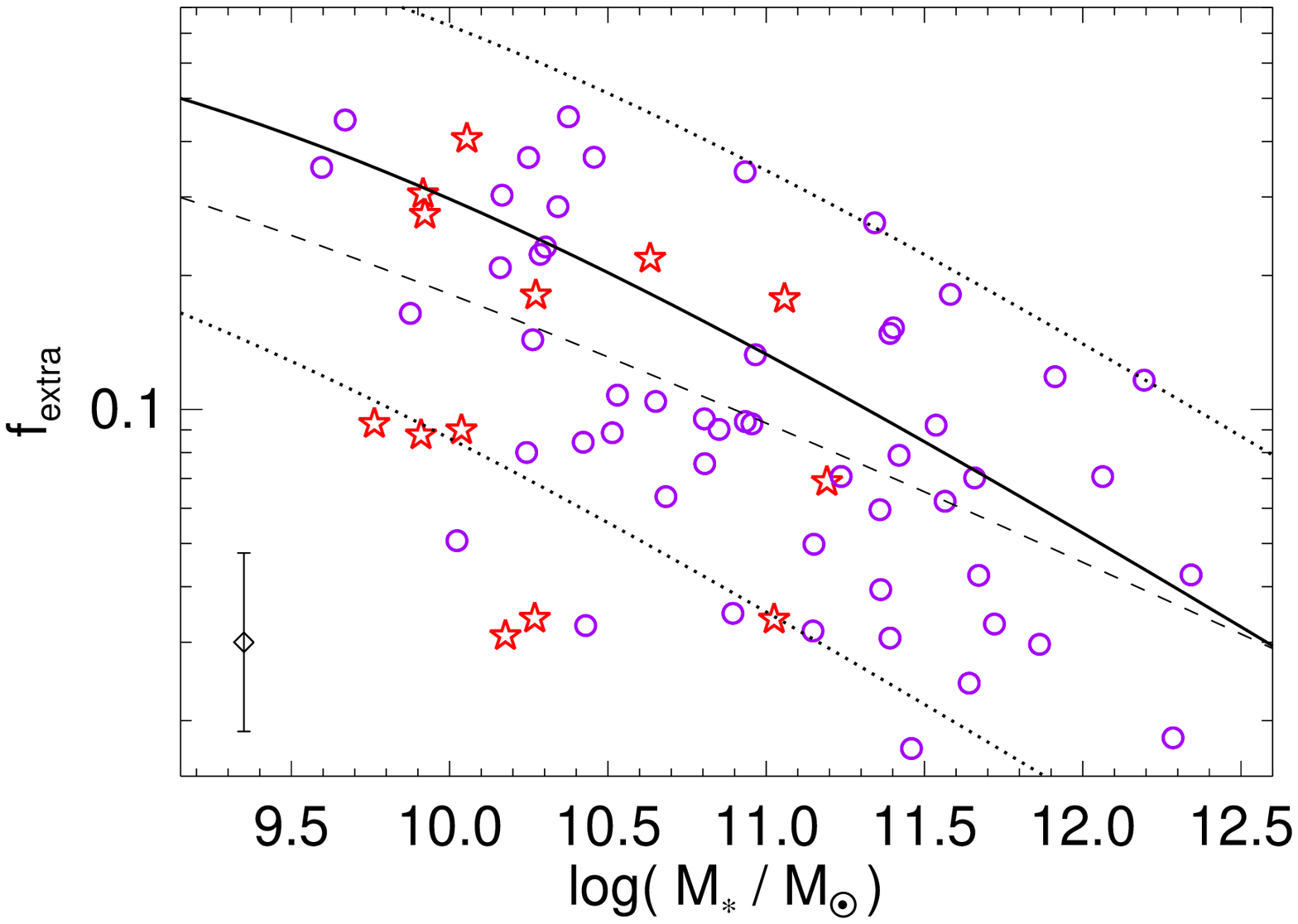}
    \caption{As Figure~\ref{fig:fgas.needed}, but showing 
    our empirically fitted $f_{\rm extra}$ as a function of 
    stellar mass. The trend of increasing dissipation 
    at lower masses is still clear and is consistent with 
    that in Figure~\ref{fig:fgas.needed}, but with an 
    expected extra factor $\sim2-3$ scatter 
    from the scatter in our purely empirical estimator. 
    Solid line shows the best-fit from Figure~\ref{fig:fgas.needed}, 
    dotted lines show the $\pm1\,\sigma$ scatter expected 
    based on the scatter in $f_{\rm extra}(f_{\rm sb})$ (see Figure~\ref{fig:check.fe.fsb}). 
    Dashed line is a fit just to these data (statistically consistent 
    with the solid line, and ruling out no dependence of dissipation on  
    mass at $>5\,\sigma$ confidence). 
    \label{fig:fgas.needed.fextra}}
\end{figure}

Admittedly, the estimation of this trend requires some comparison with our simulations, 
and one might argue that perhaps it is driven by some deficiency in them. However, 
we can repeat this exercise with the empirically fitted extra light component
$f_{\rm extra}$, and show the results in Figure~\ref{fig:fgas.needed.fextra}. 
The trend seen in $f_{\rm extra}(M_{\ast})$ is completely consistent with 
that in $f_{\rm sb}$, but with a scatter larger by a factor $\sim2$ (exactly 
what we expect, based on the predicted and observed scatter 
in $f_{\rm extra}(f_{\rm sb})$). Considering just the data in Figure~\ref{fig:fgas.needed.fextra}, 
even given its increased scatter, the trend of decreasing extra light fraction 
with mass is significant at $>5\,\sigma$. We have experimented with 
alternative, non-parametric (albeit less accurate) 
estimators based on e.g.\ the concentration 
indices or stellar populations 
of our simulations and observed systems, and obtain a similar answer. 
In short, even without reference to our simulations, 
however we derive an estimate of the
dissipational component, it is difficult to escape the conclusion that
it is more prominent in lower-mass ellipticals. This confirms 
a long-standing 
expectation of the merger hypothesis 
that if spirals are indeed the progenitors of ellipticals, more dissipation 
is required in lower mass systems in order to explain 
their densities and fundamental plane correlations 
\citep[we examine this in more detail in][]{hopkins:cusps.fp}.

This, in fact, should be expected. It is well-established that the gas
fractions of spirals are strongly decreasing functions of mass, at any
given redshift. To the extent that these are the progenitors of the
cusp ellipticals, then, the amount of dissipation involved in the
formation of ellipticals should reflect this trend. We therefore
compare in Figure~\ref{fig:fgas.needed} the range of observed gas
fractions of spirals as a function of baryonic mass, estimated at
$z=0$, $z=1$, and $z\sim2-3$. The gas fractions follow, at each
redshift, a similar trend to that we find for the 
dissipational fractions of the observed ellipticals. They do, as expected for
almost any reasonable cosmological history, increase systematically
with redshift, and the typical disk gas fractions at $z=0$ and
$z\sim2-3$ appear to roughly bracket the low and high end of the
dispersion in the inferred elliptical progenitor gas fractions.

In other words, the distribution in progenitor gas fractions implied
by the elliptical surface brightness profiles is, as a function of
mass, exactly what would be predicted if one assumes that the
progenitors were spirals, and that most of the systems were formed by
a major merger sometime between a redshift of $\sim0-3$
\footnote{Technically the post-merger 
elliptical mass is not exactly the mass of a single initial
spiral, but correcting for this amounts to a small horizontal shift of
the disk and elliptical points relative to one another in the figure
(i.e.\ the disk points should be shifted by $0.3$\,dex to larger
masses if all mergers are $1:1$, or $0.1$\,dex for more likely $3:1$
mergers), and does not change our comparison (in fact it makes the
agreement slightly better).}.  Indeed,
this is exactly what is inferred for the formation times of cusp
ellipticals from both observations of the early-type or red galaxy
mass functions \citep{bundy:mtrans,borch:mfs,fontana:highz.mfs,
hopkins:red.galaxies,hopkins:groups.ell},
from direct stellar population synthesis studies
\citep{trager:ages,thomas05:ages, gallazzi:ssps}, and 
by association of elliptical galaxy formation with the 
triggering of quasar activity
\citep[e.g.][]{hopkins:qso.all,hopkins:bol.qlf,
hopkins:groups.qso}.

\subsection{Size of the ``Extra Light'' Component}
\label{sec:properties.size}

\begin{figure}
    \centering
    \scaleup
    \plotter{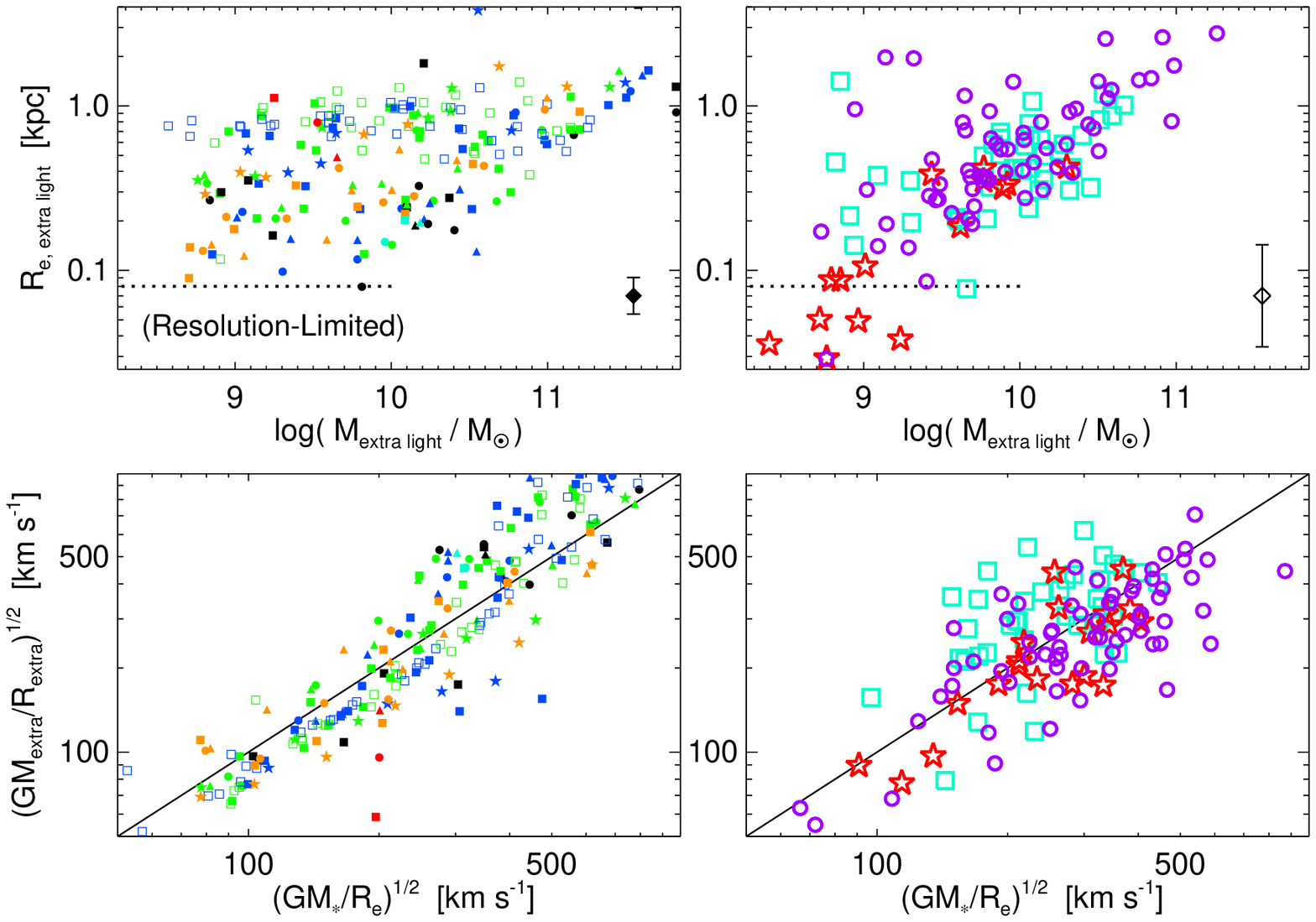}
    \caption{{\em Top:} Effective radius of the extra light component 
    ({\em not} equivalent to the radius where it breaks from the outer \Sersic\ fit) 
    as a function of extra light mass
    (points as in Figure~\ref{fig:ns.mass}). Simulations ({\em left}) and observations ({\em right}) 
    are similar, especially 
    if we restrict to simulations with initial gas fractions $\sim0.2-0.4$. Resolution limits 
    (see Appendix~\ref{sec:appendix:resolution}) prevent us from simulating systems with 
    $R_{\rm extra}\ll 100\,$pc, but this is only important for the few very lowest-mass 
    ellipticals ($L\lesssim 0.01\,L_{\ast}$; discussed in \S~\ref{sec:obs}).
    Filled diamond is typical sightline-to-sightline variance in the simulations, open 
    diamond the source-to-source (or band-to-band) scatter in observed profile fits. 
    {\em Bottom:} Effective velocity dispersion of the extra light component vs.\ that for the 
    whole galaxy. Solid line shows $(G\,M_{\rm extra}/R_{\rm extra})=(G\,M_{\ast}/R_{e})$ -- the 
    extra light collapses to the point where it is self-gravitating.
    The observed systems ({\em right}) follow a trend which agrees 
    well with the simulations ({\em left}).
    \label{fig:sizes}}
\end{figure}

The extra light components in simulations and  
cusp ellipticals also appear to follow a similar size-mass
relation, shown in Figure~\ref{fig:sizes}. The correspondence is
especially close if we consider simulations with initial gas fractions
$\sim0.2-0.4$, which tend to be the best analogs to the observed
systems. This radius is the effective radius of the fitted
inner extra light component, and is not the radius
at which the system appears to deviate from the outer \Sersic\ law (but
is more physically robust). We do not see simulations with 
extra light effective radii $\ll100\,$pc, corresponding to the smallest 
extra light components seen in the very low-mass observed 
systems, but as discussed in Appendix~\ref{sec:appendix:resolution}, 
this probably owes to our 
resolution limits. 

In \paperone\ we show that the size-mass relation is driven by the
condition that the gas collapsing into the central regions in the
final starburst becomes self-gravitating, i.e.\ that $(G\,M_{\rm
extra}/R_{\rm extra})\sim(G\,M_{\ast}/R_{e})$ in terms of the extra light
mass and effective radius. We show that the observations obey a
similar condition, with (small) scatter and dynamic range similar 
to that in our simulated mergers. That observed systems
follow a similar correlation suggests both that we are at least
roughly capturing the most relevant physics determining the scales of
extra light, and that we are not being severely biased by resolution
effects over most of the mass range of interest.

\breaker
\section{Impact of ``Extra Light'' on Galactic Structure}
\label{sec:structural.fx}

\subsection{Galaxy Sizes and the Fundamental Plane}
\label{sec:structural.fx.sizes}

\begin{figure*}
    \centering
    \scaleup
    \plotter{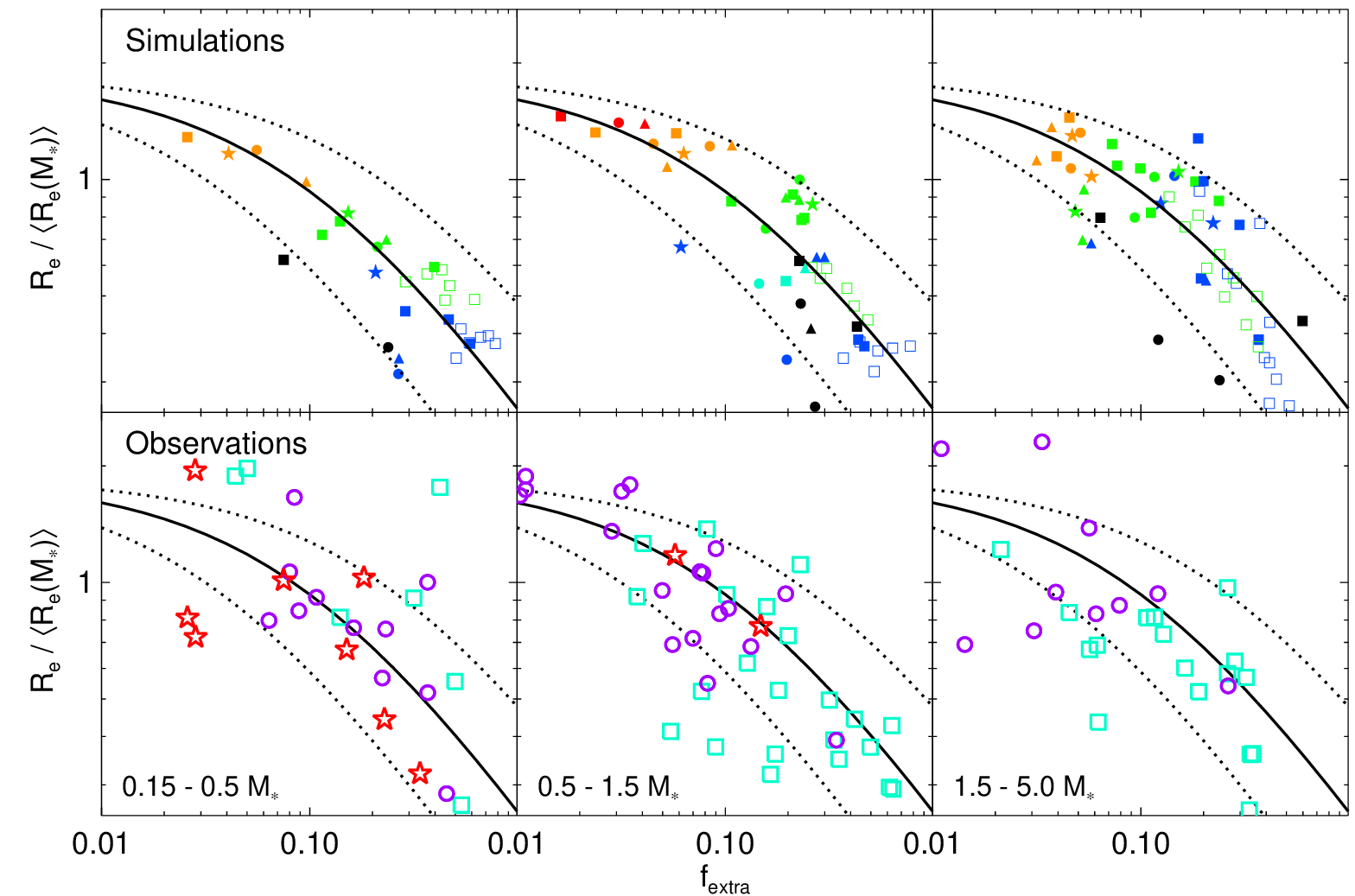}
    \caption{Effective radius $R_{e}$ relative to the 
    median value for all ellipticals of the same stellar mass, 
    as a function of our 
    fitted extra light fractions (the empirical 
    tracer of the dissipational/starburst mass fraction). 
    We compare simulated gas-rich 
    merger remnants ({\em top}) with 
    observed cusp ellipticals and gas-rich merger remnants 
    ({\em bottom}), with points as in Figure~\ref{fig:ns.mass}.
    We show this in three bins of stellar mass
    (relative to $\mstar\approx10^{11}\,\msun$, or $M_{V}^{\ast}=-21$).
    Solid (dotted) lines show the mean ($\pm1\,\sigma$) correlation, 
    following the analytic solution for dissipational mergers and 
    fits to our simulation in \citet{covington:diss.size.expectation}. 
    Simulations and observations exhibit the same 
    behavior: systems with smaller $R_{e}$ at fixed mass have 
    systematically larger extra light fractions ($>6\,\sigma$ significance 
    in the observations). 
    This implies that, at fixed mass, 
    systems are driven along the fundamental plane by the relative amount of 
    dissipation involved in their formation. 
    \label{fig:re.sigma.cusp}}
\end{figure*}

Figure~\ref{fig:re.sigma.cusp} shows how, at fixed mass, the effective radius 
(of the entire elliptical profile) scales with extra light mass. We
consider three mass bins, below, at, and above $\sim\mstar$. In each,
we plot $R_{e}$ relative to that expected
for the given stellar mass, as a function of the fitted extra light
fraction. Specifically, we determine $\langle{R_{e}(M_{\ast})}\rangle$
from the sample of \citet{shen:size.mass}, and take the ratio of 
the half mass radius of each system
(determined directly from the light profile,
or from the fits, it does not change the comparison) 
to that value.  Our mass bins are small enough,
however, that this makes little difference compared to just e.g.\
considering $R_{e}$ in a given bin. There is a strong trend: at a
given stellar mass, systems with larger extra light have
systematically smaller $R_{e}$
(they also have slightly larger velocity dispersion $\sigma$, although the 
scatter is larger there in both simulations and observations). 
In each case, the simulations and observed systems occupy a
similar locus. We can also construct this plot with the starburst mass
fraction $\fsb$ of the best-fitting simulation as the independent
variable, and find an even tighter correlation of the same nature.

This directly implies some structural 
change (some subtle non-homology, albeit not necessarily traditional 
structural or kinematic non-homology)
in the fundamental plane tilt. At
fixed mass, smaller systems are so because a larger fraction of their
mass is formed in a central dissipational starburst. 
This dissipational starburst is compact, so even though
the pre-existing stars are scattered to large radii, the effective
radius is smaller.

Given two progenitors of known size and mass, it is straightforward to 
predict the size of the remnant of a dissipationless merger, simply 
assuming energy conservation \citep[see e.g.][]{barnes:disk.halo.mergers}; 
in the case of a dissipative merger, we can very crudely model 
the results in Figure~\ref{fig:re.sigma.cusp} by assuming the 
non-extra light stars follow 
a \citet{hernquist:profile} profile with effective radius $R_{e}(\fsb=0)$, and 
the extra light is all at $r=0$. More accurately, 
\citet{covington:diss.size.expectation} use the impulse approximation to 
estimate the energy loss in the gaseous component, followed 
by collapse in a self-gravitating starburst. 
This yields a detailed approximation as a function 
of e.g.\ initial structural and orbital parameters, but if we assume typical progenitor 
disks and parabolic orbits, it reduces to the 
remarkably simple approximation
\begin{equation}
R_{e} \approx \frac{R_{e}(\fsb=0)}{1+(f_{\rm sb}/f_{0})}, 
\end{equation}
where $f_{0}\approx0.25-0.30$ and $R_{e}(\fsb=0)$ 
is the radius expected for a gas-free remnant. 
We plot this in Figure~\ref{fig:re.sigma.cusp}, with the scatter seen 
in the simulations. At all masses, in both simulated and observed 
cusp ellipticals, more dissipational ellipticals are smaller 
in the manner predicted. 
In the absence of dissipation, the stellar light of observed 
systems would follow a uniform virial relation, but 
dissipation results in smaller $R_{e}$ at fixed stellar mass, and (given the 
concentration of mass in this central starburst) therefore a higher 
baryon fraction inside $R_{e}$ -- i.e.\ changing the total $M/L$ 
(total dark matter plus stellar mass to stellar mass ratio) within 
$R_{e}$ of the stellar light \citep[see, e.g.][]{KormendyGebhardt01}.

\begin{figure}
    \centering
    \plotter{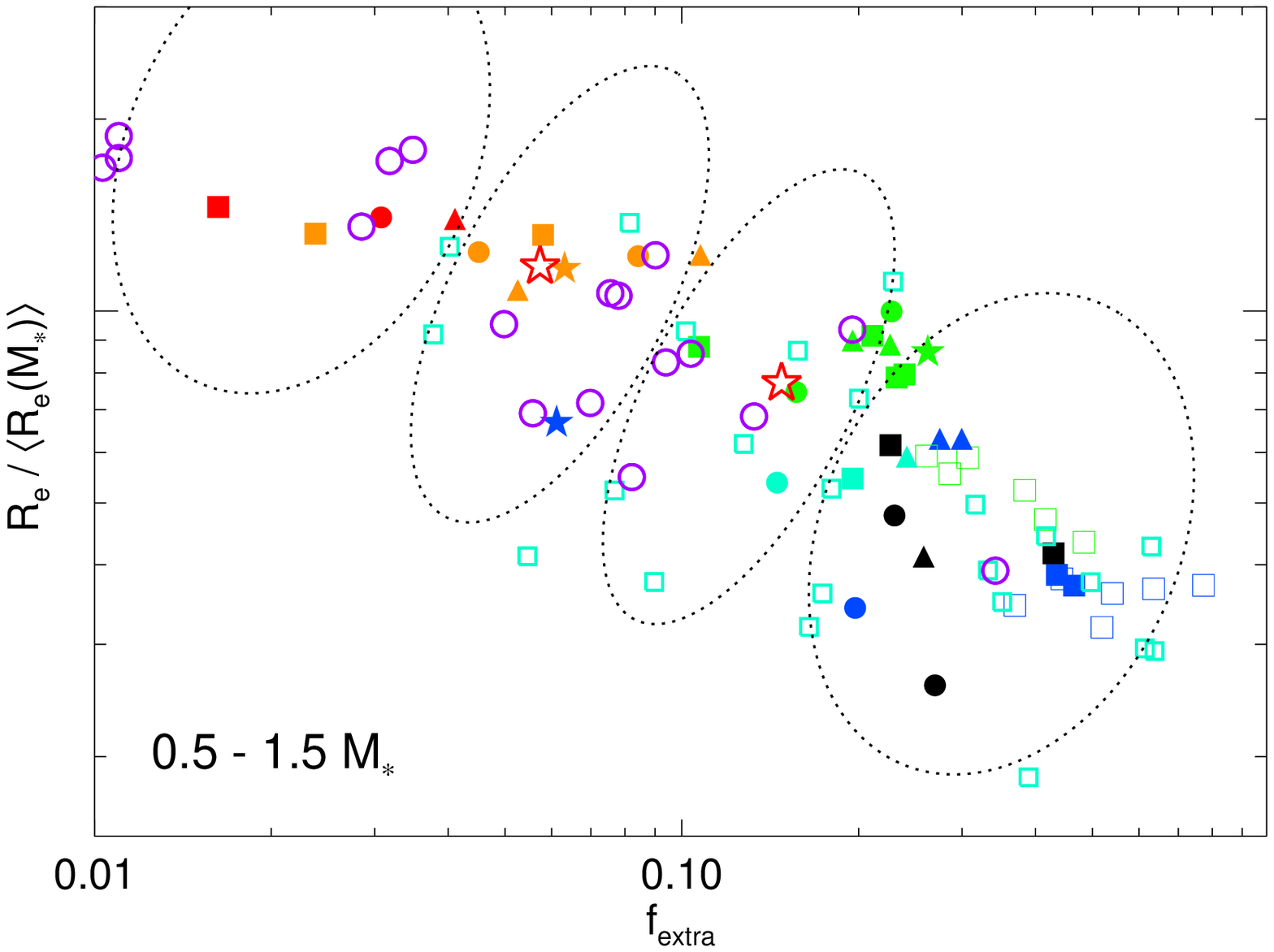}
    \caption{Center panels of Figure~\ref{fig:re.sigma.cusp}, with 
    four regions along the correlation between effective radius and extra 
    light fraction at fixed mass highlighted. The surface brightness profiles 
    in these regions are shown in Figures~\ref{fig:profile.vs.fsb.sims} \& \ref{fig:profile.vs.fsb}.
    \label{fig:re.sigma.cusp.zoom}}
\end{figure}
\breaker

\begin{figure}
    \centering
    \scaleup
    \plotter{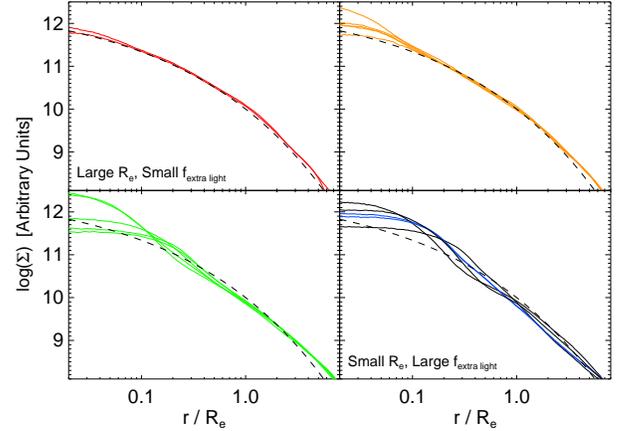}
    \caption{Light profiles of simulated systems with different $\re$ at fixed stellar mass. 
    We plot all profiles from our simulations (colors 
    denote simulation gas fractions as Figure~\ref{fig:ns.mass}) from each of the labeled regions 
    in the $\re-\fsb$ space for the $\sim\mstar$ galaxy mass bin in Figure~\ref{fig:re.sigma.cusp.zoom}, 
    in order from largest $\re$ (smallest $\fsb$) to smallest $\re$ (largest $\fsb$) (left to right, 
    top to bottom). Similar results are obtained for the other mass ranges. 
    The dashed black line shows a constant \Sersic\ profile, the same in each panel, for 
    comparison.
    There is a substantial systematic difference: smaller $\re$ systems at fixed stellar 
    mass have more prominent central mass concentrations, driven by dissipation in our 
    simulations. 
    \label{fig:profile.vs.fsb.sims}}
\end{figure}
\breaker

\begin{figure}
    \centering
    \scaleup
    \plotter{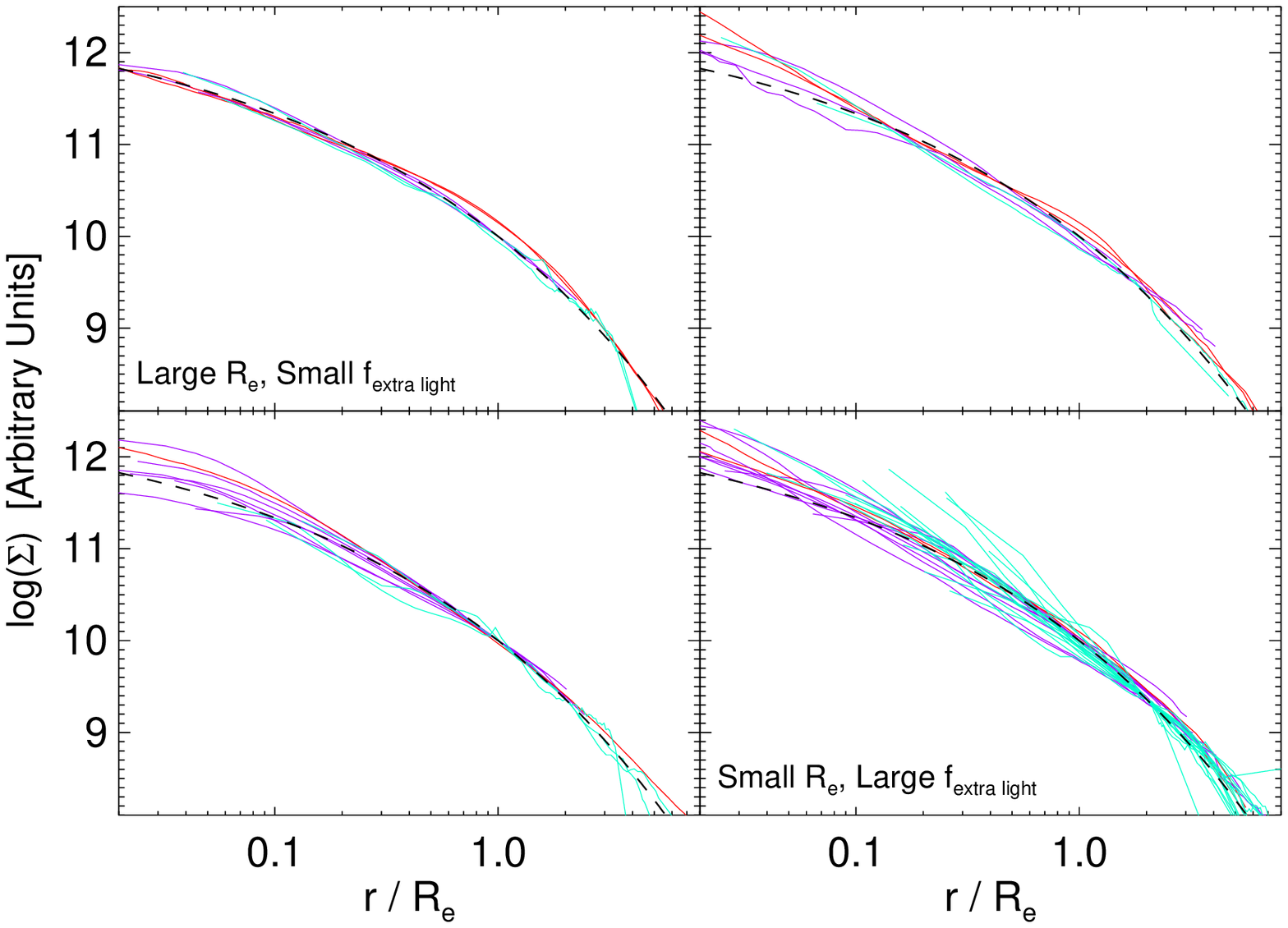}
    \caption{As Figure~\ref{fig:profile.vs.fsb}, but for the observed surface brightness profiles.
    Color denotes the observed sample as Figure~\ref{fig:ns.mass}.
    \label{fig:profile.vs.fsb}}
\end{figure}

We directly illustrate this homology-breaking in
Figures~\ref{fig:re.sigma.cusp.zoom}-\ref{fig:profile.vs.fsb}, 
by considering the light profiles of
systems in Figure~\ref{fig:re.sigma.cusp}, along the correlation
between $R_{e}$ and $f_{\rm extra}$ at fixed stellar mass. We consider the
$\sim\mstar$ mass bin (the other two give similar results, but have
fewer observed systems), and identify four regions in the $R_{e}-f_{\rm extra}$
space along the mean correlation, highlighted in Figure~\ref{fig:re.sigma.cusp.zoom}. 
We then plot the simulated (Figure~\ref{fig:profile.vs.fsb.sims}) and
observed (Figure~\ref{fig:profile.vs.fsb}) 
light profiles in each region. We also plot an $n_{s}=3$ \Sersic\
profile, which provides a good fit (in the mean) to most of the outer
\Sersic\ profiles in this mass range.

It is clear in Figure~\ref{fig:profile.vs.fsb} that both simulated and observed 
systems at fixed mass with the largest $R_{e}$ show profiles close to a 
pure \Sersic\ law, with little evidence for any extra light component in their centers 
(indeed, they have $\fsb\lesssim0.03$). The observed systems in this regime 
are still cusps, but the cusps tend to be prominent at very small radii 
and (in several cases) somewhat shallow, and contribute negligibly to the 
stellar mass. However, moving to smaller $R_{e}$, 
deviations from a \Sersic\ law at $r\ll\re$ become more prominent. That is not to say 
that these deviations are universal (that the extra light always takes the same 
shape/form), but there are increasingly prominent central 
light concentrations. If the systems were strictly homologous, there should be 
no differences in Figure~\ref{fig:profile.vs.fsb}. Since we have scaled 
each system at its $\re$, they should be identical -- instead, it appears that
a central light concentration, formed in our simulations via gas dissipation,
drives an important non-homology.

We discuss this in \citet{hopkins:cusps.fp}, where we show 
that this is sufficient to explain the tilt of the 
(stellar mass) fundamental plane. In that work, we 
demonstrate that the fitting of the central component in the light profile 
can be used as a direct observational test for the role of dissipation in 
the fundamental plane. Given a larger central light component, the 
effective radius of the stellar light profile must be smaller. Because the 
central regions of the galaxy are the most baryon-dominated, moving the 
effective radius inwards results in a different (larger) ratio of stellar to 
dynamical mass, i.e.\ driving tilt in the fundamental plane. We note 
that although this is technically non-homology (i.e.\ the profiles 
are not perfectly self-similar), it does not drive tilt in the sense 
of traditional structural or kinematic non-homology; i.e.\ the 
practical ``homology assumption'' that the true mass enclosed 
in $R_{e}$ is proportional to the dynamical mass estimator $\sigma^{2}\,R_{e}/G$ 
remains true. Rather, the non-homology induced is a subtle effect 
that provides a tracer of dissipation, which changes the physical 
ratio of baryonic to dark matter within $R_{e}$.

\subsection{Galaxy Shapes and Kinematics}
\label{sec:structural.fx.shapes}

\begin{figure*}
    \centering
    \scaleup
    \plotterr{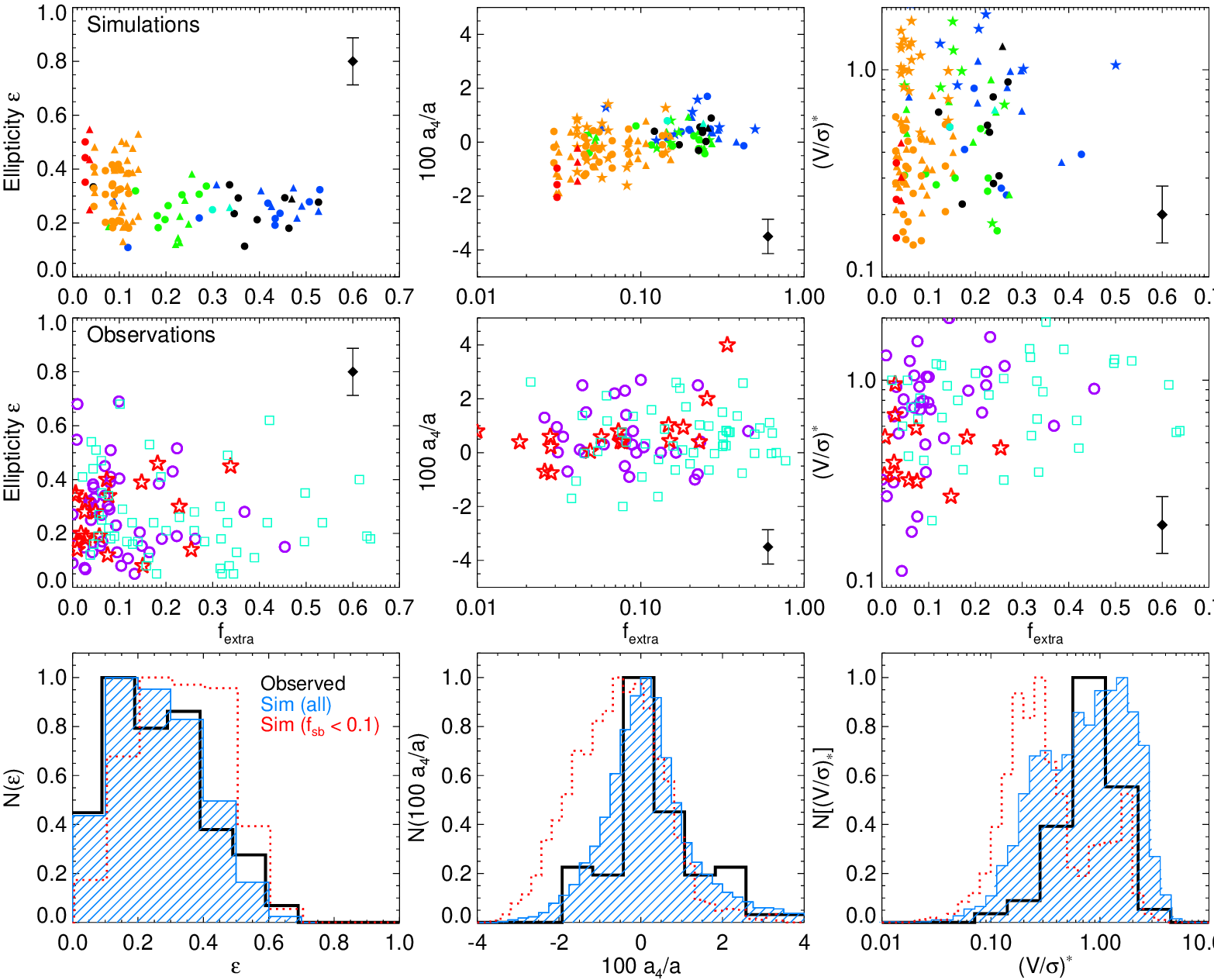}
    \caption{{\em Top:} 
    Correlation between fitted extra light fraction (in both simulations and observed 
    cuspy ellipticals, plotted as in Figure~\ref{fig:ns.mass}) and 
    global kinematic properties of the galaxy: rotation $(V/\sigma)^{\ast}$, boxy/disky-ness 
    $100\,a_{4}/a$, and ellipticity $\epsilon$. Median values across 
    sightlines are plotted. Black points with error bars show the 
    typical $\sim1\sigma$ sightline-to-sightline dispersion in each quantity 
    from our simulations. 
    {\em Bottom:} Histograms show the distribution in each property, 
    for observed systems (thick black line), dissipational simulations 
    which have similar moderate extra light fractions (blue shaded),  
    and gas-poor (nearly dissipationless) simulations ($\fsb<0.1$; red dotted). 
    The simulation distributions uniformly 
    sample each simulation in solid angle over $\sim100$ lines-of-sight (including the 
    large sightline-to-sightline dispersion in the top panels). 
    Broadly, 
    cuspy ellipticals and gas-rich merger remnants occupy a wider range in these parameters at 
    low $\fsb$, but are uniformly rapid rotators, with slightly diskier isophotal shapes, and 
    slightly rounder ellipticities at large $\fsb\gtrsim0.1$. 
    For more details, see \citet{cox:kinematics}. 
    \label{fig:cusp.vs.kinematics}}
\end{figure*}

Figures~\ref{fig:cusp.vs.kinematics} shows how the shapes and 
global kinematic properties vary with extra light. We plot 
the rotation, isophotal shapes, and ellipticity of our simulations and 
the observed systems as a function of the fitted extra light fraction. 
Here, we consider the global properties of the galaxy -- namely the 
rotation $(V/\sigma)^{\ast}$ measured within $\re$, and 
boxy/disky-ness $100\,a_{4}/a$, and 
ellipticity $\epsilon$ measured for the half-mass projected ($\re$) isophotal 
contour, for each of $\sim100$ lines-of-sight to the remnant uniformly 
sampling the unit sphere (i.e.\ representing the distribution across random 
viewing angles). The details of the fitting for our simulations are described in 
\citet{cox:kinematics}, and for the observations in \citet{bender:87.a4,
bender:88.shapes,bender:ell.kinematics,faber:catalogue,
simien:kinematics.1,simien:kinematics.6}. 
We define rotation in the standard manner, relative to that 
of an oblate isotropic rotator, with the parameter $(V/\sigma)^{\ast}$, 
defined using the maximum circular velocity $V_{\rm c}$, 
velocity dispersion within $\re$, and ellipticity as \citep{kormendy:rotation.equation} 
\begin{equation}
(V/\sigma)^{\ast} = (V/\sigma) / \sqrt{\epsilon/(1-\epsilon)}.
\end{equation}
We exclude the coplanar merger simulations 
from our comparisons here: those 
simulations are idealized perfectly coplanar prograde orbits, and as such 
produce pathological orbit and phase structure (we do, however, include some  
more representative orbits in that are not far from coplanar). 

In each sense, the simulations and observations occupy a similar
locus.  While there are no real correlations, there some broad differences 
between the distributions for systems with significant 
or insignificant extra 
light components. Systems with small extra light fractions ($\fsb\lesssim0.1$)
have a wide range of $(V/\sigma)^{\ast}$, $a_{4}/a$, and $\epsilon$ --
they range from relatively slow rotation \citep[albeit not as slow as
many core elliptical slow rotators; see][]{emsellem:sauron.rotation}
to rotationally supported objects $(V/\sigma)^{\ast}\gtrsim1$, with
both boxy and disky isophotal shapes ($-2\lesssim 100\,a_{4}/a
\lesssim 2$), and a range in ellipticity from fairly spherical to
highly flattened $\epsilon\sim0.5$.  Despite in some instances having e.g.\
slow rotation properties, these systems do not follow other
trends of massive, cored slow rotators -- they usually are not
simultaneously slowly rotating, boxy, and round. This is similar to 
several observed systems, which generally show evidence for
formation in a gas-rich merger but owing to sightline effects may
be deviant in one of these properties. 

At high extra light fractions, however, both our simulated systems and
those observed tend to be more rapid rotators
(fractionally more populating 
$(V/\sigma)^{\ast}\gtrsim 0.4$), with less boxy isophotal shapes
(typical $-0.5\lesssim 100\,a_{4}/a \lesssim2$), and slightly rounder
ellipticities ($\epsilon\lesssim0.4$; note they are 
still more elliptical than core galaxies, in agreement with observed trends). 
A more detailed analysis of the
role of central mass concentrations and dissipation in shaping the
orbital structure of gas-rich merger remnants will be the focus of
\citet{hoffman:prep}. Briefly, however, these differences are
physically expected, and follow what has been 
seen in earlier work \citep[e.g.][]{barneshernquist96,naab:gas,
cox:kinematics,onorbe:diss.fp.details,
jesseit:kinematics,burkert:anisotropy}: 
the central dissipational mass concentration in
these systems is highly concentrated, and acts effectively like a
point mass at the center of the galaxy to much of the material at
$R_{e}$. When this becomes a large fraction of the galaxy mass
($\fsb\gtrsim0.1$), the potential becomes more spherical, which
eliminates some of the phase space density that might otherwise
support boxy and radial orbits
\citep{binney:box.orbits,hernquist:phasespace,naab:gas,hoffman:prep},
and the central cusp can itself disrupt box orbits as they pass near
the center \citep{gerhard:box.orbits.w.cusp}.  This results in a
larger fraction of rotationally supported tube orbits, diskier
isophotal shapes, and (directly) a rounder overall system.

If we consider simulations with sufficiently large dissipational fractions
($\fsb\gtrsim0.1$, similar to what we see in the observed systems), then, 
our gas-rich merger remnant 
simulations match well each of the distributions of $\epsilon$, $a_{4}/a$ and 
$(V/\sigma)^{\ast}$ for cuspy ellipticals. 
This is demonstrated in much greater detail in \citet{cox:kinematics}. 
Here, we highlight 
that there is a significant difference in many of these 
properties between cusp and core ellipticals, as has been well 
established in previous studies \citep[e.g.][]{bender89,faber:ell.centers}. 
It is clear that our gas-rich merger simulations 
provide a good match to the observed distributions of 
$(V/\sigma)^{\ast}$, $a_{4}/a$, and $\epsilon$ in 
cuspy ellipticals, as we would expect, but they are not reproducing the more 
slowly rotating, boxy, round distributions characteristic of core ellipticals. 
This should also be borne in mind considering Figure~\ref{fig:cusp.vs.kinematics}: 
the lack of real correlations with extra light fraction appears to be true 
{\em within the cuspy population}; if core ellipticals are altered through
subsequent spheroid-spheroid ``dry'' mergers, 
these properties can be modified 
\citep{naab:dry.mergers,cox:remerger.kinematics,burkert:anisotropy}, 
and it is expected that these processes will bring the predicted distributions 
into better agreement with those observed in the cored population.

\subsection{Nuclear Black Hole Masses}
\label{sec:structural.fx.mbh}

\begin{figure}
    \centering
    \scaleup
    \plotter{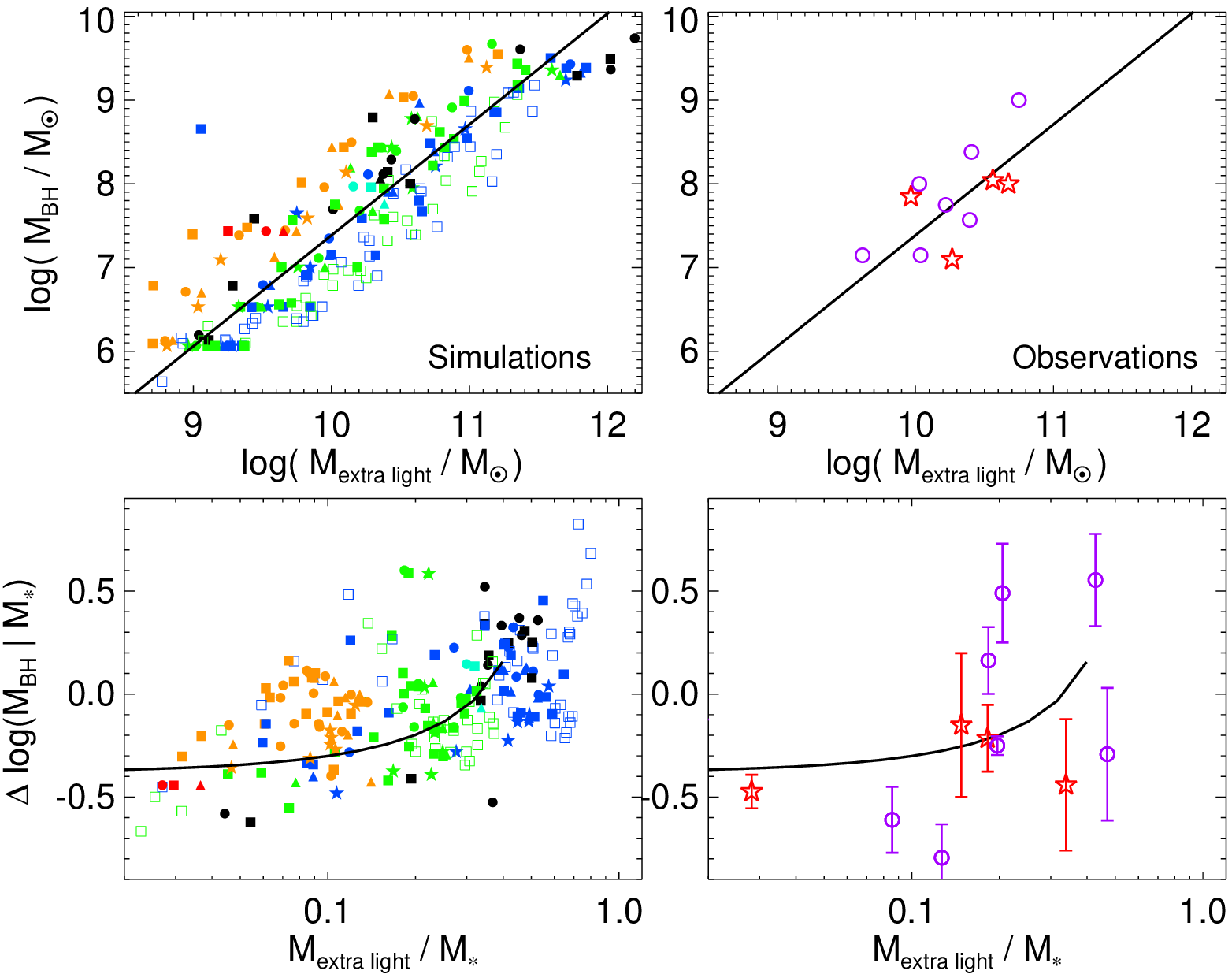}
    \caption{{\em Top:} BH mass as a function of fitted extra light mass, 
    in our simulations ({\em left}) and observed systems that have direct BH mass 
    measurements ({\em right}; only a small fraction of our sample have such 
    BH masses measured). {\em Bottom:} Residual correlation of BH mass with 
    extra light mass at fixed stellar mass (i.e.\ $M_{\rm BH}/\langle M_{\rm BH}(M_{\ast})\rangle$ 
    versus $M_{\rm extra}/M_{\ast}$). Solid line is the expectation if 
    systems obey the correlation between $\fsb$ and $\re$ from Figure~\ref{fig:re.sigma.cusp} and 
    the black hole fundamental plane \citep{hopkins:bhfp.obs}. 
    There is a significant $M_{\rm BH}-M_{\rm extra}$ correlation, but it largely 
    reflects the $M_{\rm BH}-M_{\ast}$ correlation or BHFP. 
    At fixed stellar mass, the (weak) residual trend 
    comes from the BHFP as increased dissipation builds a more compact remnant 
    with deeper potential, and therefore larger BH. 
    \label{fig:bh}}
\end{figure}

Figure~\ref{fig:bh} shows 
how the BH masses scale with extra light mass, both globally and 
at fixed stellar mass. 
We plot all of our simulations which include central BHs, and include the cusp 
ellipticals in our sample for which direct kinematic or maser measurements of a 
central BH have been possible
\citep[compiled from][]{magorrian,merrittferrarese:msigma,tremaine:msigma,
marconihunt,haringrix,aller:mbh.esph}. 
For more details, we refer to \citet{hopkins:bhfp.obs}.
There is a correlation in simulations and observations between 
BH mass and extra light mass. However, this appears to be largely driven by the 
correlation between BH mass and total spheroid stellar mass, which has 
smaller dispersion and a weaker residual dependence on 
e.g.\ gas fraction or orbital properties. This is expected: if BH mass is 
actually sensitive to the 
depth of the central potential and spheroid 
binding energy, as argued from the nature of the fundamental plane-like 
correlation for BH masses and host properties demonstrated in 
\citet{hopkins:bhfp.theory,hopkins:bhfp.obs} 
and \citet{aller:mbh.esph}, then this is better correlated with the total 
stellar mass setting the potential than the few percent of the mass in the extra light component. 

We might expect though, that at fixed total stellar mass, systems with
larger extra light components, since this formed a more compact
remnant, would have somewhat deeper potentials. Given the black hole
fundamental plane (BHFP) in terms of stellar mass and $\re$ (i.e.\
scaling of BH mass with $\re$ or, equivalently, $\sigma$ at fixed
galaxy stellar mass, as in \citet{hopkins:bhfp.theory,hopkins:bhfp.obs}),
and the scaling of $\re$ with extra light
fraction at fixed mass seen in Figure~\ref{fig:re.sigma.cusp}, we can
estimate the dependence of BH mass on extra light fraction at fixed
stellar mass. Figure~\ref{fig:bh} shows the residual of BH mass (i.e.\
BH mass relative to that expected at each stellar mass) as a function
of extra light fraction, compared to this expectation. There is
considerable scatter, but the simulations and few observed systems
broadly follow the expected trend -- namely that the cusp mass does
not primarily drive $\mbh$, but that at fixed stellar mass, increasing
dissipation builds more compact remnants which have higher binding
energies and therefore, via the BHFP, larger BH masses.

\breaker
\section{Impact of ``Extra Light'' on Stellar Populations}
\label{sec:ssp.fx}

If a significant fraction of gas dissipates to produce a central mass
concentration, and this forms stars in a rapid central starburst, we
should expect this to leave an imprint on stellar population
gradients in the remnant \citep[e.g.][]{mihos:gradients}.

\begin{figure*}
    \centering
    \plotter{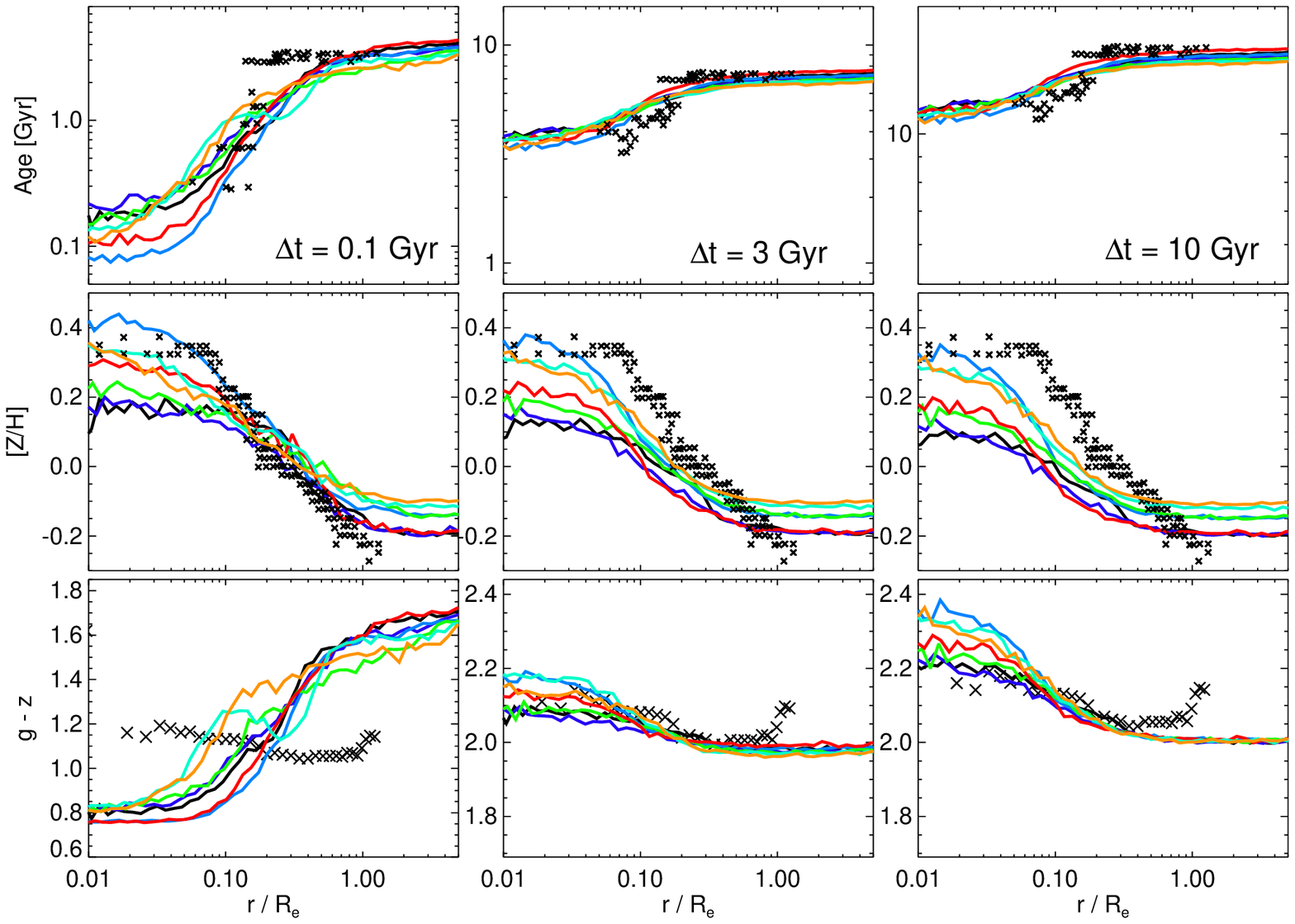}
    \caption{Comparison of the observed stellar population 
    and color gradients (points) of NGC 3377 with those in the $\sim10$ simulations (lines) which 
    most closely match its surface brightness profile (see Figure~\ref{fig:jk2}). 
    The $B$-band light weighted parameters are plotted at three different times 
    after the merger (labeled), with $\Delta t=3\,$Gyr corresponding roughly to the 
    observed mean stellar population ages in NGC 3377 (points there are plotted 
    exactly as observed). At $\Delta t = 0.1$ and $10\,$Gyr, in contrast, we have 
    added the mean age offsets ($-2.9\,$Gyr and $+7$\,Gyr), and shifted the colors 
    by the mean difference expected for passive evolution of a single stellar 
    population with solar metallicity and the quoted age; this highlights the change in 
    shape of the simulation gradients, as well as their normalizations. 
    The simulated age and metallicity gradients fade slightly with time, as the 
    difference in stellar populations becomes less prominent. At all times, however, the 
    gradients are primarily driven by the mean difference in age and metallicity between 
    the compact central starburst populations and the violently relaxed pre-merger populations. 
    At early times, young, 
    starburst populations often yield blue cores and a rising color gradient; 
    at late times, the age effect weakens, and the enhanced metallicity of the 
    starburst population dominates, reversing the color gradient (yielding 
    red cores). 
    \label{fig:grad.case.study}}
\end{figure*}

Figure~\ref{fig:grad.case.study} presents an illustrative example of
these effects.  We consider the elliptical NGC 3377, for which
detailed stellar population gradients have been measured by
\citet{sanchezblazquez:ssp.gradients}. The system is a clear cusp
galaxy and rapid rotator with disky isophotal shapes, and is included
in our \citet{jk:profiles} sample (Figure~\ref{fig:jk2}).  From our
comparison in Figure~\ref{fig:jk2}, we select the $\sim10$ simulations
with the most similar surface brightness profiles. These generally
fall within the range of those plotted in the figure, and all provide
reasonably good fits to the observed surface brightness profile. For
each simulation, we then extract the stellar populations as a function of
radius, viewed from the same projection as that which provides the
most similar surface brightness profile to NGC 3377 (although 
the sightline-to-sightline variation is weak).

We model the emission from each star particle treating it as a single
stellar population with the formation time and metallicity determined
self-consistently from the star-forming gas in our simulations, and
convolving with the stellar population models of \citet{BC03},
assuming a \citet{chabrier:imf} initial mass function.  The observed
stellar population parameters are effectively light-weighted in the
optical SED, so as a rough proxy for this we extract, in each
major-axis radial annulus, the $B$-band light weighted stellar
population age and metallicity from the simulation. The $g-z$ color is
determined directly in each annulus.  

Of course, doing this requires that we pick a definite time after the
merger to view a simulation. We therefore show results for three
representative times.  First, just $\sim10^{8}$\,yr after the final
coalescence of the two galactic nuclei, when the object would likely
be classified as a recent merger remnant. Second, $\sim3$\,Gyr after
the final merger, which is representative of younger $\sim
L_{\ast}$ cusp ellipticals. Specifically, we choose this time because
the total light-weighted age, integrated over the whole galaxy, at
this point matches that observationally inferred for NGC 3377 -- i.e.\
when our comparisons are most appropriate. Third, we show the
remnant after $\sim10\,$Gyr, i.e.\ having evolved in isolation for 
nearly a
Hubble time, comparable to the oldest observed ellipticals.  At
$t=3$\,Gyr, we directly plot the observed age, metallicity, and color
profiles from \citet{sanchezblazquez:ssp.gradients}. At the other
times, we make the lowest-order reasonable corrections to
highlight relative evolution in the shapes of the profiles: we add or
subtract the appropriate age difference uniformly from the age
profile, and likewise add or subtract the mean color difference
expected for a single stellar population of the mean observed age and
luminosity.  Again, these leave the profile shapes unchanged -- we
merely shift them by the expected mean to compare with the shape
changes predicted by our simulations.

At early times ($t=0.1\,$Gyr), the gradients are strong. The stars
formed in the central starburst are very young, and thus dominate the
central light. Unsurprisingly, then, the age at the center of the
galaxy is approximately just the time since the merger,
$\sim0.1\,$Gyr. This rises to $\sim3-5$\,Gyr in the outermost regions,
representative of the ages of the stars that were forming before the
merger proper began.  The central population is also the most metal
rich, producing a similar strong metallicity gradient. The young, blue
stellar populations of the center result in a strong color gradient
with a blue core, common among young merger remnants
\citep[e.g.][]{schweizerseitzer92,schweizer96,rj:profiles,yamauchi:ea.gradients}.

The gradients in these quantities are most pronounced
between $\sim0.05-0.1\,R_{e}$ and $\sim R_{e}$: at much smaller or
larger radii they tend to flatten. This intermediate region is the
transition between dominance of the pre-merger stellar populations and
central cusp -- the gradients are primarily driven by the mean
difference in stellar population parameters between the central
starburst and outer old stellar components. This, in
principle, allows arbitrarily strong gradients at the radii typically
observed, but at larger radii, for example, violent relaxation mixes
the old stars, washing out initial gradients and leading to a
flattening in the total gradient. As predicted, these trends are
typical in observed stellar population gradients
\citep[e.g.][]{mehlert:ssp.gradients,
sanchezblazquez:ssp.gradients,reda:ssp.gradients}.

By $\sim3\,$Gyr after the merger, the age and metallicity gradients
have weakened slightly. The apparent age gradient,
quantified as e.g.\ ${\rm d}\log{({\rm age})}/{\rm d}\log{(r)}$ is
much weaker at this time, but largely for artificial reasons -- adding
a uniform $\sim3$\,Gyr to the age of the system accounts for most of
this, since it makes the difference in $\log{(\rm age)}$ smaller.  In
terms of ${\rm d}{\rm (age)}/{\rm d}\log{(r)}$, the effects are more
subtle, comparable to what is seen in the metallicity gradient. The
gradients do, at this level, weaken slightly.  This is because the now
older central stellar populations have a lower $L/M$, more comparable
to the old stars with which they are mixed. There is therefore
slightly more mixing between the pre-merger and starburst
populations. This can be seen in the metallicity gradients, which we
would expect (to lowest order) to remain constant with time. Still,
the effect is clearly second-order. These trends continue with time,
as can be seen at the time $\sim10\,$Gyr after the merger.

The color gradients evolve significantly with time, however. The inner regions are younger, 
which at early times (when they are very young, $\lesssim0.5$\,Gyr) typically 
results in blue cores and a color gradient which becomes redder at larger radii 
(note, however, that at times very close to the merger, dust can reverse this trend). 
However, they are also more metal-rich, 
which tends to make them redder. After $\sim1\,$Gyr, the effects of the age difference 
are much less prominent, and the color gradient becomes dominated by the 
metallicity gradient, resulting in {\em red} cores, and reversing the 
sense of the gradient. At even later times, the strength of this gradient (towards redder 
central regions) becomes stronger, although it remains relatively weak. 
We therefore expect that most ellipticals, even those which are relatively young, 
will have weak color gradients with red cores, as observed \citep{faber:catalogue,
bender:ell.kinematics.a4,trager:ages,cote:virgo,ferrarese:profiles}. 
This should caution strongly against inferring too much from an observed color 
gradient, since the physical meanings and sense of the color gradients in 
typical simulations not only quantitatively depend on time and relative 
age, metallicity, and abundance gradients, but in fact usually reverse 
their meaning and behavior with time. 

Because the star formation history depends on spatial location, 
mergers and dissipation can also induce gradients in e.g.\ the 
chemical abundance patterns and $\alpha$-enrichment of ellipticals. However, we 
do not show these explicitly for two reasons.  First, our simulation 
code does not currently track separate enrichment by different species, so 
our estimates of such are based on crude analytic estimates (taking e.g.\ the 
star formation history in radial bins and estimating the 
$\alpha$-enrichment based on a closed-box model). Second, and more 
important, the effects are much more sensitive to the initial 
conditions and cosmological merger history. For example, one can 
imagine a situation where stars form over a 
fairly extended time period in the initial disks, leading to relatively high 
$[{\rm Fe/Mg}]$ at large radii. The merger then forms a rapid starburst in the 
center, which will preferentially be $\alpha$-enriched but not have time 
to self-enrich in heavier elements, leading to a decreasing $\alpha$-enrichment 
gradient. However, to the extent that {\em most} of the metal content of the 
central starburst often comes from the previous history of enrichment of 
that gas in the pre-merger disks (i.e.\ gas that enriched over long timescales), 
the additional effect of the starburst will be quite weak, leading to 
gradients of only $\sim0.1$ in $[{\rm Fe/Mg}]$ over $\sim0.01-10\,R_{e}$. 

Furthermore, one can also imagine a scenario in which the progenitor 
disks rapidly exhaust their gas in a short timescale leading to 
high $\alpha$-enrichment, retaining 
a relatively poor gas reservoir for the final merger. Because this gas is 
retained over long time periods, it enriches in ${\rm Fe}$-elements from 
the older disk stellar populations, 
and after forming stars in the central starburst creates a gradient with 
the opposite sense (i.e.\ decreasing $\alpha$-enrichment with radius). 
Again, our estimates suggest the effect will be weak.

In either case, 
because the magnitude of the induced $\alpha$-enrichment gradients 
are comparable to those observed in spiral galaxies 
\citep[e.g.][]{moorthy:spiral.ssp.grad,ganda:spiral.ssp.grad}, 
the final prediction is much more sensitive to the initial 
conditions (unlike e.g.\ the induced total metallicity gradients, which 
are typically stronger than those in comparable disks). 
The mean or total $\alpha$-enrichment, on the other hand, depends of 
course primarily on the timescale for the formation of most of the 
stellar mass -- which in almost all the cases of interest here means the 
pre-merger disks, which we are not attempting to model. 
Further complicating matters, cosmological infall or accretion of 
new gas and minor mergers will play a (fractionally) larger role. 
For these reasons, a more comprehensive set of models 
is needed to say anything quantitative about 
$\alpha$-enrichment gradients, but it is 
nevertheless reassuring that 
the magnitude and sense (both positive and negative) of these 
possible induced gradients are in fact qualitatively similar to 
what is observed \citep[e.g.][]{reda:ssp.gradients,sanchezblazquez:ssp.gradients}. 

\begin{figure*}
    \centering
    \plotter{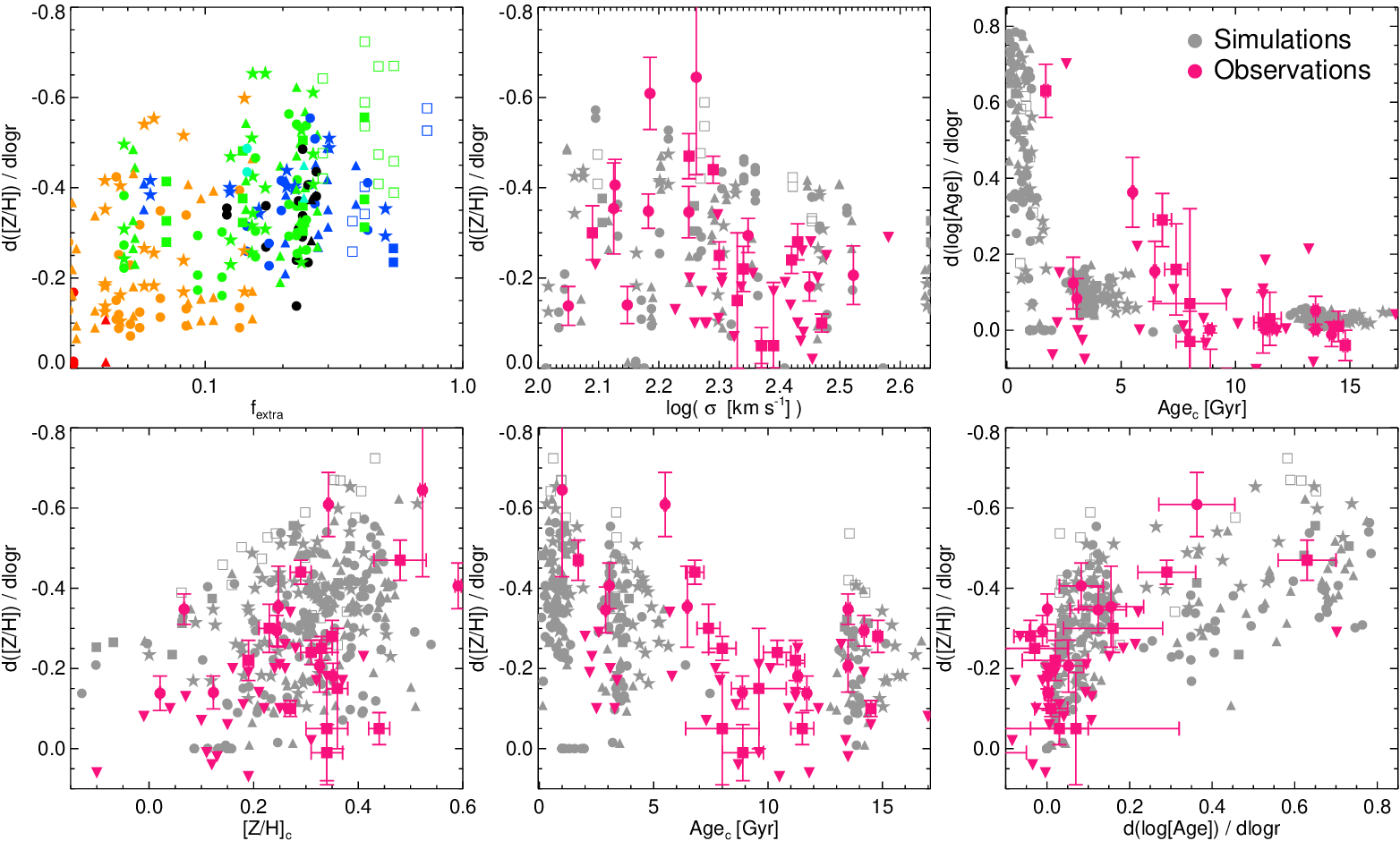}
    \caption{{\em Top Left:} Correlation between metallicity gradient 
    (${\rm d[Z/H]/d}\log{r}$)
    and extra light fraction ($f_{\rm extra}$) 
    in our simulations (points as in Figure~\ref{fig:ns.mass}). 
    {\em Remaining Panels:} Various correlations 
    between metallicity gradients (${\rm d[Z/H]/d}\log{r}$), 
    age gradients (${\rm d\log{(Age)}/d}\log{r}$), 
    central velocity dispersion $\sigma$, 
    central light-weighted stellar population age (${\rm Age}_{\rm c}$), 
    and central light-weighted stellar population metallicity 
    (${\rm [Z/H]}_{\rm c}$). For clarity, the simulations are plotted in these 
    panels as light grey points, and compared with observed ellipticals 
    (magenta points; triangles, squares, and diamonds are from the samples of 
    \citet{mehlert:ssp.gradients,reda:ssp.gradients,sanchezblazquez:ssp.gradients}, respectively). 
    The correlation between gradient strength and the excess light fraction 
    ({\em top left}) reflects the fact that the central light drives most of the gradients. 
    This gives rise to the associated correlations shown in other panels. 
    In each case, the observations occupy a similar 
    locus to the simulations. 
    \label{fig:grad.compare}}
\end{figure*}

Figure~\ref{fig:grad.case.study} is instructive, but it is difficult to compare 
with all systems. We therefore globally compare gradient strengths by fitting the 
mean gradient in metallicity and age over the range 
$\sim0.1-1\,R_{e}$ in each of our simulations, comparable to the 
range typically used observationally \citep[e.g.][and references therein]{reda:ssp.gradients}. 
We consider the results extracted 
at a series of times after the merger similar to those in Figure~\ref{fig:grad.case.study}. 

Figure~\ref{fig:grad.compare} shows how these gradient strengths 
compare to those in observed ellipticals \citep[from][]{mehlert:ssp.gradients,
reda:ssp.gradients,sanchezblazquez:ssp.gradients}. We 
consider gradient strength as a function of 
extra light fraction, velocity dispersion, central stellar population age 
(averaged within an aperture of $R_{e}/8$, again comparable to typical 
observations), and central metallicity. 
There is a noticeable correlation between gradient strength and the excess light fraction; 
we expect this, since as noted above this central light drives most of the gradients. 
This also gives rise to a correlation between 
e.g.\ metallicity gradient and central metallicity, age gradient and central age, and 
different gradients themselves (stronger metallicity gradients tend to 
accompany stronger age gradients). There is not a strong dependence of 
gradient on mass, however (what there is is mostly driven by the mean dependence of 
extra light fraction on stellar mass). In each case, the observations occupy a similar 
locus to the simulations. 

There have been some claims that the metallicities
\citep{naab:no.merger.ell} and metallicity gradients
\citep{forbes:grad.vs.mass} in ellipticals are too large/strong for
them to be formed from mergers of present-day spirals. However, these
arguments fail to account for the role of dissipation in increasing
the central metallicity and giving rise to strong metallicity
gradients in merger remnants.  The left panels of
Figure~\ref{fig:grad.compare} demonstrate that, for moderate
gas fractions, our simulations occupy a similar locus in
both central metallicity and metallicity gradients to observed
ellipticals. The metallicities typically
quoted in stellar population studies are {\em central} metallicities,
measured within $\sim R_{e}/8$ (or within a central fiber in automated
surveys such as SDSS). It therefore requires only a small amount of
material to rapidly self-enrich in the central regions to explain
these metallicities -- our simulations even with $\sim5\%$ excess
light fractions are able to do so (reaching ${\rm [Z/H]}_{c}\sim0.5$ 
and ${\rm d[Z/H]/d}\log{(r)}\sim-0.6$). Since strong central metallicities
often accompany strong metallicity gradients, 
the actual total metal mass of the galaxy (estimated from
the combination of the surface density profile and metallicity
profile) can easily be a factor $\sim3$ smaller than the metal mass
that would be inferred if the system had a constant metallicity equal
to the quoted central values.

We emphasize that the pre-merger stars in our simulations
are initialized to have a low uniform metallicity with no initial
gradients (so that we can be sure any gradients we see are the result
of the simulation dynamics, not artifacts of our initial
conditions). Since more realistic initial disks will already have high
metallicities $\sim Z_{\sun}$ \citep[similar, in fact, to ellipticals of the 
same mass; see e.g.][]{gallazzi:ssps}, and metallicity gradients in the same
sense as those produced in our simulations, 
the metallicities and metallicity gradients in our
simulations are only lower bounds to the true values that can arise in
dissipational mergers. Furthermore, recycling of metal-rich gas 
by stellar evolution \citep[which we do not model here, but see e.g.][]{ciottiostriker:recycling}
will subsequently enrich the system.

\begin{figure}
    \centering
    \plotter{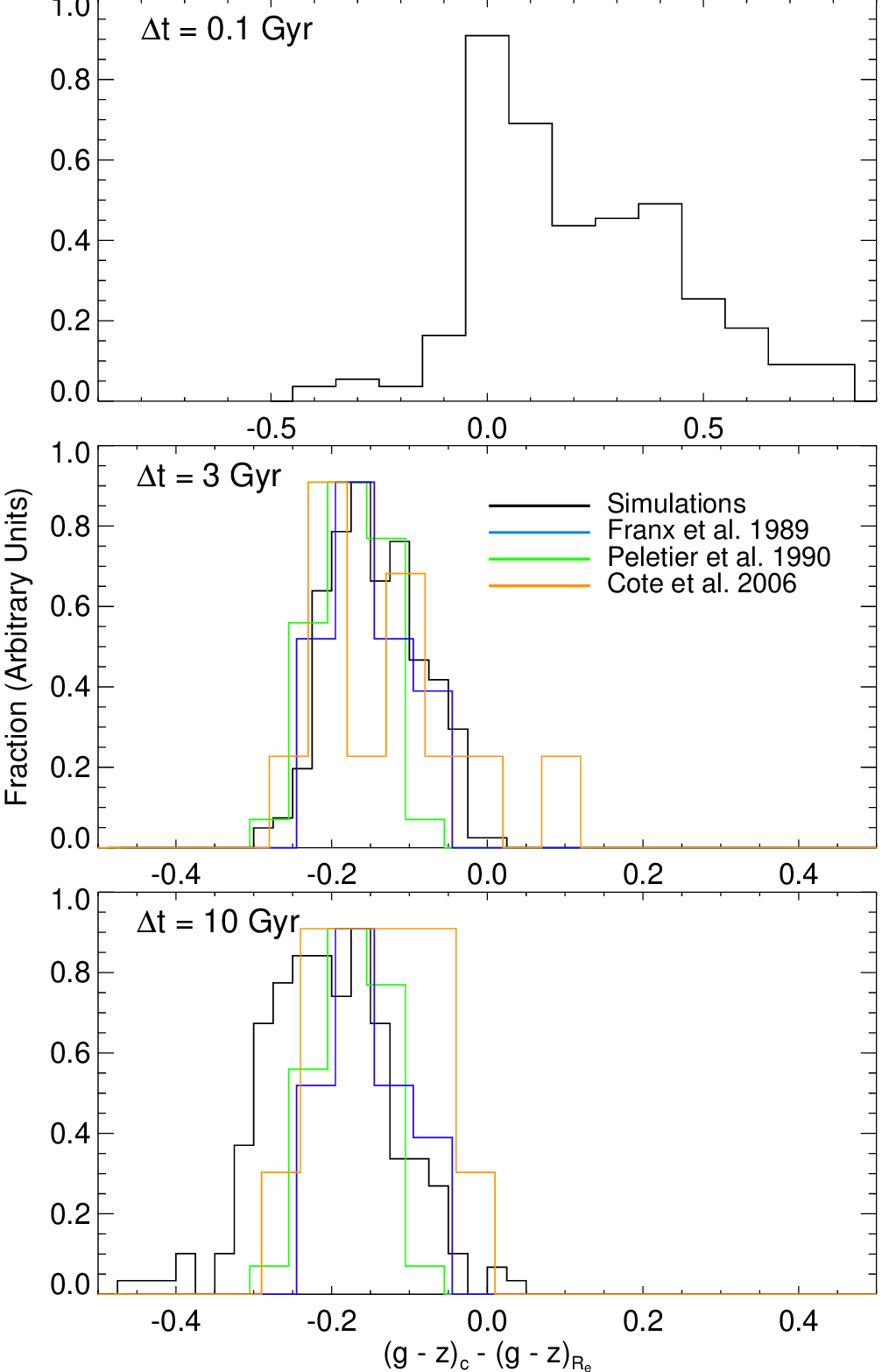}
    \caption{Color gradients (here the difference in mean color 
    within $R_{e}/8$ and $R_{e}$) in simulations  
    (black lines) at different times after the merger (labeled). We compare to 
    the distribution for observed cusp elliptical samples (for the \citet{cote:virgo} 
    objects, we show the distribution for ellipticals $<6\,$Gyr 
    [{\em middle}] and $>6$\,Gyr [{\em bottom}] old). 
    Positive values indicate blue cores. For ages typical of cusp ellipticals, 
    the remnants have relaxed to where metallicity gradients dominate the overall 
    color gradient and yield red cores, albeit with generally weak gradients, 
    with distributions similar to observed ellipticals of comparable ages. 
    \label{fig:color.grad}}
\end{figure}

Since the behavior of the color gradients is somewhat more complex, we 
try and reduce it to the key qualitative element. 
Figure~\ref{fig:color.grad} shows the distribution in color difference at $R_{e}/8$ and 
at $R_{e}$, a rough proxy for the color gradients, at various times after the merger. 
Observed distributions from samples of 
cusp ellipticals, with typical ages $\sim3-8$\,Gyr, 
are also shown. The transition from initial blue cores shortly after the merger to 
red cores in relaxed ellipticals is obvious, as is the weakness of the color 
differences. 
The simulations and observations trace 
broadly the same distribution for post-merger times of $\sim3-10$ Gyr. 
We find a similar agreement comparing to the $V-I$ color gradients in 
\citet{lauer:centers}, and the $V-R$ and $V-I$ color gradients 
in \citet{bender:data}.

\begin{figure}
    \centering
    \plotter{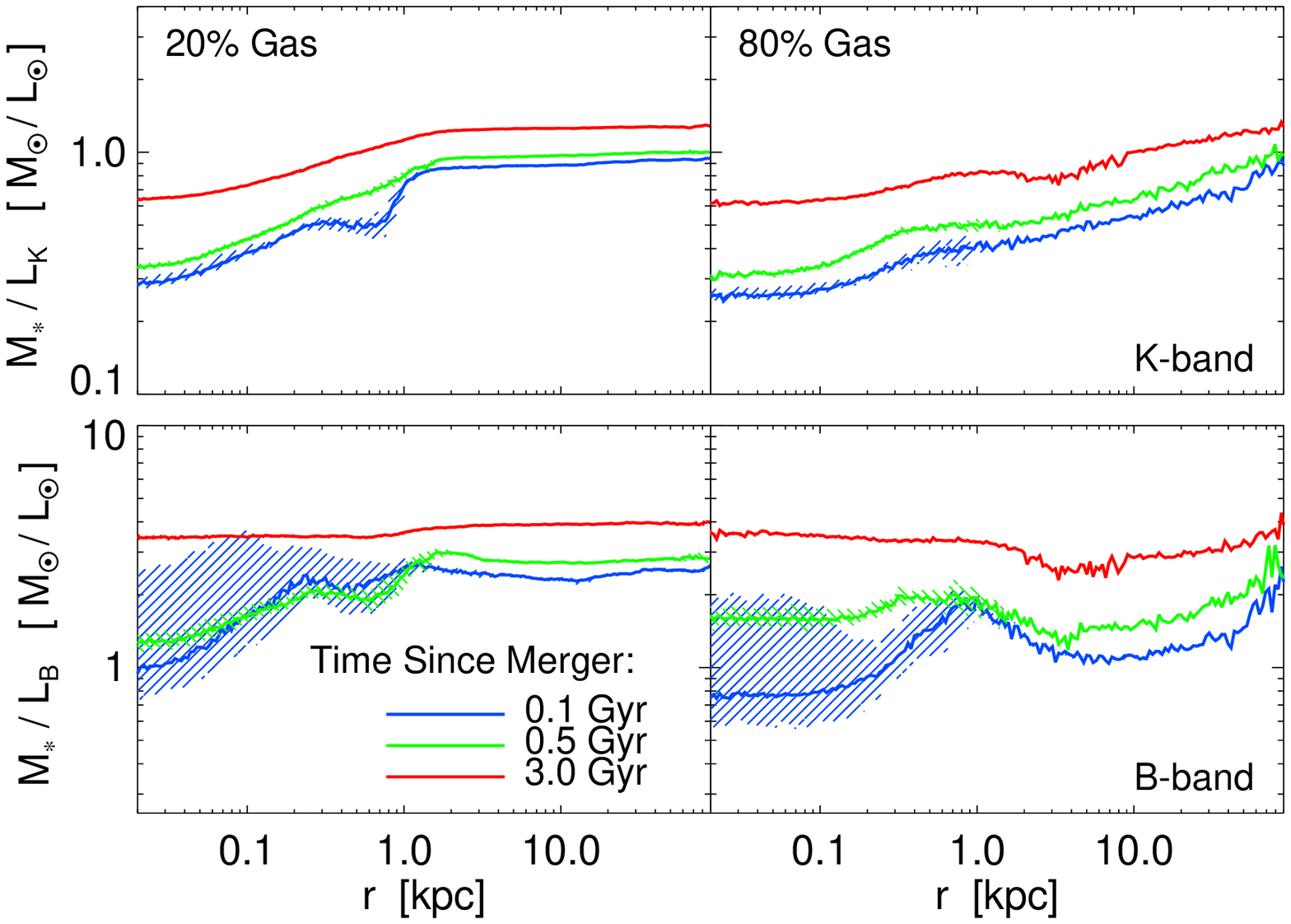}
    \caption{Mass-to-light ratio ($M/L$) 
    as a function of radius and time elapsed since the end of the 
    merger-induced starburst, in $K$-band ({\em top}) and $B$-band ({\em bottom}), 
    for relatively low gas fraction ({\em left}) and high gas fraction ({\em right}), 
    otherwise equal ``typical'' ($\sim\lstar$) merger simulations. Shaded range shows 
    the $\sim1\sigma$ range in $M/L$ across different sightlines. 
    At early times ($\lesssim1\,$Gyr), young central stellar populations lead to a 
    significant gradient (related to the strong ``blue cores'' in Figure~\ref{fig:color.grad}). 
    By $\gtrsim3\,$Gyr (relevant for almost all the ellipticals in our sample), 
    there is no significant dependence in optical bands, and only weak 
    dependence in the near infrared (owing to the remaining metallicity 
    gradients; for detailed discussion see \paperone).
    \label{fig:m.l.r}}
\end{figure}

The weak color gradients in observed ellipticals are often used to justify the 
assumption of a constant {\em stellar} mass-to-light ratio as a function of 
radius, an assumption we have used in comparing our simulations and 
observed systems. In Figure~\ref{fig:m.l.r}, we test this 
assumption directly in our simulations. We consider two $\sim\lstar$ galaxy mergers, 
one relatively gas poor (a case which happens to provide a reasonable 
match to many of the observed ellipticals), and another otherwise equal but 
gas rich merger, both of which have typical color and stellar population gradients. We 
directly calculate the stellar mass-to-light ratios in narrow major-axis radial annuli 
in both $K$-band and $B$-band, at different times after the peak of the 
merger-induced starburst. 

The $M/L$ gradients reflect what is expected from the color gradients, 
and supports our assumption of nearly constant $M/L$ with radius. At 
early times $\ll 1\,$Gyr, the central stellar populations are bright (associated 
with blue cores), and in $B$-band, nuclear dust obscuration makes the central $M/L$ 
sensitive to viewing angle. This is a relevant for the recent merger remnants 
studied in \paperone\ (although they are observed in the $K$-band), 
and we discuss the implications for the youngest systems (with very large 
apparent $f_{\rm extra}\gtrsim0.5$) therein. By $\sim2\,$Gyr after the merger, 
however, the remnants have typical weak, red cores and $M/L$ 
in optical bands such as $B$ and $V$ is usually constant as a function of radius to within 
$\lesssim20\%$. At these ages, the parameters derived by fitting to 
optical profiles are identical to those obtained directly fitting the stellar mass profiles. 
The gradients in age (younger at the center) and metallicity (more metal rich 
at the center) have opposite effects in the optical, yielding negligible dependence of 
$M/L$ on radius (in $K$ band, the effects do not cancel; however 
among our observed samples only the recent 
merger remnants from \citet{rj:profiles} are observed in $K$-band, so we refer to 
\paperone\ for more details). 
Reassuringly, this is similar to a purely empirical estimate based on the 
color-dependent $M/L$ from \citet{bell:mfs} and the observed color gradients. 
All of the ellipticals we study have estimated stellar population ages 
older than this threshold (and weak color gradients) and are observed in these 
optical bands, so our results should be robust to stellar population effects.

\begin{figure}
    \centering
    \scaleup
    \plotter{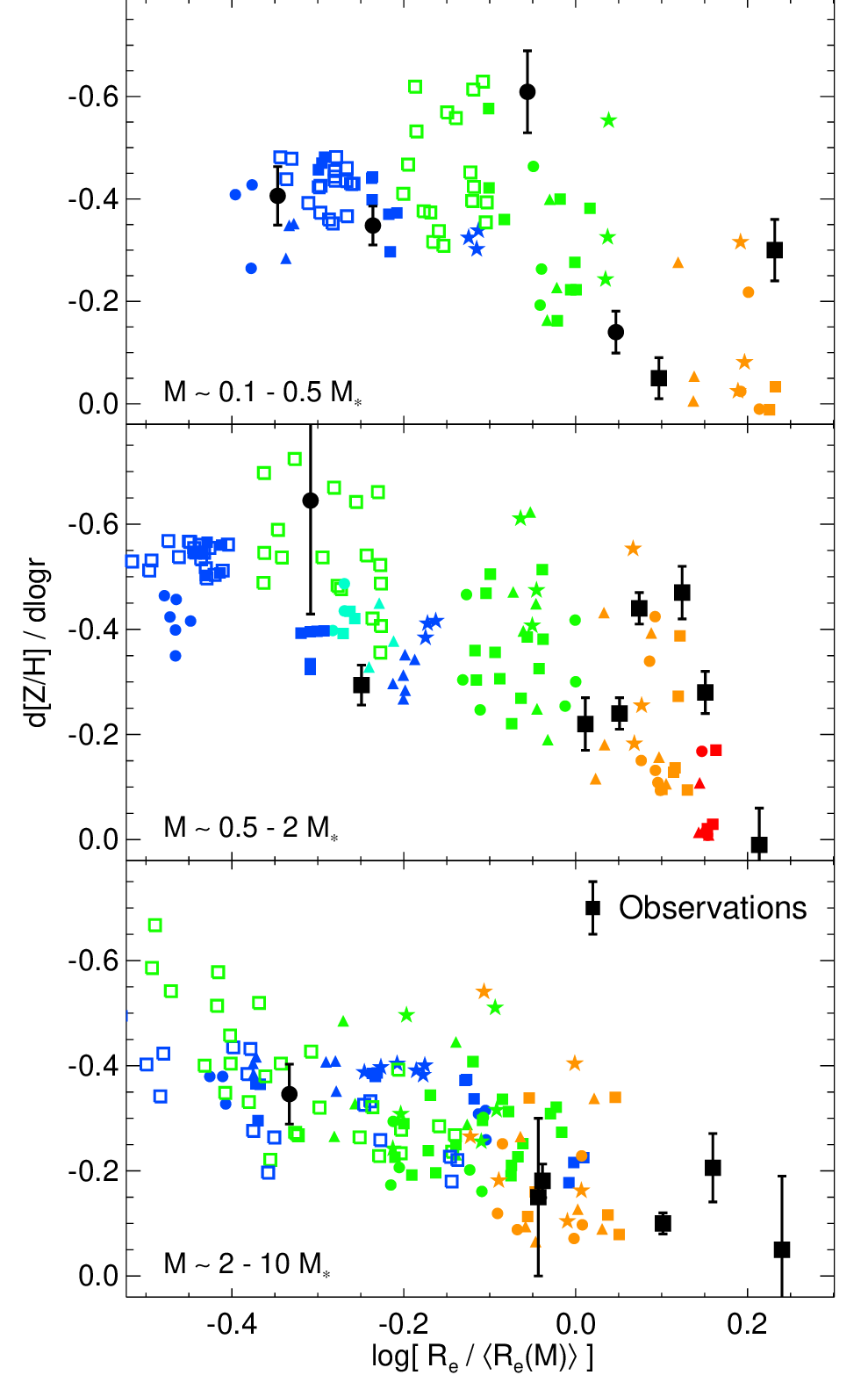}
    \caption{Correlation between metallicity gradient strength and 
    effective radius (relative to the mean at each stellar mass), at 
    fixed stellar mass (in the mass bins labeled). Our simulations are 
    shown (point style as in Figure~\ref{fig:ns.mass}) along with the observed 
    systems from Figure~\ref{fig:grad.compare} 
    (black circles and squares from \citet{sanchezblazquez:ssp.gradients} and 
    \citet{reda:ssp.gradients}, respectively). In the simulations, increased 
    dissipation yields more compact remnants, with stronger stellar population 
    gradients (unfortunately the color and age gradients are strongly sensitive 
    to subsequent time evolution effects, so are less robust tracers of dissipation), 
    predicting a correlation of the nature shown. Observations tentatively suggest the 
    same, but more are needed to confirm this prediction of different degrees of dissipation 
    along the fundamental plane. 
    \label{fig:re.ssp.grad}}
\end{figure}

Given the dependence of both 
gradient strength and effective radius (Figure~\ref{fig:re.sigma.cusp}) on 
dissipation, we would expect that systems with smaller $R_{e}$
at fixed stellar mass (an indicator of greater dissipation) 
should exhibit somewhat stronger stellar population
gradients. Figure~\ref{fig:re.ssp.grad} shows the predicted gradient
strength versus effective radius from simulations
and observed systems. We focus on the metallicity
gradient: the color and age 
gradients evolve strongly with time, introducing much larger 
scatter and possibly even reversing the predicted trends here. 
There is a significant trend predicted, and the observations show some
tentative evidence for such behavior, but more data are needed to
determine this robustly (because there is a mean offset in the
range of metallicity gradients at different masses, 
it is important to consider narrow ranges in observed 
stellar mass). To the extent that the strength of the metallicity
gradients is an indicator of the degree of dissipation in the
elliptical formation event,
this prediction potentially provides a completely non-parametric means
to search for non-homology (i.e.\ differential degrees of dissipation
as a function of radius at fixed mass) along the fundamental plane.

\section{Discussion}
\label{sec:discuss}

We have studied the formation and properties of extra light and cusps
in elliptical galaxies using a large library of both numerical
simulations and local observed cuspy ellipticals.  We demonstrate the
relation of these observed cusps to the prediction of
\citet{mihos:cusps} with our improved numerical models: namely, that
tidal torques in major mergers of gas-rich disks channel gas into the
central regions of the galaxy, where it forms a dense central
starburst. The starburst leaves a central light excess with a high
phase space density, making the remnant more compact and reconciling
the densities of disks and elliptical galaxies. Improvements in 
spatial resolution (the ``extra light'' in previous numerical studies 
has generally been entirely unresolved), 
numerical treatments (see Appendix~\ref{sec:appendix:resolution}), and models for 
star formation allow us to make detailed comparisons 
and for the first time demonstrate good agreement between the extra light and 
cusps in simulated gas-rich merger remnants and observed cusp 
ellipticals. 

We argue that stars in these cuspy ellipticals (specifically, 
we intend this to represent those ellipticals formed 
immediately in gas-rich major mergers; not those which may have had subsequent gas-poor 
re-mergers) should be separated into at least two distinct populations. 
First, stars which are formed in the disks (or otherwise in extended distributions in 
progenitor galaxies) before the final merger and coalescence 
of the two galaxies. The final merger scatters these stellar orbits and they 
undergo violent relaxation. They dominate the light, even in highly gas-rich 
merger remnants, outside of $\sim0.5-1\,$kpc, and form a \Sersic-law profile 
owing to their partial violent relaxation. Second, 
the starburst or dissipational population, 
formed in the central gas concentration in the final merger. 
This component is compact, and 
dominates the light inside a small radius $\lesssim0.5-1\,$kpc. 
These stars {\em do not} undergo 
significant violent relaxation, but form in a nearly fixed background potential 
set by the dissipationless component of the galaxy. The size of the 
dissipational component is set primarily by the radius at which it becomes 
self-gravitating; the gas is then generally stable against further collapse (even with 
cooling) and rapidly forms stars.\footnote{There is also a third 
component present in simulations but not prominent in light profile fitting: 
gas moved to large radii temporarily either by feedback or tidal effects, which 
settles into the relaxed remnant and re-forms small rotationally supported 
components \citep[embedded disks, kinematically decoupled cores, etc.; e.g.][]{hernquist:kinematic.subsystems,hopkins:disk.survival}.}

\subsection{Comparing Simulations and Observations:
Empirical Decomposition of Light Profiles}
\label{sec:discuss:separating}

Observations and simulations have, in recent years, been driven to convergence 
on this multi-component description of elliptical light profiles, at least 
within the ``cusp'' population 
\citep{kormendy99,jk:profiles,cote:virgo,cote:smooth.transition,ferrarese:profiles}. 
In particular, the combination of ground-based and high-resolution {\em HST} photometry
has allowed observers, from a purely phenomenological perspective, to 
recognize central extra light components 
beyond the inward extrapolation of an outer profile. The interpretation proposed 
herein has been advanced as the explanation for these central profile 
shapes, supported in a purely empirical manner by 
gradients in e.g.\ stellar populations, isophotal shapes, and stellar 
kinematics. 

By applying observational methods to a large library of simulations, 
we can directly compare simulations with different degrees of dissipation 
to observed light profiles, in order to determine what is required
in a merger-driven scenario in order to 
reproduce the observations. We find that in all 
cases of ellipticals of moderate masses $\gtrsim 0.1\,\mstar$,
we have simulations which provide excellent matches to the observed
systems (down to our resolution limits $\sim30-50$\,pc), 
comparable to the typical point-to-point variance inherent
in the simulation surface brightness profiles
($\dmu\lesssim0.1$). We can therefore further use this ensemble of simulations 
to test different means of decomposing observed 
profiles in order to calibrate a method which reliably and robustly 
recovers the appropriate {\em physical} decomposition in those 
simulations. 
We demonstrate (Figures~\ref{fig:demo.fit.danger} \&\ \ref{fig:check.fe.fsb}) 
that an appropriately designed 
parametric fit to the total light profile recovers, on average, the correct 
(known physically in the simulations) decomposition of the 
light -- the mass fraction and effective radius of both 
the physical starburst component and, separately, 
the outer dissipationless component. 

The details of our methodology are discussed in \S~\ref{sec:fits}: 
essentially, we fit the surface brightness profiles to the sum 
of two \Sersic\ components, an inner and outer profile. 
The quantitative details of the decomposition are, of course, 
sensitive to the methodology and assumed functional forms; 
for this reason observations have often disagreed over the 
quantitative scales of extra light even for similar profiles 
\citep[compare e.g.][]{cote:virgo,jk:profiles}. 
Fitting the entire light profile to 
a single \Sersic\ or a core-\Sersic\ law, for example, and comparing the 
central regions with the fit can yield physically meaningless 
values for the ``extra light,'' even though the result is, in a global sense, 
a formally good fit to the profile. 
Physically motivated models and simulated light profiles of the sort 
studied here are therefore necessary in order to make quantitative 
statements about extra light. 
Furthermore, a wide dynamic range 
is necessary; we find that the extra light is typically 
$\sim1-10\%$ of the galaxy light and becomes important within 
$\sim500\,$pc ($\sim0.1\,R_{e}$). The extra light component,
as it becomes larger, typically blends in more smoothly with the outer
profile, and does not necessarily appear as a sharp departure from the
outer light profile. Therefore, simultaneous resolution of small scales 
$\sim100\,$pc where the extra light dominates, and large scales 
$\sim20-100\,$kpc where the shape of the outer, dissipationless 
component can be robustly determined, is a critical observational 
ingredient enabling these comparisons. 

We apply this decomposition to our simulations and to a number of
observed samples: including the combination of detailed elliptical
surface brightness profiles over a wide dynamic range in
\citet{jk:profiles} and \citet{lauer:bimodal.profiles}, and a large
sample of gas-rich merger remnants observed in \citet{rj:profiles}. 
The results from 
our fits are included in Tables~\ref{tbl:cusp.fits} \&\ \ref{tbl:mgr.fits} 
(see Figures~\ref{fig:jk1}-\ref{fig:lauerpp1} and Appendix~\ref{sec:appendix:jk}). 
For the observed
ellipticals, we consider only those which are observed to have central
cusps in high-resolution HST imaging, as opposed to those with central
cores or mass deficits, as the latter are commonly believed to form in
gas-poor spheroid-spheroid 
remergers, which we do not model here. We instead undertake a
comparison of those objects and simulations of both gas-rich and
gas-poor mergers in \citet{hopkins:cores}. 

Applying these methods to observations, then, provides a powerful new 
diagnostic of the formation history of galaxies. 
In the physical 
context of the models considered here, 
it is not only possible to 
estimate (from the existence of some central component) whether or 
not the formation of a given galaxy requires some dissipation, 
but {\em to empirically, quantitatively estimate the degree of dissipation -- i.e.\ ``how 
gas-rich'' was the history or progenitor of a given elliptical}. 
We extend this to develop a number of new 
probes of galaxy merger history, and 
corresponding 
predictions and tests of the merger hypothesis.

\subsection{Predictions and Observations}
\label{sec:discuss:predictions}

{\bf (1)} From either fitted extra light/dissipational component masses, or
a direct comparison of simulations and observations, we find that
{\em there is a clear trend of increasing dissipation in the profiles of
less massive ellipticals} (Figures~\ref{fig:fgas.needed}-\ref{fig:fgas.needed.fextra}). 
The required initial gas fractions in
mergers to form profiles that resemble the observed systems are a
decreasing function of mass, and at all masses from $M_{\ast}\sim
10^{9}-10^{12}\,\msun$ span a range bracketed by the typical observed
gas fractions of spiral galaxies at the same mass, at $z=0$
(bracketing the low end of the required gas fractions) and $z\sim2-3$
(bracketing the high end). In terms of the mass fraction in the final
dissipational or starburst component, the trend is even more clear
(significance $\gg5\,\sigma$), and is reflected in the observed extra
light components. The trend is given by Equation~(\ref{eqn:fgas.m}): 
\begin{eqnarray}
\nonumber & & \langle f_{\rm starburst} \rangle \sim {\Bigl[}1+{\Bigl(}\frac{M_{\ast}}{10^{9.15}\,\msun}{\Bigr)}^{0.4}{\Bigr]}^{-1}.
\end{eqnarray}
We are able to, for the first time, observationally directly infer the degree of dissipation 
from the surface brightness profiles of observed ellipticals 
and use it to show that, in line with predictions from e.g.\ their fundamental plane 
correlations \citep{carlberg:phase.space,gunn87,lake:merger.remnant.phase.space,
kormendy:dissipation,hernquist:phasespace}, 
dissipation is more important in lower mass ellipticals,
reflecting the increasingly gas-rich nature of lower-mass progenitor
spirals.
 \\

{\bf (2)} At a given mass, the degree of dissipation strongly influences the
size of the remnant. In both observations and simulations {\em we
demonstrate a tight correlation between effective (half-light) radius
at a given stellar mass and the inferred dissipational/extra light
fraction} (Figure~\ref{fig:re.sigma.cusp}). 
This owes to the compact nature of the central dissipational
component -- increasing the 
mass fraction in this component means that the half-light radius must be
smaller.  This is also reflected as an increasing velocity dispersion
with extra light fraction at fixed mass, but this effect is 
weaker (because of the role of dark matter in setting the total mass
and velocity dispersion). 

This has important implications for the fundamental plane of
elliptical galaxies -- namely that dissipation is a key driver of
systems on the fundamental plane (in terms of e.g.\ stellar mass, in
optical bands there are obviously also substantial stellar population
effects).  Looking at $\sim\lstar$ ellipticals along the sequence of
increasing dissipational component and decreasing effective radius, we
directly show how this relates to homology-breaking in ellipticals 
(Figures~\ref{fig:profile.vs.fsb.sims}-\ref{fig:profile.vs.fsb}). Those
with large effective radii and little dissipation at a given mass
are well-described by a \Sersic\ law (with our mean Sersic index
$n\sim2.5-3$) over the range $\sim0.01-5\,R_{e}$. Those with smaller
effective radii and large dissipational components typically show an
excess at small radii, reflecting the concentration of starburst light
owing to dissipation. This grows with smaller radii and larger
dissipational components, from prominence only at small radii $\ll
0.1\,R_{e}$ to $\sim0.5\,R_{e}$ in extreme systems.

We caution that there is not an obvious sense of the shape of this
extra component -- in some cases the whole profile could still be
well-fitted by a single \Sersic\ law (albeit with a higher index
$n_{s}$). In others, there is only a small excess in the $\mu-r$
plane, but it extends to large radii $\sim0.5\,R_{e}$ and therefore
contributes considerably to the mass.  In yet others it is similar to
more obvious extra light, and to that visible in our most extreme
simulations, as a sharp departure from the outer \Sersic\ law at small
radii. In any case, however, the tight correlation between dissipational
component strength and size of the elliptical can be directly seen to
relate to a subtle non-homology in the central surface brightness of the
galaxies, in both the observations and our simulations.
\\

{\bf (3)} We find that the outer \Sersic\ indices of 
cusp ellipticals are nearly constant as a function of
stellar mass or any other properties (with median 
$\sim2.5-3$ and scatter $\sim0.7$; Figures~\ref{fig:ns.mass}-\ref{fig:ns.distrib.cleaned}). 
We emphasize that, given our
two component decomposition, this outer \Sersic\ index is only
meaningful in the sense of reflecting those violently relaxed stellar
populations which were present in the progenitors before the final
merger (fitting the entire profile to a single \Sersic\ index can yield a 
very different result). 
This is because the dissipationless component is relaxed under the
influence of gravity, and is therefore roughly self-similar across
scales.

We note this constant \Sersic\ index distribution is 
relevant for {\em gas-rich merger remnants}. 
As we demonstrate in \citet{hopkins:cores}, subsequent gas-poor 
re-mergers will systematically raise the outer \Sersic\ indices, 
and pseudobulges or other low-mass bulges may 
have systematically lower indices (and together
these may drive a systematic dependence of \Sersic\ index on mass, owing
to the cosmological dependence of formation history on mass). 
This has been seen in the Virgo cusp elliptical sample of \citet{jk:profiles}, 
and each of our observed samples independently confirms
our prediction, and together they show no dependence of \Sersic\ index 
over more than three orders of
magnitude in stellar mass. Especially restricting to a sample of
ellipticals which are observed in multiple bands and from multiple
ground-based sources and considering only those results for which the
fits (from different bands and instruments) do not disagree, we
confirm both the predicted 
\Sersic\ index distribution and lack of dependence on other galaxy properties.
This agreement further suggests that cusp ellipticals are the direct remnants 
of gas-rich major mergers, without substantial subsequent re-merging.
\\

{\bf (4)} Extra light fraction does not, in any predictive sense, correlate with the global 
kinematic properties (namely, ellipticity, isophotal shape $a_{4}$, and 
rotation $(V/\sigma)^{\ast}$) of ellipticals in our simulations or in the observations. However, 
there is in a broad sense a change in these properties for large versus small 
extra light fractions (Figure~\ref{fig:cusp.vs.kinematics}). 
At small extra light fractions $\lesssim5-10\%$, remnants are 
more similar to the known properties of dissipationless disk-disk mergers 
\citep[e.g.][]{barnes:disk.halo.mergers}, with a wide range in 
rotation properties, isophotal shapes, and ellipticities. 
At substantial extra light fractions, however, the systems are 
(on average) somewhat rounder, 
and are more uniformly disky 
and rapidly rotating. 
The dissipational component itself makes the remnant potential more 
spherical, and is closely related to kinematic subsystems formed from gas 
that survives the merger which contribute to the rotation and 
isophotal shapes \citep[see e.g.][]{cox:kinematics,
robertson:fp,dekelcox:fp,naab:gas,burkert:anisotropy}. 
A distribution of dissipational/extra light fractions like that
observed yields remnants with a similar distribution in each of these 
properties to observed cusp ellipticals, in striking contrast with dissipationless simulations 
that have historically not been able to match these properties. 
\\

{\bf (5)} Black hole mass appears to be roughly correlated with the mass in the 
extra light component, in both simulations and observations, but we show that this 
is largely a reflection of the tighter correlation between black hole mass and total 
host spheroid stellar or dynamical mass (Figure~\ref{fig:bh}). 
This is not surprising: if black hole mass 
most tightly follows the central potential or binding energy of the bulge, 
as argued from recent observations favoring a black hole fundamental plane 
\citep{hopkins:bhfp.theory,hopkins:bhfp.obs,aller:mbh.esph}, 
then this is largely set by the total mass of 
the system. The extra light mass represents a perturbation to the binding energy/velocity 
dispersion the system would have if dissipationless, and there is in simulations 
a weak residual trend along the expected lines, but it is sufficiently weak that 
the present observations are inconclusive. 
\\

{\bf (6)} The dissipation of gas into the central regions of ellipticals in a central starburst 
gives rise to significant gradients in the stellar populations (Figure~\ref{fig:grad.case.study}). 
Since they are formed at
the end of the merger, from gas channeled to the center of the galaxy after being 
enriched by star formation throughout the progenitor disks, the central stellar 
populations tend to be younger and more metal rich. The detailed gradient structure 
in our simulations corresponds well to what is typically observed -- with metallicity 
increasing and age decreasing relatively smoothly from $\sim R_{e}$ to 
$\sim 0.03-0.1\,R_{e}$. We do occasionally find anomalous behavior resembling 
various observed systems -- e.g.\ non-monotonic trends in the stellar populations -- 
but these are generally rare, as is also observed. 

At large radii, observations find generally weaker gradients -- our
simulations suggest that these are not driven by dissipation in the
merger, but rather reflect pre-existing gradients in the initial
disks. To be conservative, we have not included them here, but
experiments suggest that they are not completely destroyed: it is well
known, for example, that particles roughly preserve their ordering in
binding energy in mergers \citep{barnes:disk.disk.mergers}, so given
the trend in spirals towards less metal enrichment and older ages in
their outer regions, we expect a smeared out version of these
gradients to continue weakly into the outer regions of merger
remnants, where the stars are primarily those violently relaxed from
the initial disks. Over the range where observed gradients are
typically fit, we quantify the gradient strengths in our simulations
and find they are similar to those observed, as a function of system
mass, age, and metallicity (Figure~\ref{fig:grad.compare}). 
Previous claims that mergers would only
yield weak gradients in ellipticals ignored the importance of
dissipation, which allows us to form systems with metallicity
gradients comparable to the most extreme observed (${\rm
d[Z/H]/d}\log{r}\sim-0.8$).

In general, the gradients are not strongly dependent on mass, but do
depend somewhat on the degree of dissipation (Figure~\ref{fig:grad.compare}). 
They also evolve
significantly with stellar population age, as suggested by observations -- 
not only do the stars
become uniformly older, but as they age after the merger, the
difference in mass to light ratio between young stars formed in the
starburst and older stars becomes less significant, so in any
light-weighted stellar population the younger starburst stars become
(relatively) less prominent, and the gradient weakly washes out (Figure~\ref{fig:grad.case.study}). 
Of course, being driven by the central dissipational population, the
gradient strengths are also correlated with central metallicity and
with one another, albeit with large scatter (and for e.g.\ uniformly
high metallicities or weak gradients, these correlations break down).
All of these trends are in good agreement with present-day observed
ellipticals.

Furthermore, the central stellar metallicities of ellipticals (both
their mean and distribution about that mean) as a function of e.g.\
mass and velocity dispersion, are reproduced in our simulations, given
only the assumption that the initial spirals have metallicities
appropriate for the observed disk galaxy mass-metallicity relation --
in fact, when the initial disk gas fractions are moderate ($\gtrsim20\%$), we
obtain this result from the self-enrichment tracked in our code (which
dominates the final total central metallicity) even if the initial
gas and stars have zero metallicity (Figure~\ref{fig:grad.compare}). 
Given that we simultaneously
reproduce the central metallicities, metallicity gradients, and
surface brightness profiles of observed cusp ellipticals, the claim
that these systems have too much mass in metals to be produced from
the merger of local spirals \citep{naab:no.merger.ell} does not stand
up, at least insofar as we are restricting to {\em cuspy} ellipticals.

We also reproduce the typical observed color gradients of these
ellipticals (Figures~\ref{fig:grad.case.study} \&\ \ref{fig:color.grad}).  
We caution that while time evolution of even
physical quantities such as the metallicity and age gradients is
non-trivial, the color gradients are even more complex. Generally,
though, systems have blue cores (are more blue towards their centers) for
a short time after the merger, commonly taken as a signature of a
recent merger or central starburst, owing to the young central
starburst population. However, this population is also more metal
rich, so the gradient rapidly weakens and after a short period,
$\sim$\,Gyr, the metallicity difference becomes more important and the
systems have red cores (are redder towards their centers).  The
gradients are fairly weak, a difference of $\sim0.1-0.2$ mag between
the color within $R_{e}/8$ and $R_{e}$, and the simulated distribution
of color gradient strengths agrees well with that observed in samples
of local ellipticals (especially if we restrict broadly to similar
mean stellar ages).
\\

{\bf (7)} Since increasing the amount of dissipation at a given total mass tends both to yield 
smaller effective radii and stronger stellar population gradients, we predict that the 
two should be correlated (again, at fixed stellar mass - considering too wide a range 
in stellar mass will wash out this correlation, since the two scale differently in the 
mean with stellar mass; Figure~\ref{fig:re.ssp.grad}). 
The trend is not predicted to be dramatic, but should be 
observable, and comparison with recent observations does suggest that systems with 
smaller effective radii for their mass do have uniformly higher metallicity gradient strengths. 
Future observations should be able to test this more robustly, using metallicity gradients 
(we caution that the age gradients and, especially, color gradients are too age-sensitive 
to be constraining in this sense).  

To the extent that gas-rich progenitor disks (at least at low-moderate redshifts) 
are 
so because they have more extended star formation histories, this should also 
be reflected in the integrated stellar populations of the ellipticals. 
This is well known in an integrated sense -- low mass ellipticals (which form from 
low-mass, gas-rich disks) tend to have more extended star formation histories with 
less $\alpha$-element enrichment \citep[e.g.][]{trager:ages,thomas05:ages}. 
At fixed mass, however, 
if systems undergo mergers over roughly the same time period 
(which is generally true for systems of the same mass), then those with smaller 
effective radii (and lower $M_{\rm dyn}/M_{\ast}$) should have formed 
with more dissipation from more gas rich disks, which presumably had more extended 
star formation histories. This should yield younger pre-merger stellar population ages 
and less $\alpha$-enriched stellar populations, relative to progenitors of the same 
mass which had shorter star formation histories and exhausted their 
gas supply before the merger. There is one important caveat -- systems of the same mass 
might also have had more gas-rich progenitors because they underwent mergers 
at very early times (making them older and more $\alpha$-enriched), but 
cosmological estimates \citep[e.g.][]{hopkins:groups.ell} suggest that systems 
with such early mergers will usually have multiple subsequent mergers 
at later times, so they will grow significantly in mass and have 
their effective radii substantially modified by these additional 
processes (and may reflect the same trends as more typical systems by $z\sim0$). 
In any case, this latter process is not dominant at a given stellar mass. 
We study this in greater detail in \citet{hopkins:cusps.evol}.

\subsection{Synopsis}
\label{sec:discuss:summary}

We have developed a paradigm in which to understand the
structure of (cusp or extra light) ellipticals, in which there are
fundamentally two stellar components: the relic of a
a dissipational central
starburst and a more extended violently relaxed component 
\citep[introduced and discussed in e.g.][]{mihos:cusps,kormendy99,cox:kinematics,jk:profiles}.
We have shown that these components can be separated with observations
of sufficient quality, and can be used to constrain the formation histories of
ellipticals and infer physically meaningful properties in a hierarchical 
formation scenario. This allows
us to demonstrate that dissipation is critical in
understanding the properties of ellipticals, including (but not
limited to) the structure of their surface brightness profiles, their
sizes, ellipticities, isophotal shapes and rotation, age, color, and
metallicity gradients (and their evolution), and the gas content and
properties of their progenitors.

We have studied these properties and identified robust trends across a 
large library of simulations, in which we vary e.g.\ the galaxy masses, initial gas 
fractions, concentrations, halo masses, presence or absence of bulges, presence or 
absence of black holes, feedback parameters from supernovae and stellar winds, 
orbital parameters and disk inclinations, and mass ratios of the merging galaxies. 
This range of parameters allows us to identify the most important physics. Most of these 
choices, for example, affect the surface brightness profile, extra light mass and 
radius of the extra light, concentration and effective radius of the remnant, 
and even its ellipticity and isophotal shape only indirectly. Ultimately, what 
determines the structure of the remnant (insofar as the properties we have considered) 
is, to first order, how much mass is in the dissipationless (violently relaxed) 
component versus the dissipational/starburst component at the time of the final merger(s). 
Therefore, changing something like the orbital parameters or initial galaxy structure 
can alter the remnant substantially, but predominantly only insofar as it affects
the amount of gas which will be available at the time of the final coalescence of the 
galaxy nuclei \citep[i.e.\ how much mass ends up in the starburst component, as opposed to 
being violently relaxed in this final merger; see also][]{cox:feedback}. 
Moreover, merger-induced starbursts may not be the only source of 
dissipation \citep[for example, stellar mass loss may replenish the gas supply and 
lead to new dissipational bursts, see e.g.][]{ciottiostriker:recycling}, 
and the merger history and series of induced dissipational events may be 
more complex than a single idealized major merger 
\citep[see e.g.][]{kobayashi:pseudo.monolithic,naab:etg.formation}, but for 
our purposes, all dissipational star formation will appear similar when observed 
and have the same effects (we are essentially measuring the integrated amount of 
dissipation). 

We have demonstrated that this makes predictions for how fundamental
plane scalings arise, which we study further in \citet{hopkins:cusps.fp}. Given 
these decompositions, we can attempt to observationally test whether sufficient 
dissipation, as a function of stellar mass, 
could have occurred in the inner regions of ellipticals to 
explain the discrepancies between their central densities and those of their 
proposed progenitor spirals. We
make a wide range of new predictions for the distributions of these
properties and how they scale with the degree of dissipation, and how
they should scale with each other and various other observational
proxies for this degree of dissipation (which we define herein).
We have predicted and shown (given these proxies) that dissipation is
indeed more important (contributing a larger mass fraction) in
low-mass ellipticals, in line with expectations based on how gas
fractions are known to scale with disk mass.  Testing all of these
with better observations should be possible in the near future, with
well-defined samples of ellipticals and continued improvements in
mapping e.g.\ the surface brightness profiles, stellar populations and
their gradients, and structural properties of ellipticals over a wide
dynamic range.

To the extent that their parameter correlations 
and theoretical formation scenarios are similar, 
these conclusions should also 
generalize to ``classical'' bulges in disk-dominated galaxies 
(as opposed to ``pseudo-bulges'' formed in secular processes, which we 
do not model herein, and which may or may not show similar structure). 
We have excluded them from our observational study in this paper because the 
presence of a large disk greatly increases the uncertainties in profile 
fitting (and makes it difficult to robustly identify multi-component 
structure in the bulge), but emphasize that re-analyzing the S0 and early-type 
spiral galaxies in our observational samples demonstrates that they are 
in all cases consistent with our conclusions. Indeed, 
studies of the central profiles of classical bulges in S0-Sbc galaxies, 
in those cases with sufficient resolution and dynamic range and without 
much obstruction from the outer disk, 
have begun to see evidence 
for central two-component bulge structure with 
extra light components similar to those predicted here \citep[consistent 
in their profile shapes, sizes, mass fractions, kinematics, 
isophotal shapes, stellar populations, and colors; see e.g.][]{balcells:bulge.xl,
peletier:spiral.maps}. We emphasize that these 
are distinct from the nuclear star clusters seen in some bulges, discussed 
in Appendix~\ref{sec:appendix:nuclei}. 

It is important to extend this study to central core ellipticals,
increasingly believed to be shaped by subsequent gas-poor
(spheroid-spheroid) re-mergers, and we consider these objects in
\citet{hopkins:cores}. However, since cuspy ellipticals dominate the
$\sim\lstar$ population, most of the mass density in present-day
ellipticals is contained in those cuspy objects that we study herein.
We also strongly emphasize that, if core populations are indeed the
re-mergers of initially cuspy ellipticals, then the extra
light or dissipational components are {\em not} destroyed in these
mergers. Indeed, most simulations suggest that the dense, central
components are sufficiently tightly bound that they feel relatively little
disturbance in a dry merger. Although black hole scouring may scatter
stars from the central region and create such a core
\citep[e.g.][]{milosavljevic:core.mass}, these stars represent only a
small fraction of the extra light population. In other words, scouring
will flatten the central power-law like behavior of cusps of extra
light, but not fundamentally remove the $\sim500$\,pc central
light concentrations remaining from a dissipational starburst.
Furthermore, the impact of scouring
will be restricted to mass scales of order the black hole mass,
which is only $\sim 10^{-3}$ the stellar mass, much smaller than
the typically masses of the extra light components inferred through
our analysis \citep[e.g.][]{hopkins:cores}. Given this, it should be 
possible to generalize our modeling and constraints to much 
more complex merger histories than the idealized single major 
merger scenarios considered in our simulation study, provided 
we recognize that the dissipational and dissipationless components 
are really integral sums over the dissipational and dissipationless 
events in the formation history of a given elliptical. 
Of course, subsequent gas poor 
mergers may have a number of other effects on the
structural properties of the galaxies, modifying many of the
properties we have argued are initially set by or correlated with the
degree of dissipation -- therefore we have excluded core 
ellipticals from our
observational samples, and wish to emphasize the importance of doing
so in subsequent observational comparisons.

As we noted in \paperone, 
there is considerable room for progress in modeling the 
extra light component itself and structure of nuclear profiles 
at small radii ($\ll100\,$pc), where observations are making rapid progress; however, 
modeling these radii in a meaningful sense requires not just improved numerical 
resolution but also the inclusion of new physics that will allow simulations to 
self-consistently form the structures that would be resolved (giant molecular 
clouds, star forming regions and star clusters, individual 
supernova remnants, the multiple gas-phases of 
the ISM and their exchange, etc.). It is an important and ambitious goal 
that the next generation of studies move beyond the sub-resolution prescriptions 
necessary when modeling larger scales, and include realistic ISM structure important 
in galactic nuclei. 

Here, and in previous work
\citep[e.g.][]{hopkins:qso.all,hopkins:red.galaxies,
hopkins:groups.ell,hopkins:groups.qso} we have developed a conceptual
framework for understanding the origin of starbursts, quasar activity,
supermassive black holes, and elliptical galaxies.  In this paradigm,
the various objects and phenomena we have examined are connected in an
evolutionary sequence that is ultimately triggered by mergers between
gas-rich disk galaxies. 
Previously, from observed correlations between supermassive black
holes and properties of their hosts, we have argued that elliptical
galaxies must have originated through a process that blends together
gas- and stellar-dynamics.  This is motivated by simple physical
considerations implying that supermassive black holes are mainly
assembled by gas accretion \citep[e.g.][]{lynden-bell69} in order to
satisfy the \citet{soltan82} constraint relating black hole growth to
quasar activity, and the notion that elliptical galaxies were put
together by violent relaxation \citep[e.g.][]{lynden-bell67}
acting on stars.  The results of our analysis applied to recent
merger products \citep{hopkins:cusps.mergers}, cuspy ellipticals (this
paper), and elliptical galaxies with cores \citep{hopkins:cores}
indicate that the same blend of gas- and stellar-dynamics is further
essential for understanding the structural properties of the galaxies
that harbor supermassive black holes. 
Together, this provides further support to the idea that elliptical galaxies and
supermassive black holes originate via a common physical process
(mergers of gas-rich galaxies) and, given the \citet{soltan82}
constraint and the discovery of ULIRGs, quasars and starbursts,
respectively, as well.

\acknowledgments We thank Sandy Faber, Marijn Franx, 
and Barry Rothberg for helpful discussions and contributed data sets 
used in this paper. We also thank the anonymous referee whose 
comments greatly improved the presentation of the manuscript. 
This work
was supported in part by NSF grants ACI 96-19019, AST 00-71019, AST
02-06299, and AST 03-07690, and NASA ATP grants NAG5-12140,
NAG5-13292, and NAG5-13381. 
JK's work was supported in part by NSF grant AST 06-07490.
Support for 
TJC was provided by the W.~M.\ Keck 
Foundation. 

\bibliography{/Users/phopkins/Documents/lars_galaxies/papers/ms}

\clearpage
\longtabler
\begin{\tableset}{lccccccccccccc}
\tablecolumns{14}
\tabletypesize{\scriptsize}
\tablecaption{Fits to Cusp Ellipticals\label{tbl:cusp.fits}}
\tablewidth{0pt}
\tablehead{
\colhead{Name} &
\colhead{Ref.} &
\colhead{${\rm N}_{\rm phot}$} &
\colhead{$M_{\ast}$} &
\colhead{$M_{V}$} &
\colhead{$\sigma$} &
\colhead{$R_{e}$} &
\colhead{$\epsilon$} &
\colhead{$100\,a_{4}/a$} &
\colhead{$(v/\sigma)^{\ast}$} &
\colhead{$n_{s}$ (fit)} &
\colhead{$n_{s}$ (sim)} &
\colhead{$f_{e}$ (fit)} & 
\colhead{$f_{sb}$ (sim)} \\
\colhead{(1)} &
\colhead{(2)} &
\colhead{(3)} &
\colhead{(4)} &
\colhead{(5)} &
\colhead{(6)} &
\colhead{(7)} &
\colhead{(8)} &
\colhead{(9)} &
\colhead{(10)} &
\colhead{(11)} &
\colhead{(12)} &
\colhead{(13)} &
\colhead{(14)} 
}
\startdata
NGC 0596 & 2 &  4 & $10.89$ & $-20.90$ & $164$ & $3.24$ & $0.07$ & $ 1.30$ & $0.67$ & $7.50^{+0.59}_{-0.74}$ & $3.29^{+1.38}_{-0.47}$ & $0.035^{+0.009}_{-0.005}$ & $0.185^{+0.125}_{-0.097}$ \\
NGC 0636 & 3 &  3 & $10.80$ & $-20.83$ & $156$ & $3.81$ & $0.13$ & $ 0.80$ & $1.04$ & $3.39^{+0.18}_{-0.31}$ & $3.07^{+2.05}_{-0.63}$ & $0.090^{+0.005}_{-0.005}$ & $0.159^{+0.105}_{-0.061}$ \\
NGC 0821$^{\ast}$ & 2 &  2 & $11.36$ & $-21.71$ & $209$ & $3.63$ & $0.38$ & $ 2.50$ & $0.70$ & $3.38^{+1.57}_{-1.57}$ & $3.02^{+0.80}_{-0.14}$ & $0.228^{+0.193}_{-0.193}$ & $0.154^{+0.182}_{-0.064}$ \\
NGC 1172 & 2 &  1 & $10.46$ & $-20.13$ & $113$ & $4.37$ & $0.09$ & -- & -- & $29.92$ & $2.64^{+0.70}_{-0.11}$ & $0.001$ & $0.106^{+0.166}_{-0.035}$ \\
NGC 1199$^{\ast}$ & 3 &  3 & $10.95$ & $-21.49$ & $207$ & $3.23$ & $0.23$ & -- & $0.49$ & $1.96^{+0.06}_{-0.06}$ & $2.58^{+1.60}_{-0.80}$ & $0.106^{+0.012}_{-0.015}$ & $0.105^{+0.126}_{-0.061}$ \\
NGC 1400 & 3 &  2 & $9.88$ & $-18.45$ & $250$ & $0.67$ & $0.13$ & $ 0.00$ & -- & $1.91^{+0.06}_{-0.06}$ & $2.82^{+8.00}_{-2.82}$ & $0.164^{+0.001}_{-0.001}$ & $0.232^{+0.114}_{-0.095}$ \\
NGC 1426 & 2 &  3 & $10.83$ & $-20.78$ & $153$ & $3.63$ & $0.31$ & $ 0.01$ & -- & $5.26^{+0.11}_{-0.35}$ & $3.51^{+0.31}_{-0.63}$ & $0.001^{+0.001}_{-0.001}$ & $0.186^{+0.159}_{-0.096}$ \\
NGC 1427 & 2 &  2 & $10.83$ & $-20.79$ & $170$ & $3.16$ & $0.29$ & -- & -- & $2.12^{+1.31}_{-1.31}$ & $3.32^{+0.10}_{-0.31}$ & $0.251^{+0.199}_{-0.199}$ & $0.149^{+0.159}_{-0.061}$ \\
NGC 1439 & 2 &  3 & $10.85$ & $-20.82$ & $159$ & $4.07$ & $0.15$ & -- & $0.32$ & $3.47^{+1.11}_{-1.56}$ & $3.30^{+0.39}_{-0.64}$ & $0.114^{+0.142}_{-0.057}$ & $0.126^{+0.205}_{-0.050}$ \\
NGC 2314$^{\ast}$ & 3 &  2 & $11.34$ & $-21.95$ & $290$ & $3.37$ & $0.18$ & -- & $1.17$ & $2.02^{+0.59}_{-0.59}$ & $5.32^{+3.07}_{-2.64}$ & $0.269^{+0.059}_{-0.059}$ & $0.194^{+0.152}_{-0.097}$ \\
NGC 2434 & 2 &  1 & $11.14$ & $-21.33$ & $229$ & $3.63$ & $0.07$ & -- & -- & $4.72$ & $2.76^{+1.13}_{-0.12}$ & $0.001$ & $0.138^{+0.136}_{-0.048}$ \\
NGC 2534$^{\ast}$ & 3 &  1 & $10.43$ & $-20.25$ & -- & $22.67$ & -- & -- & -- & $1.98$ & $4.67^{+0.25}_{-1.73}$ & $0.033$ & $0.054^{+0.063}_{-0.017}$ \\
NGC 2693$^{\ast}$ & 3 &  2 & $11.58$ & $-22.59$ & $279$ & $6.65$ & -- & -- & -- & $1.77^{+0.02}_{-0.02}$ & $4.57^{+0.11}_{-2.61}$ & $0.181^{+0.001}_{-0.001}$ & $0.084^{+0.098}_{-0.048}$ \\
NGC 2768$^{\ast}$ & 3 &  2 & $11.01$ & $-21.56$ & $198$ & $6.78$ & $0.68$ & -- & $0.28$ & $2.63^{+0.21}_{-0.21}$ & $2.55^{+5.91}_{-0.89}$ & $0.011^{+0.004}_{-0.004}$ & $0.037^{+0.066}_{-0.007}$ \\
NGC 2778 & 2 &  1 & $9.67$ & $-18.75$ & $166$ & $1.62$ & $0.20$ & $ 3.50$ & $2.00$ & $2.15$ & $4.62^{+0.50}_{-1.83}$ & $0.447$ & $0.230^{+0.125}_{-0.115}$ \\
NGC 2974 & 2,3 &  5 & $11.00$ & $-21.09$ & $143$ & $4.07$ & $0.30$ & $ 0.50$ & $1.54$ & $4.06^{+0.77}_{-0.48}$ & $3.06^{+1.12}_{-0.63}$ & $0.049^{+0.007}_{-0.034}$ & $0.158^{+0.142}_{-0.103}$ \\
NGC 3193 & 3 &  2 & $10.51$ & $-20.30$ & $205$ & $1.81$ & -- & $ 0.30$ & $0.80$ & $2.73^{+0.21}_{-0.21}$ & $4.15^{+0.97}_{-1.97}$ & $0.089^{+0.010}_{-0.010}$ & $0.205^{+0.130}_{-0.097}$ \\
NGC 3309$^{\ast}$ & 3 &  3 & $11.39$ & $-22.24$ & $266$ & $6.67$ & $0.15$ & -- & -- & $1.97^{+0.46}_{-0.46}$ & $2.66^{+1.92}_{-0.70}$ & $0.148^{+0.031}_{-0.031}$ & $0.077^{+0.104}_{-0.046}$ \\
NGC 3311$^{\ast}$ & 3 &  3 & $11.91$ & $-23.14$ & $210$ & $12.15$ & $0.11$ & -- & -- & $1.36^{+0.00}_{-0.58}$ & $4.47^{+0.27}_{-2.56}$ & $0.086^{+0.073}_{-0.002}$ & $0.047^{+0.055}_{-0.001}$ \\
NGC 3377 & 1,2,3 &  6 & $10.42$ & $-20.07$ & $141$ & $2.24$ & $0.50$ & $ 0.50$ & $0.72$ & $1.96^{+0.67}_{-0.08}$ & $3.27^{+0.73}_{-0.50}$ & $0.133^{+0.064}_{-0.051}$ & $0.221^{+0.125}_{-0.115}$ \\
NGC 3605 & 2 &  3 & $10.16$ & $-19.61$ & $120$ & $1.58$ & $0.34$ & -$ 0.90$ & $0.74$ & $1.31^{+0.29}_{-0.66}$ & $3.04^{+2.08}_{-0.95}$ & $0.175^{+0.150}_{-0.041}$ & $0.283^{+0.176}_{-0.165}$ \\
NGC 3610 & 2,3 &  6 & $10.93$ & $-21.16$ & $163$ & $1.45$ & $0.52$ & $ 2.50$ & $1.10$ & $4.78^{+7.11}_{-3.60}$ & $3.04^{+0.40}_{-0.87}$ & $0.342^{+0.099}_{-0.012}$ & $0.273^{+0.133}_{-0.126}$ \\
NGC 3818$^{\ast}$ & 3 &  2 & $10.24$ & $-19.69$ & $206$ & $1.61$ & $0.39$ & $ 2.30$ & $0.93$ & $2.81^{+0.06}_{-0.06}$ & $3.54^{+0.92}_{-1.13}$ & $0.080^{+0.005}_{-0.005}$ & $0.205^{+0.132}_{-0.096}$ \\
NGC 3894$^{\ast}$ & 3 &  1 & $11.15$ & $-21.70$ & -- & $4.63$ & -- & -- & -- & $2.07$ & $3.20^{+0.86}_{-0.47}$ & $0.050$ & $0.036^{+0.069}_{-0.008}$ \\
NGC 3904$^{\ast}$ & 3 &  2 & $10.93$ & $-21.07$ & $215$ & $3.05$ & -- & $ 0.00$ & $0.36$ & $2.34^{+0.17}_{-0.17}$ & $6.06^{+2.84}_{-4.04}$ & $0.094^{+0.009}_{-0.009}$ & $0.109^{+0.140}_{-0.019}$ \\
NGC 3923$^{\ast}$ & 3 &  3 & $11.36$ & $-22.00$ & $216$ & $5.92$ & -- & -- & -- & $2.90^{+0.12}_{-0.26}$ & $2.76^{+3.31}_{-2.76}$ & $0.057^{+0.007}_{-0.002}$ & $0.037^{+0.070}_{-0.007}$ \\
NGC 3962$^{\ast}$ & 3 &  1 & $10.85$ & $-21.04$ & $211$ & $10.31$ & -- & -- & -- & $3.74$ & $4.72^{+0.08}_{-2.22}$ & $0.090$ & $0.105^{+0.122}_{-0.049}$ \\
NGC 4125 & 3 &  2 & $11.39$ & $-22.53$ & $229$ & $4.97$ & -- & $ 0.95$ & $0.93$ & $2.33^{+0.09}_{-0.09}$ & $3.65^{+0.24}_{-1.31}$ & $0.031^{+0.003}_{-0.003}$ & $0.032^{+0.069}_{-0.004}$ \\
NGC 4318 & 1 &  1 & $9.83$ & $-16.26$ & $101$ & $0.59$ & $0.34$ & $ 0.40$ & -- & $1.12$ & $2.36^{+0.42}_{-0.17}$ & $0.077$ & $0.336^{+0.168}_{-0.146}$ \\
NGC 4382$^{\ast}$ & 1,2,3 &  4 & $11.26$ & $-22.43$ & $196$ & $8.43$ & $0.19$ & $ 0.59$ & $0.33$ & $3.89^{+1.11}_{-1.35}$ & $3.11^{+0.63}_{-0.80}$ & $0.034^{+0.008}_{-0.010}$ & $0.090^{+0.086}_{-0.054}$ \\
NGC 4387 & 1,2,3 &  5 & $9.95$ & $-19.11$ & $84$ & $1.46$ & $0.43$ & -$ 1.00$ & $0.70$ & $2.04^{+0.11}_{-0.22}$ & $2.42^{+0.97}_{-2.42}$ & $0.007^{+0.002}_{-0.002}$ & $0.287^{+0.148}_{-0.147}$ \\
NGC 4434 & 1 &  1 & $10.59$ & $-19.55$ & $118$ & $1.44$ & $0.08$ & $ 0.44$ & -- & $2.62$ & $2.87^{+0.05}_{-0.46}$ & $0.150$ & $0.323^{+0.137}_{-0.111}$ \\
NGC 4458 & 1 &  1 & $10.36$ & $-18.89$ & $85$ & $1.34$ & $0.12$ & $ 0.41$ & $0.32$ & $2.72$ & $2.87^{+0.84}_{-0.09}$ & $0.075$ & $0.276^{+0.128}_{-0.135}$ \\
NGC 4459 & 1 &  1 & $10.56$ & $-20.88$ & $168$ & $3.30$ & $0.18$ & $ 0.22$ & $0.96$ & $3.09$ & $3.68^{+0.14}_{-0.67}$ & $0.028$ & $0.194^{+0.152}_{-0.096}$ \\
NGC 4464 & 1 &  1 & $10.26$ & $-18.40$ & $120$ & $0.55$ & $0.30$ & $ 0.40$ & -- & $2.11$ & $2.36^{+0.69}_{-0.17}$ & $0.228$ & $0.368^{+0.170}_{-0.134}$ \\
NGC 4467 & 1 &  1 & $9.71$ & $-16.87$ & $67$ & $0.39$ & $0.32$ & $ 0.60$ & $0.35$ & $1.89$ & $2.36^{+1.26}_{-0.17}$ & $0.028$ & $0.355^{+0.152}_{-0.140}$ \\
NGC 4473 & 1 &  1 & $11.13$ & $-20.82$ & $192$ & $3.14$ & $0.39$ & $ 1.03$ & $0.28$ & $3.60$ & $2.92^{+0.39}_{-0.27}$ & $0.148$ & $0.216^{+0.120}_{-0.101}$ \\
NGC 4476$^{\ast}$ & 3 &  3 & $10.25$ & $-19.85$ & $41$ & $1.57$ & $0.28$ & -- & $0.60$ & $1.19^{+0.26}_{-0.37}$ & $2.59^{+1.89}_{-0.76}$ & $0.355^{+0.114}_{-0.066}$ & $0.233^{+0.113}_{-0.096}$ \\
NGC 4478 & 1,3 &  3 & $10.29$ & $-19.77$ & $149$ & $1.15$ & $0.19$ & -$ 0.80$ & $0.95$ & $2.21^{+0.00}_{-0.18}$ & $2.89^{+0.90}_{-1.29}$ & $0.223^{+0.003}_{-0.195}$ & $0.296^{+0.206}_{-0.130}$ \\
NGC 4486a & 1 &  1 & $10.36$ & $-18.84$ & $154$ & $0.50$ & $0.45$ & $ 4.00$ & -- & $2.40$ & $2.41^{+0.99}_{-0.08}$ & $0.338$ & $0.368^{+0.158}_{-0.120}$ \\
NGC 4486b & 1 &  1 & $9.90$ & $-17.70$ & $200$ & $0.19$ & $0.40$ & $ 0.80$ & $0.59$ & $2.18$ & $2.41^{+0.69}_{-0.08}$ & $0.073$ & $0.325^{+0.135}_{-0.144}$ \\
NGC 4494 & 2,3 &  7 & $11.24$ & $-21.50$ & $155$ & $3.72$ & $0.15$ & $ 0.30$ & $1.24$ & $1.97^{+0.63}_{-0.07}$ & $3.06^{+1.31}_{-0.64}$ & $0.071^{+0.007}_{-0.018}$ & $0.117^{+0.187}_{-0.027}$ \\
NGC 4515 & 1 &  1 & $10.03$ & $-18.52$ & $90$ & $0.87$ & $0.14$ & $ 2.00$ & $0.47$ & $3.87$ & $2.48^{+0.79}_{-0.76}$ & $0.254$ & $0.336^{+0.133}_{-0.087}$ \\
NGC 4551 & 1,2 &  5 & $10.02$ & $-19.08$ & $100$ & $1.25$ & $0.32$ & -$ 0.70$ & $0.55$ & $1.85^{+0.24}_{-0.58}$ & $2.79^{+0.59}_{-1.19}$ & $0.040^{+0.025}_{-0.014}$ & $0.300^{+0.155}_{-0.182}$ \\
NGC 4564 & 2,3 &  7 & $10.53$ & $-19.66$ & $153$ & $2.75$ & $0.45$ & $ 2.20$ & $1.05$ & $1.62^{+0.75}_{-0.47}$ & $3.13^{+1.60}_{-2.04}$ & $0.108^{+0.087}_{-0.029}$ & $0.141^{+0.235}_{-0.050}$ \\
NGC 4621 & 1,2,3 &  7 & $11.26$ & $-21.41$ & $237$ & $6.51$ & $0.34$ & $ 1.50$ & $0.81$ & $3.23^{+1.51}_{-0.75}$ & $3.06^{+1.70}_{-0.18}$ & $0.053^{+0.005}_{-0.044}$ & $0.140^{+0.168}_{-0.050}$ \\
NGC 4660 & 2 &  2 & $10.46$ & $-19.41$ & $191$ & $1.05$ & $0.40$ & $ 2.70$ & $1.04$ & $1.88^{+0.41}_{-0.41}$ & $2.87^{+0.56}_{-0.37}$ & $0.402^{+0.161}_{-0.161}$ & $0.323^{+0.179}_{-0.140}$ \\
NGC 4697 & 2,3 &  4 & $11.23$ & $-21.49$ & $165$ & $5.13$ & $0.41$ & $ 1.40$ & $0.78$ & $3.51^{+0.38}_{-2.05}$ & $2.87^{+0.92}_{-0.74}$ & $0.001^{+0.020}_{-0.001}$ & $0.103^{+0.152}_{-0.068}$ \\
NGC 4742 & 2,3 &  4 & $10.30$ & $-19.79$ & $93$ & $1.24$ & -- & $ 0.41$ & $1.62$ & $2.06^{+9.33}_{-0.08}$ & $3.28^{+3.09}_{-0.61}$ & $0.232^{+0.012}_{-0.045}$ & $0.249^{+0.100}_{-0.066}$ \\
NGC 5018$^{\ast}$ & 3 &  2 & $11.36$ & $-22.29$ & $223$ & $4.17$ & $0.25$ & -- & -- & $4.07^{+0.15}_{-0.15}$ & $2.72^{+0.57}_{-2.72}$ & $0.040^{+0.005}_{-0.005}$ & $0.106^{+0.124}_{-0.036}$ \\
NGC 5127$^{\ast}$ & 3 &  1 & $11.15$ & $-21.87$ & -- & $8.32$ & -- & -- & -- & $2.90$ & $2.30^{+1.76}_{-0.20}$ & $0.032$ & $0.046^{+0.057}_{-0.016}$ \\
NGC 5444$^{\ast}$ & 3 &  3 & $11.20$ & $-21.91$ & $221$ & $5.46$ & -- & -- & -- & $2.94^{+2.03}_{-0.11}$ & $3.13^{+1.58}_{-0.57}$ & $0.078^{+0.005}_{-0.078}$ & $0.105^{+0.117}_{-0.068}$ \\
NGC 5576 & 3 &  3 & $10.81$ & $-20.82$ & $187$ & $3.52$ & $0.30$ & -$ 0.50$ & $0.22$ & $3.90^{+1.06}_{-1.06}$ & $3.06^{+8.90}_{-0.56}$ & $0.082^{+0.018}_{-0.018}$ & $0.200^{+0.113}_{-0.098}$ \\
NGC 5638$^{\ast}$ & 3 &  3 & $10.65$ & $-20.60$ & $159$ & $1.99$ & $0.08$ & $ 0.20$ & $0.72$ & $1.33^{+0.58}_{-0.58}$ & $4.01^{+2.06}_{-1.66}$ & $0.169^{+0.070}_{-0.070}$ & $0.129^{+0.178}_{-0.038}$ \\
NGC 5812 & 3,4 &  4 & $10.97$ & $-21.12$ & $204$ & $2.63$ & $0.05$ & $ 0.00$ & $0.52$ & $3.17^{+0.08}_{-0.64}$ & $3.86^{+0.94}_{-2.20}$ & $0.133^{+0.022}_{-0.004}$ & $0.221^{+0.106}_{-0.106}$ \\
NGC 5831 & 3 &  3 & $10.68$ & $-20.42$ & $166$ & $2.56$ & $0.17$ & -- & $0.19$ & $3.47^{+0.30}_{-1.06}$ & $ 3.40^{+4.40}_{-2.00}$ & $0.054^{+0.021}_{-0.001}$ & $0.205^{+0.140}_{-0.103}$ \\
NGC 5845 & 3 &  3 & $10.38$ & $-19.46$ & $251$ & $0.51$ & $0.15$ & $ 0.80$ & $0.91$ & $1.18^{+0.34}_{-0.02}$ & $2.59^{+6.31}_{-1.91}$ & $0.455^{+0.027}_{-0.077}$ & $0.404^{+0.123}_{-0.117}$ \\
NGC 6482$^{\ast}$ & 3 &  2 & $11.57$ & $-22.76$ & $287$ & $4.51$ & $0.27$ & -- & -- & $2.13^{+0.05}_{-0.05}$ & $3.57^{+0.90}_{-1.07}$ & $0.062^{+0.001}_{-0.001}$ & $0.084^{+0.120}_{-0.055}$ \\
NGC 6487$^{\ast}$ & 3 &  3 & $11.54$ & $-23.11$ & -- & $12.91$ & -- & -- & -- & $2.26^{+0.32}_{-0.02}$ & $4.95^{+3.44}_{-3.05}$ & $0.101^{+0.002}_{-0.018}$ & $0.089^{+0.046}_{-0.043}$ \\
NGC 7562 & 3 &  2 & $11.42$ & $-22.27$ & $243$ & $6.01$ & $0.29$ & -- & -- & $2.49^{+0.10}_{-0.10}$ & $2.95^{+0.33}_{-0.48}$ & $0.079^{+0.009}_{-0.009}$ & $0.046^{+0.073}_{-0.017}$ \\
NGC 7626 & 3 &  3 & $11.67$ & $-22.62$ & $234$ & $8.81$ & $0.13$ & $ 0.11$ & $0.12$ & $2.94^{+0.20}_{-0.04}$ & $3.39^{+1.00}_{-1.43}$ & $0.042^{+0.001}_{-0.007}$ & $0.070^{+0.036}_{-0.038}$ \\
VCC 1199 & 1 &  1 & $8.97$ & $-15.52$ & $69$ & $0.17$ & $0.28$ & $ 0.05$ & -- & $1.91$ & $2.42^{+0.68}_{-0.34}$ & $0.049$ & $0.322^{+0.120}_{-0.149}$ \\
VCC 1440 & 1 &  1 & $9.13$ & $-16.75$ & $59$ & $0.70$ & $0.17$ & $ 0.60$ & -- & $3.43$ & $2.65^{+0.40}_{-0.34}$ & $0.009$ & $0.306^{+0.100}_{-0.102}$ \\
VCC 1627 & 1 &  1 & $9.11$ & $-16.43$ & -- & $0.30$ & $0.20$ & $ 0.40$ & -- & $2.26$ & $2.42^{+0.68}_{-0.34}$ & $0.018$ & $0.346^{+0.158}_{-0.142}$ \\
VCC 1871 & 1 &  1 & $9.27$ & $-17.33$ & $51$ & $0.61$ & $0.14$ & $ 0.80$ & -- & $1.90$ & $2.36^{+1.26}_{-0.17}$ & $0.010$ & $0.403^{+0.239}_{-0.155}$ \\
UGC 10638 & 2 &  1 & $11.90$ & $-22.66$ & -- & $14.79$ & -- & -- & -- & $4.83$ & $2.53^{+0.97}_{-0.36}$ & $0.001$ & $0.088^{+0.093}_{-0.042}$ \\
ESO 462-15 & 2 &  2 & $11.99$ & $-22.83$ & $289$ & $7.24$ & $0.27$ & -- & -- & $2.65^{+1.14}_{-1.14}$ & $3.13^{+1.47}_{-0.61}$ & $0.141^{+0.140}_{-0.140}$ & $0.099^{+0.127}_{-0.067}$ \\
IC 2738 & 2 &  1 & $11.64$ & $-22.21$ & -- & $23.09$ & -- & -- & -- & $6.33$ & $3.09^{+0.15}_{-0.18}$ & $0.024$ & $0.090^{+0.092}_{-0.029}$ \\
A0147-M1$^{\ast}$ & 2 &  1 & $11.86$ & $-22.60$ & -- & $11.34$ & -- & -- & -- & $3.86$ & $3.54^{+0.15}_{-0.47}$ & $0.030$ & $0.078^{+0.063}_{-0.017}$ \\
A0160-M1$^{\ast}$ & 2 &  1 & $12.19$ & $-23.18$ & -- & $21.72$ & -- & -- & -- & $1.54$ & $2.64^{+0.57}_{-0.11}$ & $0.116$ & $0.047^{+0.058}_{-0.002}$ \\
A0189-M1 & 2 &  1 & $11.46$ & $-21.89$ & -- & $9.33$ & -- & -- & -- & $3.69$ & $2.59^{+0.02}_{-0.27}$ & $0.017$ & $0.090^{+0.099}_{-0.054}$ \\
A0261-M1 & 2 &  1 & $12.06$ & $-22.95$ & -- & $22.91$ & -- & -- & -- & $6.85$ & $3.09^{+0.15}_{-0.92}$ & $0.071$ & $0.090^{+0.078}_{-0.043}$ \\
A0419-M1 & 2 &  1 & $11.40$ & $-21.79$ & -- & $8.13$ & -- & -- & -- & $2.11$ & $2.59^{+0.02}_{-0.27}$ & $0.153$ & $0.098^{+0.096}_{-0.061}$ \\
A0912-M1 & 2 &  3 & $11.66$ & $-22.24$ & -- & $9.55$ & -- & -- & -- & $3.08^{+0.24}_{-0.24}$ & $2.59^{+1.10}_{-0.25}$ & $0.065^{+0.011}_{-0.011}$ & $0.105^{+0.116}_{-0.045}$ \\
A1308-M1$^{\ast}$ & 2 &  1 & $12.34$ & $-23.44$ & -- & $26.66$ & -- & -- & -- & $4.60$ & $3.09^{+0.15}_{-0.56}$ & $0.043$ & $0.077^{+0.058}_{-0.030}$ \\
A1836-M1$^{\ast}$ & 2 &  2 & $12.29$ & $-23.34$ & -- & $22.05$ & -- & -- & -- & $4.23^{+0.13}_{-0.13}$ & $2.82^{+0.45}_{-0.45}$ & $0.018^{+0.001}_{-0.001}$ & $0.073^{+0.045}_{-0.027}$ \\
A1983-M1$^{\ast}$ & 2 &  1 & $11.72$ & $-22.35$ & -- & $7.41$ & -- & -- & -- & $3.79$ & $3.06^{+0.83}_{-0.05}$ & $0.033$ & $0.105^{+0.120}_{-0.051}$ \\
\enddata
\tablenotetext{ \, }{{\footnotesize Compiled and fitted parameters for the confirmed cusp ellipticals 
in our observed samples. Columns show: (1) Object name. (2) Source for surface brightness 
profiles, where $1=$\citet{jk:profiles}, $2=$\citet{lauer:bimodal.profiles}, $3=$\citet{bender:data}, 
$4=$\citet{rj:profiles}. 
(3) Total number of different surface brightness profiles in our combined samples for 
the given object. (4) Stellar mass [$\log{M_{\ast}/M_{\sun}}$]. (5) $V$-band absolute magnitude. 
(6) Velocity dispersion [km/s]. (7) Effective (half-light) radius of the 
{\em total} light profile [kpc]. (8) Ellipticity. (9) Boxy/diskyness. (10) Rotation. 
(11) Outer \Sersic\ index $n_{s}$ of the two-component best-fit profile. Where multiple 
profiles are available for the same object, we show the median and minimum/maximum 
range of fitted $n_{s}$ values. 
(12) Range of outer \Sersic\ indices fit in the same manner to the best-fit simulations, 
at $t\approx1-3$\,Gyr after the merger when the system has relaxed.
(13) Fraction of light in the inner or ``extra light'' component of the fits. 
Where multiple 
profiles are available for the same object, we show the median and minimum/maximum 
range of fitted values. 
(14) Fraction of light from stars produced in the central, merger-induced starburst 
in the best-fit simulations ($\pm$ the approximate interquartile range allowed). 
This list includes all systems morphologically 
classified as ellipticals in \citet{jk:profiles} (all are E0-E4), or (where not in that sample)  
\citet{lauer:bimodal.profiles} (all are E or E/BCG). \\
}}
\tablenotetext{\tableast}{{\footnotesize Systems with ambiguous or uncertain cusp status. 
These are systems for which different sources disagree on their cusp/core status, 
or for which observations of the central regions are unavailable/ambiguous but 
for which some other evidence (e.g.\ stellar populations, gas/dust content, or kinematics) suggest a 
gas-rich merger origin. We include them here for completeness, but our 
conclusions are insensitive to their inclusion/exclusion, and they are not generally shown 
in our analysis.}}
\end{\tableset}

\clearpage

\begin{\tableset}{lccccccccccccc}
\tabletypesize{\scriptsize}
\tablecaption{Fits to Recent Merger Remnants\label{tbl:mgr.fits}}
\tablecolumns{14}
\tablewidth{0pt}
\tablehead{
\colhead{Name} &
\colhead{Ref.} &
\colhead{${\rm N}_{\rm phot}$} &
\colhead{$M_{\ast}$} &
\colhead{$M_{K}$} &
\colhead{$\sigma$} &
\colhead{$R_{e}$} &
\colhead{$\epsilon$} &
\colhead{$100\,a_{4}/a$} &
\colhead{$(v/\sigma)^{\ast}$} &
\colhead{$n_{s}$ (fit)} &
\colhead{$n_{s}$ (sim)} &
\colhead{$f_{e}$ (fit)} & 
\colhead{$f_{sb}$ (sim)} \\
\colhead{(1)} &
\colhead{(2)} &
\colhead{(3)} &
\colhead{(4)} &
\colhead{(5)} &
\colhead{(6)} &
\colhead{(7)} &
\colhead{(8)} &
\colhead{(9)} &
\colhead{(10)} &
\colhead{(11)} &
\colhead{(12)} &
\colhead{(13)} &
\colhead{(14)} 
}
\startdata
NGC 34 & 4 & 1 & $11.09$ & $-24.61$ & $201$ & $0.84$ & $0.11$ & $ 1.36$ & $1.41$ & $3.49$ & $4.62^{+0.50}_{-1.83}$ & $0.389$ & $0.388^{+0.120}_{-0.098}$ \\
NGC 455 & 4 & 1 & $11.10$ & $-24.64$ & $234$ & $3.33$ & $0.21$ & -$ 0.41$ & $1.08$ & $2.02$ & $3.20^{+1.61}_{-0.10}$ & $0.200$ & $0.194^{+0.105}_{-0.089}$ \\
NGC 828 & 4 & 1 & $11.40$ & $-25.36$ & -- & $3.51$ & $0.41$ & $ 1.70$ & -- & $3.60$ & $3.65^{+0.24}_{-1.31}$ & $0.189$ & $0.231^{+0.154}_{-0.128}$ \\
NGC 1210 & 4 & 1 & $10.72$ & $-23.72$ & $247$ & $2.43$ & $0.08$ & $ 0.25$ & $0.46$ & $1.92$ & $3.32^{+0.10}_{-0.31}$ & $0.158$ & $0.231^{+0.165}_{-0.108}$ \\
NGC 1614 & 4 & 1 & $11.15$ & $-24.74$ & $146$ & $1.69$ & $0.13$ & $ 0.76$ & $1.91$ & $2.29$ & $3.30^{+0.39}_{-0.64}$ & $0.351$ & $0.300^{+0.107}_{-0.110}$ \\
NGC 2418 & 4 & 1 & $11.38$ & $-25.31$ & $288$ & $4.81$ & $0.16$ & -$ 0.25$ & $0.66$ & $2.10$ & $3.29^{+1.38}_{-0.47}$ & $0.128$ & $0.168^{+0.169}_{-0.077}$ \\
NGC 2623 & 4 & 1 & $10.93$ & $-24.22$ & $191$ & $1.32$ & $0.24$ & $ 2.39$ & $0.37$ & $4.88$ & $5.32^{+3.07}_{-2.64}$ & $0.173$ & $0.300^{+0.107}_{-0.095}$ \\
NGC 2655 & 4 & 1 & $10.72$ & $-23.70$ & $169$ & $1.14$ & $0.19$ & $ 0.05$ & $1.00$ & $2.44$ & $2.76^{+1.13}_{-0.12}$ & $0.055$ & $0.322^{+0.126}_{-0.110}$ \\
NGC 2744 & 4 & 1 & $10.36$ & $-22.83$ & -- & $3.44$ & $0.51$ & -$ 6.42$ & -- & $1.75$ & $2.36^{+0.69}_{-0.17}$ & $0.050$ & $0.128^{+0.208}_{-0.052}$ \\
NGC 2782 & 4 & 1 & $10.77$ & $-23.83$ & $196$ & $3.30$ & $0.26$ & $ 1.03$ & $0.85$ & $1.74$ & $4.67^{+0.25}_{-1.73}$ & $0.229$ & $0.210^{+0.126}_{-0.101}$ \\
NGC 2914 & 4 & 1 & $10.64$ & $-23.51$ & $186$ & $1.39$ & $0.35$ & $ 0.68$ & $1.26$ & $0.62$ & $2.53^{+0.97}_{-0.36}$ & $0.498$ & $0.244^{+0.113}_{-0.107}$ \\
NGC 3256 & 4 & 1 & $11.14$ & $-24.72$ & $241$ & $1.79$ & $0.19$ & $ 1.35$ & $0.41$ & $1.85$ & $4.57^{+0.11}_{-2.61}$ & $0.090$ & $0.299^{+0.117}_{-0.163}$ \\
NGC 3310$^{\dagger}$ & 4 & 1 & $10.04$ & $-22.07$ & -- & $0.70$ & $0.22$ & -$ 2.00$ & -- & $1.59$ & $2.55^{+5.91}_{-0.89}$ & $0.078$ & $0.323^{+0.132}_{-0.129}$ \\
NGC 3597 & 4 & 1 & $10.72$ & $-23.72$ & $174$ & $0.83$ & $0.40$ & $ 1.11$ & $0.95$ & $0.64$ & $3.06^{+1.12}_{-0.63}$ & $0.614$ & $0.358^{+0.169}_{-0.123}$ \\
NGC 3656 & 4 & 1 & $10.72$ & $-23.70$ & $132$ & $2.55$ & $0.12$ & -$ 1.70$ & -- & $3.45$ & $2.94^{+2.18}_{-0.42}$ & $0.038$ & $0.195^{+0.132}_{-0.093}$ \\
NGC 3921 & 4 & 1 & $11.31$ & $-25.13$ & $222$ & $3.45$ & $0.21$ & $ 0.99$ & $1.02$ & $2.48$ & $4.15^{+0.97}_{-1.97}$ & $0.261$ & $0.297^{+0.120}_{-0.085}$ \\
NGC 4004$^{\dagger}$ & 4 & 1 & $10.38$ & $-22.89$ & $33$ & $3.17$ & $0.62$ & $ 2.58$ & $0.47$ & $1.50$ & $2.66^{+1.92}_{-0.70}$ & $0.422$ & $0.194^{+0.143}_{-0.104}$ \\
NGC 4194 & 4 & 1 & $10.51$ & $-23.21$ & $116$ & $0.57$ & $0.24$ & $ 0.91$ & $1.24$ & $1.57$ & $3.64^{+1.10}_{-1.74}$ & $0.535$ & $0.355^{+0.149}_{-0.091}$ \\
NGC 4441 & 4 & 1 & $10.42$ & $-22.98$ & $139$ & $1.53$ & $0.17$ & $ 0.87$ & $0.94$ & $2.47$ & $3.27^{+0.73}_{-0.50}$ & $0.140$ & $0.194^{+0.133}_{-0.081}$ \\
NGC 5018 & 4 & 1 & $11.32$ & $-25.15$ & $222$ & $2.62$ & $0.25$ & $ 1.17$ & $0.74$ & $3.14$ & $2.53^{+2.59}_{-0.44}$ & $0.062$ & $0.194^{+0.126}_{-0.096}$ \\
NGC 6052$^{\dagger}$ & 4 & 1 & $10.65$ & $-23.55$ & $80$ & $4.82$ & $0.44$ & $ 1.38$ & $0.60$ & $0.85$ & $3.54^{+0.92}_{-1.13}$ & $0.044$ & $0.118^{+0.190}_{-0.050}$ \\
NGC 6598 & 4 & 1 & $11.46$ & $-25.51$ & -- & $6.08$ & $0.16$ & -$ 0.03$ & -- & $3.39$ & $3.20^{+0.86}_{-0.47}$ & $0.045$ & $0.103^{+0.102}_{-0.067}$ \\
NGC 7135 & 4 & 1 & $10.82$ & $-23.95$ & $277$ & $4.36$ & $0.18$ & $ 0.39$ & $0.82$ & $4.13$ & $6.06^{+2.84}_{-4.04}$ & $0.081$ & $0.166^{+0.085}_{-0.077}$ \\
NGC 7252 & 4 & 1 & $11.19$ & $-24.84$ & $166$ & $2.53$ & $0.07$ & $ 0.24$ & $1.42$ & $1.27$ & $2.76^{+3.31}_{-2.76}$ & $0.316$ & $0.249^{+0.146}_{-0.124}$ \\
NGC 7585 & 4 & 1 & $11.24$ & $-24.98$ & $211$ & $4.45$ & $0.29$ & $ 0.33$ & $0.21$ & $2.41$ & $4.72^{+0.08}_{-2.22}$ & $0.107$ & $0.141^{+0.178}_{-0.043}$ \\
NGC 7727 & 4 & 1 & $10.93$ & $-24.23$ & $231$ & $2.28$ & $0.24$ & -$ 1.63$ & $1.18$ & $2.63$ & $3.02^{+0.80}_{-0.14}$ & $0.127$ & $0.249^{+0.159}_{-0.120}$ \\
UGC 6 & 4 & 1 & $10.84$ & $-24.01$ & $220$ & $1.40$ & $0.19$ & -$ 0.04$ & $0.56$ & $2.10$ & $2.92^{+0.39}_{-0.27}$ & $0.630$ & $0.345^{+0.124}_{-0.115}$ \\
UGC 2238$^{\dagger}$ & 4 & 1 & $11.08$ & $-24.58$ & -- & $1.42$ & $0.53$ & $ 2.60$ & -- & $1.09$ & $2.87^{+0.05}_{-0.46}$ & $0.165$ & $0.315^{+0.122}_{-0.150}$ \\
UGC 4079 & 4 & 1 & $10.75$ & $-23.78$ & -- & $3.65$ & $0.54$ & -$ 1.06$ & -- & $2.28$ & $2.87^{+0.84}_{-0.09}$ & $0.040$ & $0.103^{+0.124}_{-0.066}$ \\
UGC 4635 & 4 & 1 & $11.13$ & $-24.71$ & $251$ & $2.48$ & $0.34$ & $ 0.76$ & $0.65$ & $2.90$ & $3.68^{+0.14}_{-0.67}$ & $0.077$ & $0.270^{+0.133}_{-0.142}$ \\
UGC 5101 & 4 & 1 & $11.46$ & $-25.50$ & $287$ & $1.07$ & $0.18$ & $ 0.62$ & $1.29$ & $4.60$ & $2.36^{+1.26}_{-0.17}$ & $0.318$ & $0.359^{+0.149}_{-0.058}$ \\
UGC 8058$^{\ast}$ & 4 & 1 & $12.31$ & $-27.55$ & -- & $0.82$ & $0.05$ & -$ 0.30$ & -- & $4.60$ & $3.51^{+0.31}_{-0.63}$ & $0.769$ & $0.447^{+0.200}_{-0.152}$ \\
UGC 9829$^{\dagger}$ & 4 & 1 & $11.24$ & $-24.96$ & $134$ & $6.61$ & $0.41$ & $ 2.62$ & $1.00$ & $1.37$ & $2.59^{+1.89}_{-0.76}$ & $0.021$ & $0.117^{+0.157}_{-0.054}$ \\
UGC 10607 & 4 & 1 & $11.34$ & $-25.20$ & $211$ & $1.59$ & $0.24$ & $ 0.26$ & $0.99$ & $1.69$ & $2.36^{+0.42}_{-0.17}$ & $0.332$ & $0.302^{+0.132}_{-0.102}$ \\
UGC 10675 & 4 & 1 & $11.17$ & $-24.80$ & $177$ & $1.46$ & $0.18$ & $ 0.72$ & $0.57$ & $2.18$ & $3.11^{+0.63}_{-0.80}$ & $0.638$ & $0.368^{+0.139}_{-0.105}$ \\
UGC 11905 & 4 & 1 & $11.05$ & $-24.51$ & $222$ & $1.89$ & $0.24$ & -$ 0.76$ & $0.64$ & $1.21$ & $2.42^{+0.97}_{-2.42}$ & $0.417$ & $0.300^{+0.113}_{-0.089}$ \\
AM 0318-230 & 4 & 1 & $11.29$ & $-25.09$ & -- & $3.64$ & $0.17$ & -$ 0.04$ & -- & $2.27$ & $3.54^{+0.15}_{-0.47}$ & $0.283$ & $0.264^{+0.104}_{-0.135}$ \\
AM 0612-373 & 4 & 1 & $11.52$ & $-25.65$ & $303$ & $4.71$ & $0.11$ & -$ 0.64$ & $0.80$ & $1.55$ & $2.64^{+0.57}_{-0.11}$ & $0.163$ & $0.102^{+0.081}_{-0.065}$ \\
AM 0956-282 & 4 & 1 & $9.39$ & $-20.50$ & -- & $2.18$ & -- & -- & -- & $2.25$ & $2.59^{+0.02}_{-0.27}$ & $0.210$ & $0.221^{+0.106}_{-0.085}$ \\
AM 1158-333 & 4 & 1 & $10.26$ & $-22.61$ & -- & $1.41$ & $0.30$ & $ 1.50$ & -- & $2.99$ & $3.09^{+0.15}_{-0.92}$ & $0.316$ & $0.307^{+0.114}_{-0.086}$ \\
AM 1255-430 & 4 & 1 & $11.22$ & $-24.93$ & $243$ & $5.18$ & $0.28$ & -$ 0.65$ & $0.33$ & $0.87$ & $2.59^{+0.02}_{-0.27}$ & $0.261$ & $0.248^{+0.168}_{-0.139}$ \\
AM 1300-233 & 4 & 1 & $11.11$ & $-24.65$ & -- & $4.28$ & $0.68$ & -$ 0.90$ & -- & $1.92$ & $2.59^{+1.10}_{-0.25}$ & $0.102$ & $0.118^{+0.122}_{-0.044}$ \\
AM 1419-263 & 4 & 1 & $11.23$ & $-24.94$ & $260$ & $3.61$ & $0.27$ & $ 0.38$ & $0.43$ & $3.15$ & $3.09^{+0.15}_{-0.18}$ & $0.057$ & $0.114^{+0.121}_{-0.037}$ \\
AM 2038-382 & 4 & 1 & $11.13$ & $-24.70$ & $257$ & $1.77$ & $0.17$ & -$ 0.03$ & $1.21$ & $2.51$ & $3.09^{+0.15}_{-0.56}$ & $0.497$ & $0.355^{+0.112}_{-0.106}$ \\
AM 2055-425 & 4 & 1 & $11.29$ & $-25.08$ & $185$ & $2.09$ & $0.05$ & $ 0.75$ & $0.88$ & $0.75$ & $2.82^{+0.45}_{-0.45}$ & $0.344$ & $0.248^{+0.137}_{-0.133}$ \\
AM 2246-490 & 4 & 1 & $11.47$ & $-25.52$ & $267$ & $4.16$ & $0.05$ & $ 0.31$ & $0.54$ & $2.82$ & $3.06^{+0.83}_{-0.05}$ & $0.320$ & $0.197^{+0.122}_{-0.092}$ \\
Arp 156 & 4 & 1 & $11.59$ & $-25.81$ & $288$ & $6.95$ & $0.17$ & $ 1.59$ & $1.20$ & $3.14$ & $3.13^{+1.47}_{-0.61}$ & $0.116$ & $0.168^{+0.065}_{-0.082}$ \\
Arp 187 & 4 & 1 & $11.36$ & $-25.25$ & -- & $4.37$ & $0.47$ & $ 2.11$ & -- & $3.93$ & $3.07^{+2.05}_{-0.63}$ & $0.062$ & $0.141^{+0.093}_{-0.064}$ \\
Arp 193 & 4 & 1 & $11.00$ & $-24.40$ & $172$ & $1.58$ & $0.49$ & $ 0.82$ & $0.65$ & $1.17$ & $3.02^{+0.80}_{-0.14}$ & $0.332$ & $0.296^{+0.120}_{-0.142}$ \\
Arp 230 & 4 & 1 & $9.91$ & $-21.75$ & -- & $1.08$ & $0.35$ & $ 9.68$ & -- & $1.20$ & $2.64^{+0.70}_{-0.11}$ & $0.063$ & $0.335^{+0.167}_{-0.141}$ \\
IC 5298 & 4 & 1 & $11.22$ & $-24.92$ & $193$ & $1.91$ & $0.08$ & $ 0.81$ & $0.36$ & $1.62$ & $2.57^{+1.61}_{-0.79}$ & $0.335$ & $0.299^{+0.119}_{-0.106}$ \\
Mrk 1014$^{\ast}$ & 4 & 1 & $12.31$ & $-28.16$ & -- & $1.00$ & -- & -- & -- & $3.86$ & $2.82^{+8.00}_{-2.82}$ & $0.674$ & $0.263^{+0.129}_{-0.155}$ \\
\enddata
\tablenotetext{\,}{{\footnotesize As Table~\ref{tbl:cusp.fits}, but 
for the recent merger remnant sample of \citet{rj:profiles}. Note 
that we show the $K$-band as opposed to $V$-band absolute 
magnitudes. Systems marked ($\ast$) are excluded from our comparison in this 
paper owing to contamination from a central AGN. Systems marked ($\dagger$) 
should be regarded with caution, as unrelaxed or prominent disk/bar 
features make our fits unreliable. Some of the best-fit simulation parameters ($n_{s}$(sim) and 
$f_{sb}$(sim)) are slightly different from those in \paperone\ owing to an expanded 
set of simulations, but it makes no difference for our comparisons.}}
\end{\tableset}

\clearpage

\begin{appendix}

\section{Fits to the Sample of Kormendy et.\ al.\ 2008}
\label{sec:appendix:jk}

\begin{figure*}
    \centering
    \plotter{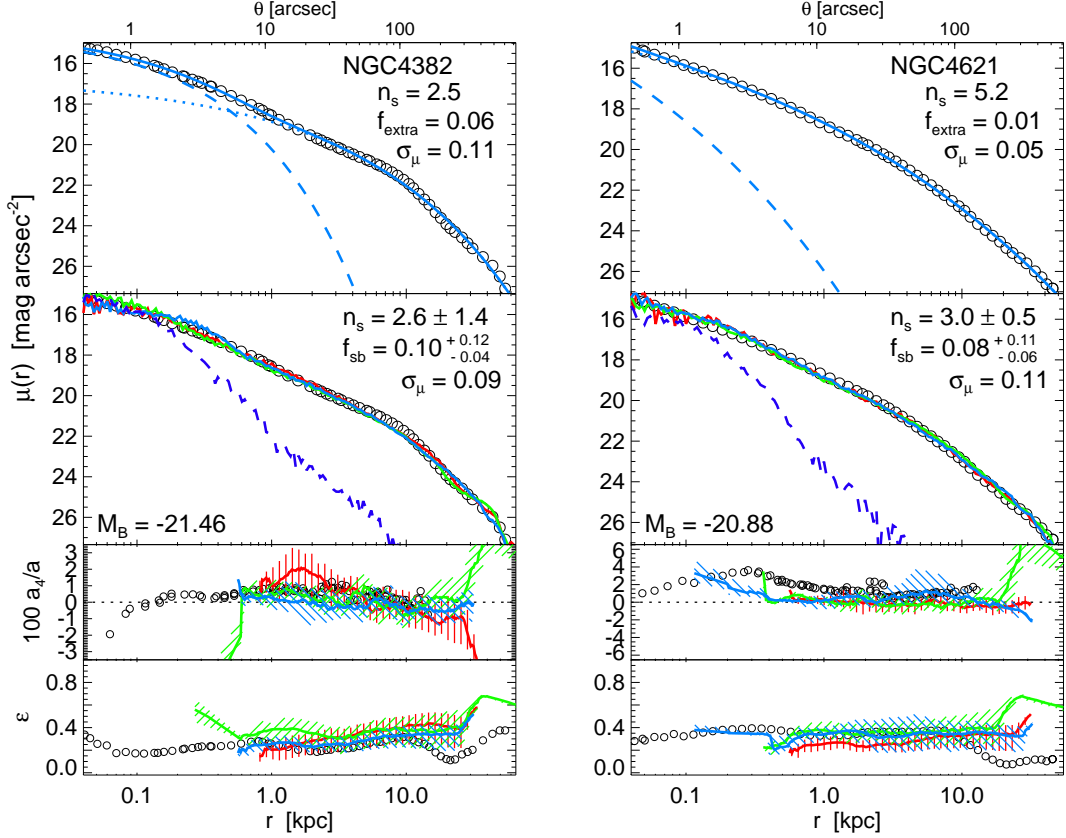}
    \caption{As Figure~\ref{fig:jk1}, but in log-log space. 
    Surface brightness profiles are shown for cuspy ellipticals in the 
    Virgo cluster.
    Open circles show the observations, from \citet{jk:profiles}. 
    These are the highest-mass cusp or extra light ellipticals in Virgo$^{\ref{foot:4382}}$
    ($\sim2\,\mstar$).
    {\em Top:} Observed V-band surface brightness profile with our 
    two component best-fit model (solid, dashed, and dotted lines show the 
    total, inner/extra light component, and outer/pre-starburst component). 
    The best-fit outer \Sersic\ index, extra light fraction, and variance about the 
    fit are shown.
    {\em  Middle:} Colored lines show the corresponding surface brightness 
    profiles from the three simulations in our library which correspond 
    most closely to the observed system (shown outside the gravitational 
    softening length, $\sim30$\,pc). Dashed line shows the 
    profile of the starburst light in the best-matching simulation. 
    The range of outer \Sersic\ indices in the simulations (i.e.\ across sightlines for 
    these objects) and range of starburst mass fractions which match the 
    observed profile are shown, with the variance of the observations about the 
    best-fit simulation$^{\ref{foot:explainfits}}$. 
    {\em Bottom:} Observed disky/boxy-ness ($a_{4}$) and ellipticity profiles, 
    with the median (solid) and $25-75\%$ range (shaded) corresponding profile 
    from the best-fitting simulations above. Note that these are not fitted for in any sense. 
    Figures~\ref{fig:jk2.log}-\ref{fig:jk10.log}
    show the other cusp ellipticals in the sample, ranked from most to least massive.
    \label{fig:jk1.log.b}}
\end{figure*}

\begin{figure*}
    \centering
    \plotter{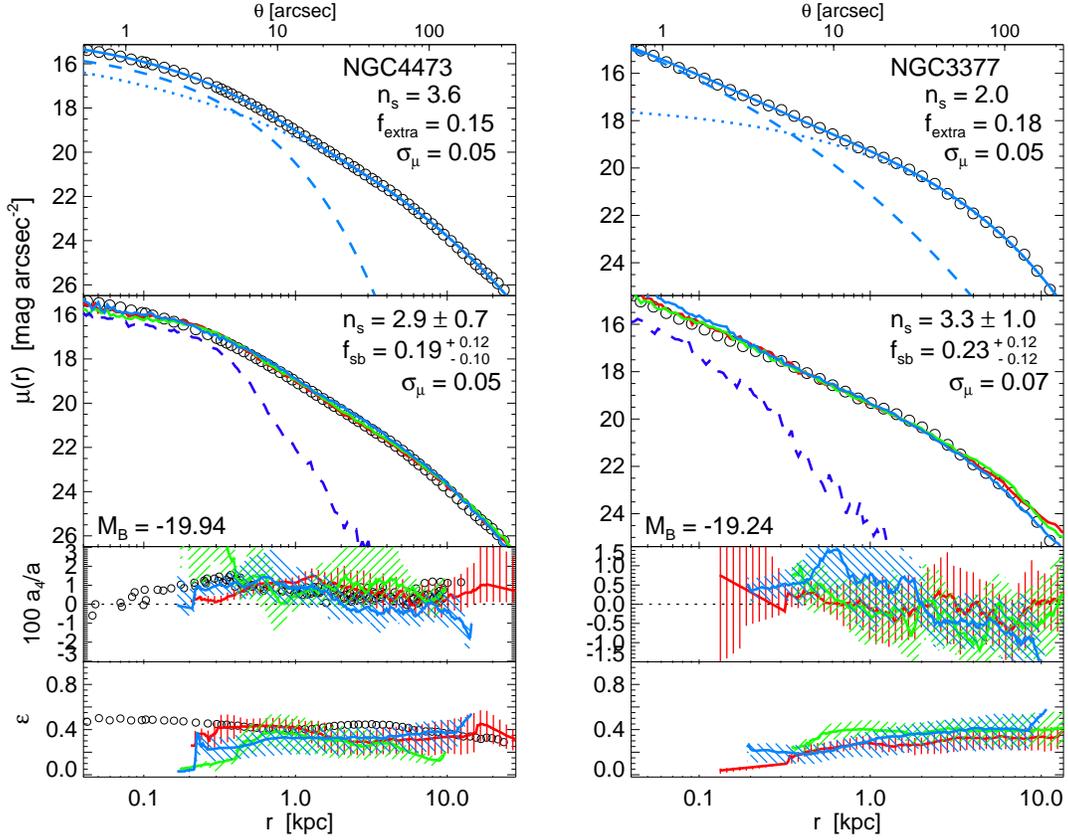}
    \caption{The next most massive cusp ellipticals ($\sim1\,\mstar$). Note that NGC 3377 is not 
    a Virgo member. (As Figure~\ref{fig:jk2}, but in log-log space.)
    \label{fig:jk2.log}}
\end{figure*}

\begin{figure*}
    \centering
    \plotter{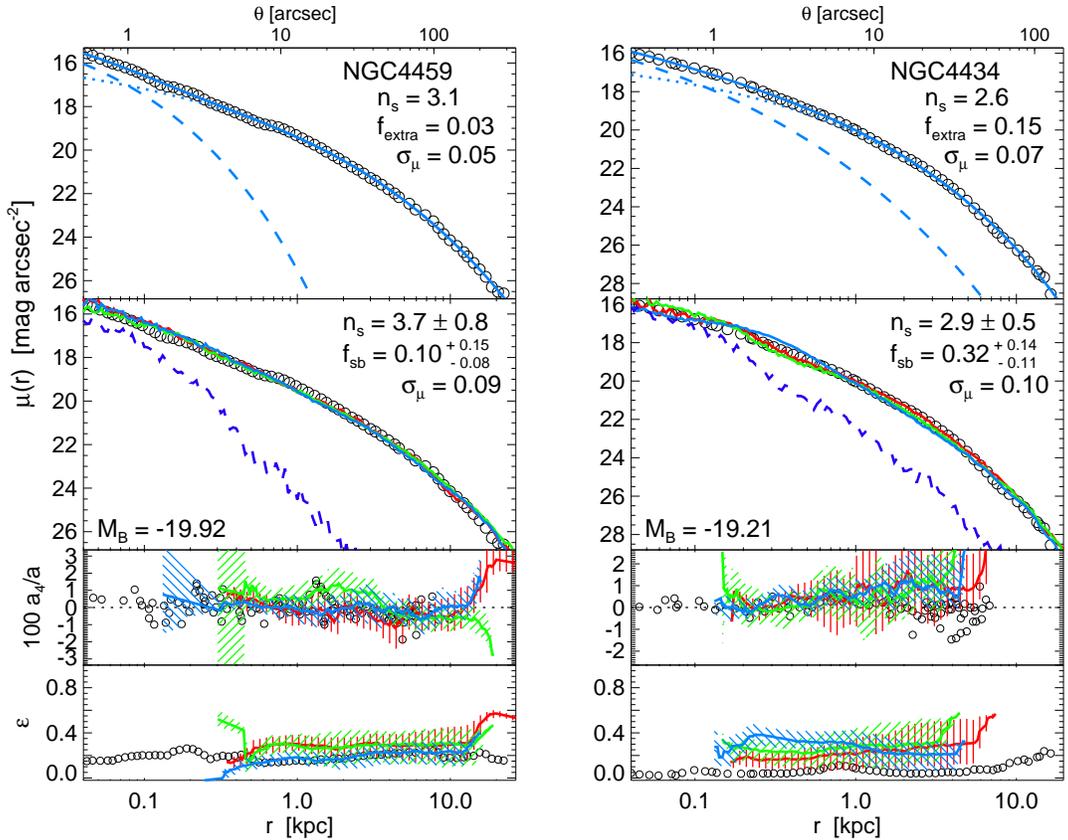}
    \caption{The next most massive cusp ellipticals ($\sim0.5\,\mstar$).
    (As Figure~\ref{fig:jk3}, but in log-log space.) 
    \label{fig:jk3.log}}
\end{figure*}

\begin{figure*}
    \centering
    \plotter{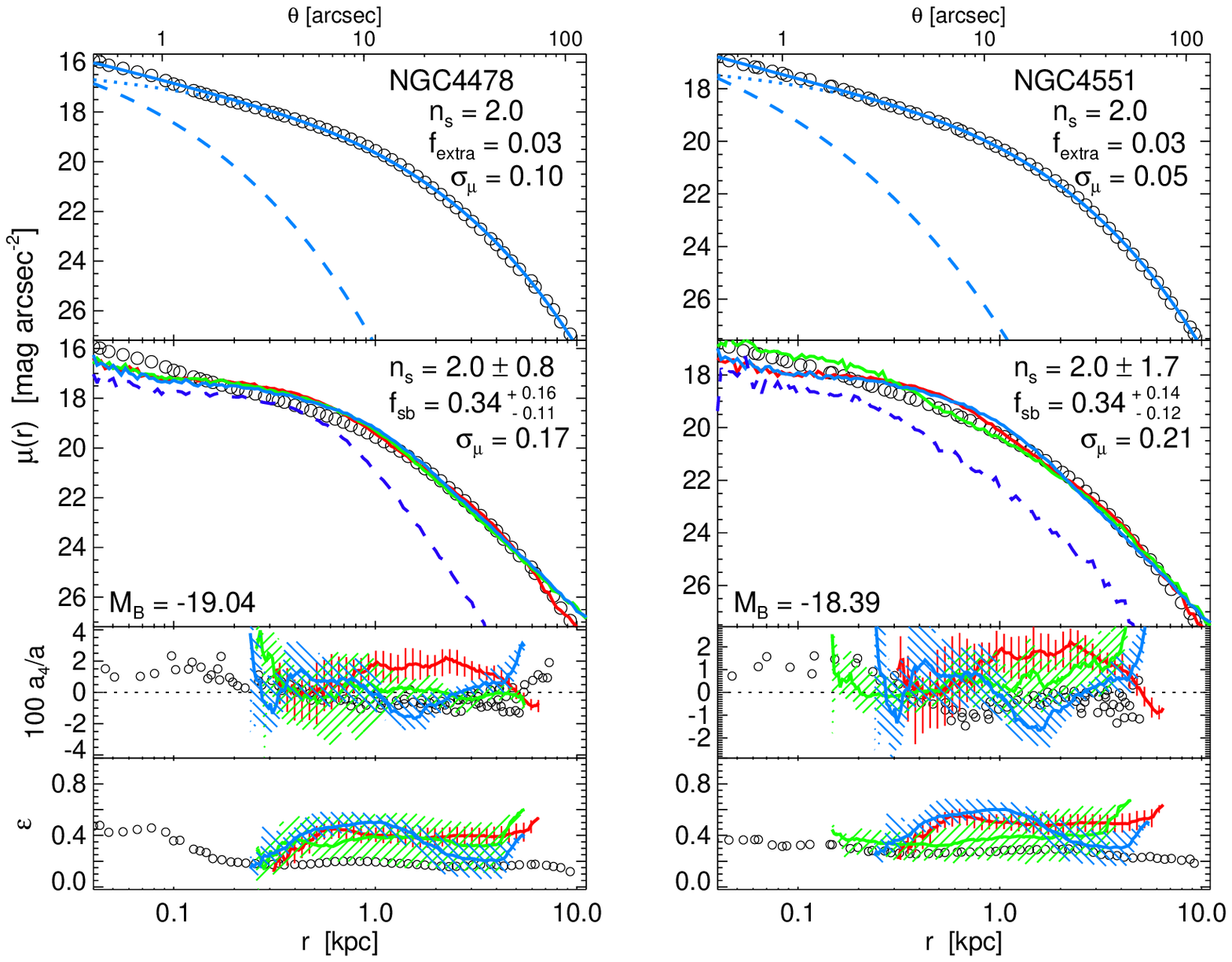}
    \caption{Lower-mass cusp ellipticals ($\sim0.2-0.3\,\mstar$). Our simulations
    reproduce the observed outer profiles and kinematic properties of such galaxies, but
    do not resolve the stellar cluster nuclei at small radii. The extra light recovered by our 
    two-component fits therefore can be misleading at such low mass.     
    (As Figure~\ref{fig:jk4}, but in log-log space.)
    \label{fig:jk4.log}}
\end{figure*}

\begin{figure*}
    \centering
    \plotter{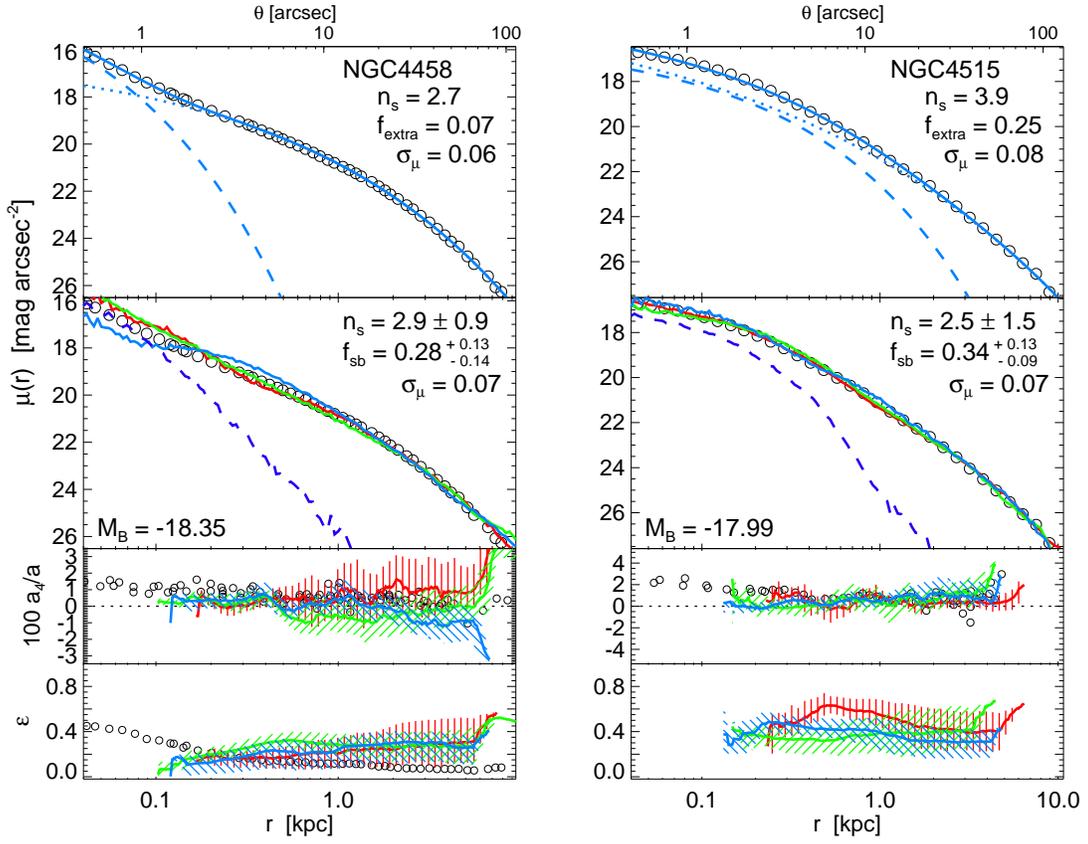}
    \caption{Additional low-mass ($\sim0.2\,\mstar$) cusp ellipticals. Our 
    fits perform better in this case.
    (As Figure~\ref{fig:jk5}, but in log-log space.)
    \label{fig:jk5.log}}
\end{figure*}

\begin{figure*}
    \centering
    \plotter{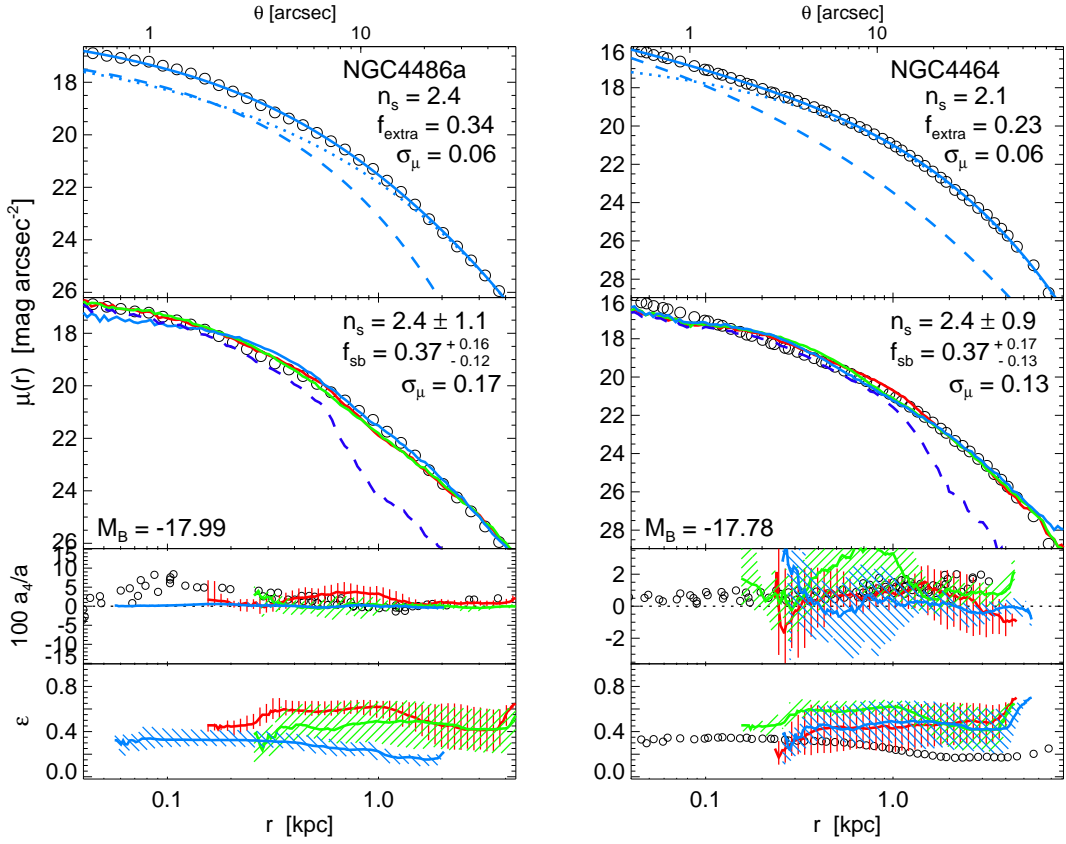}
    \caption{Additional low-mass ($\sim0.1-0.2\,\mstar$) cusp ellipticals, but 
    in this case without prominent stellar clusters in their nuclei. In this case 
    our parameterized fitting is not misled and we recover similar starburst 
    fractions to our simulations.
    (As Figure~\ref{fig:jk6}, but in log-log space.)
    \label{fig:jk6.log}}
\end{figure*}

\begin{figure*}
    \centering
    \plotter{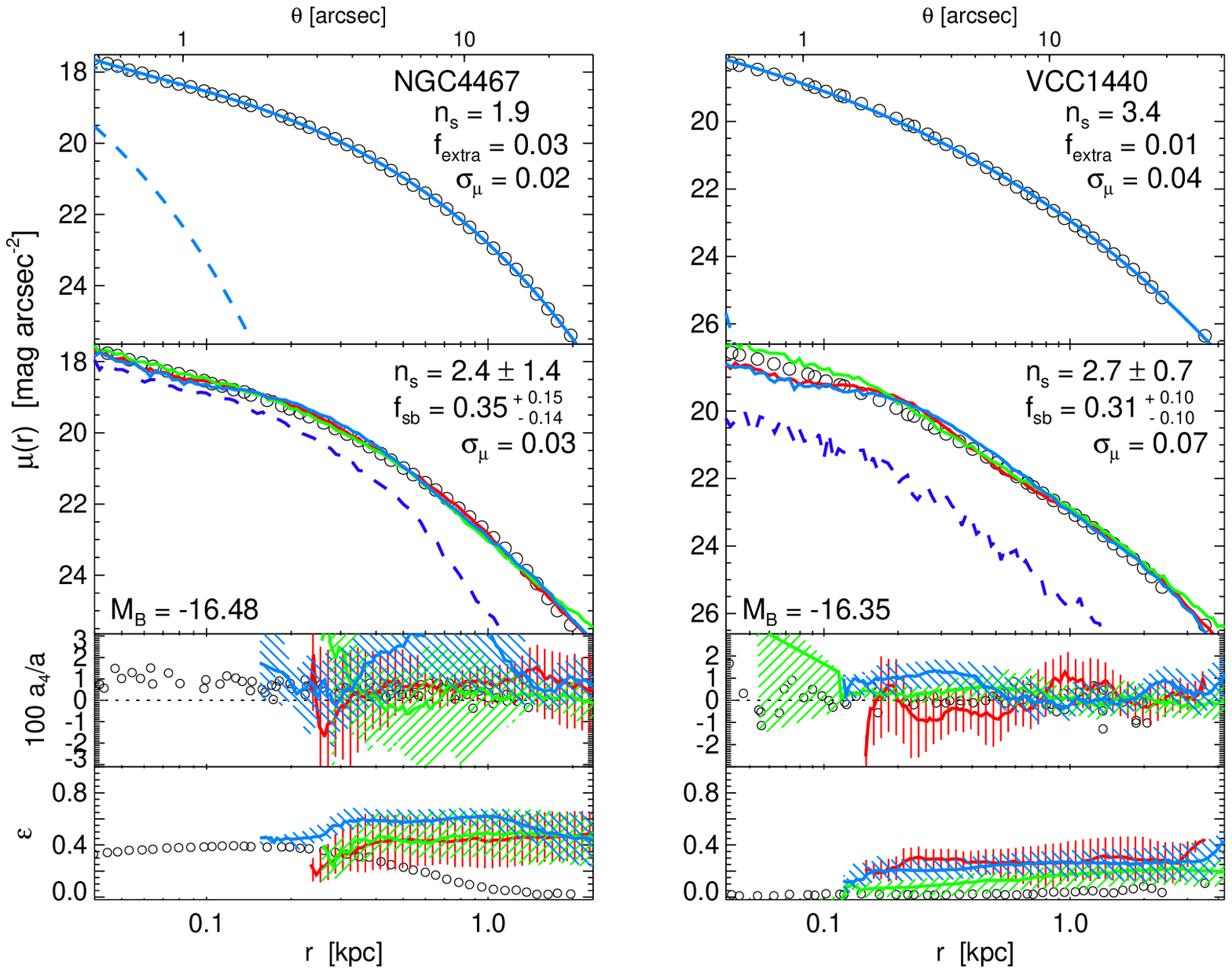}
    \caption{Very low-mass cusp ellipticals ($\sim0.03-0.1\,\mstar$). 
    Our simulations provide less good matches at 
    these luminosities, where dwarf galaxies dominate the spheroid 
    population (ellipticals at these masses are very rare). 
    Robustly resolving the extra light in these 
    very small systems probably requires $\lesssim10\,$pc 
    spatial resolution.
    (As Figure~\ref{fig:jk7}, but in log-log space.)
    \label{fig:jk7.log}}
\end{figure*}

\begin{figure*}
    \centering
    \plotter{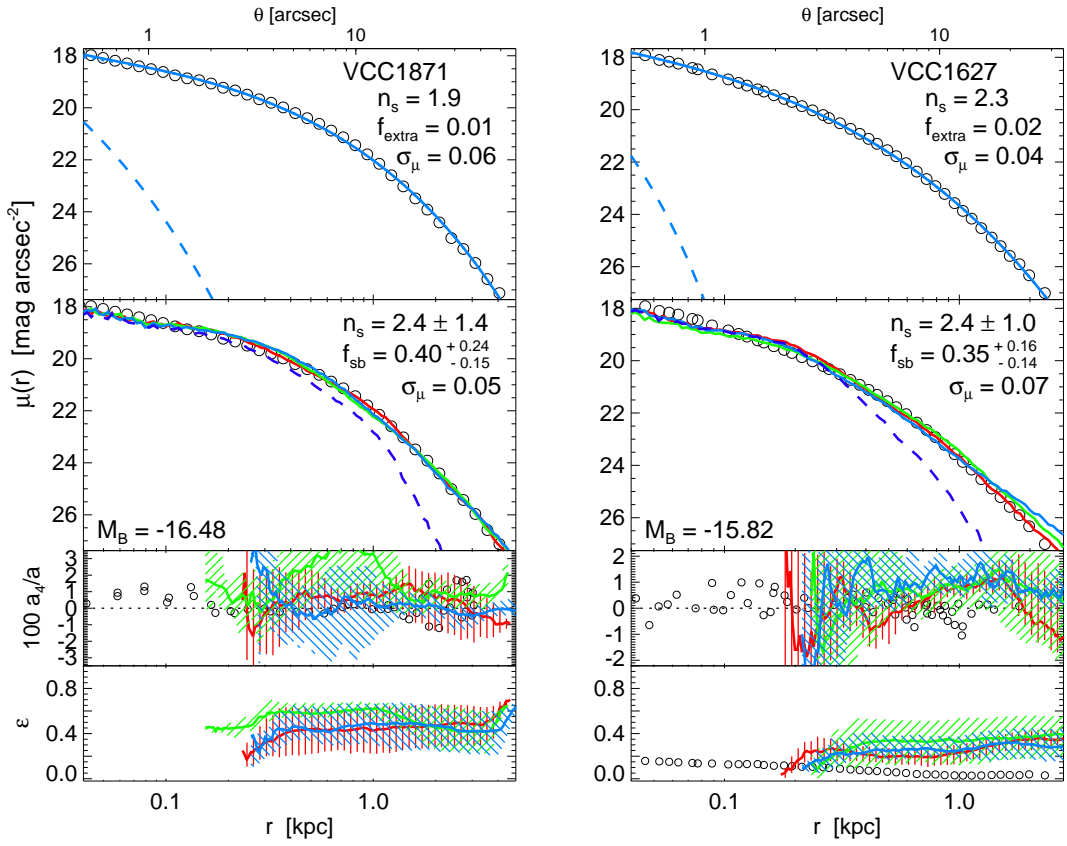}
    \caption{The lowest-luminosity cusp ellipticals in 
    Virgo ($\sim0.01\,\mstar$). The comparison with our 
    simulations is similar to Figure~\ref{fig:jk7.log}. 
    (As Figure~\ref{fig:jk8}, but in log-log space.)
    \label{fig:jk8.log}}
\end{figure*}

\begin{figure*}
    \centering
    \plotter{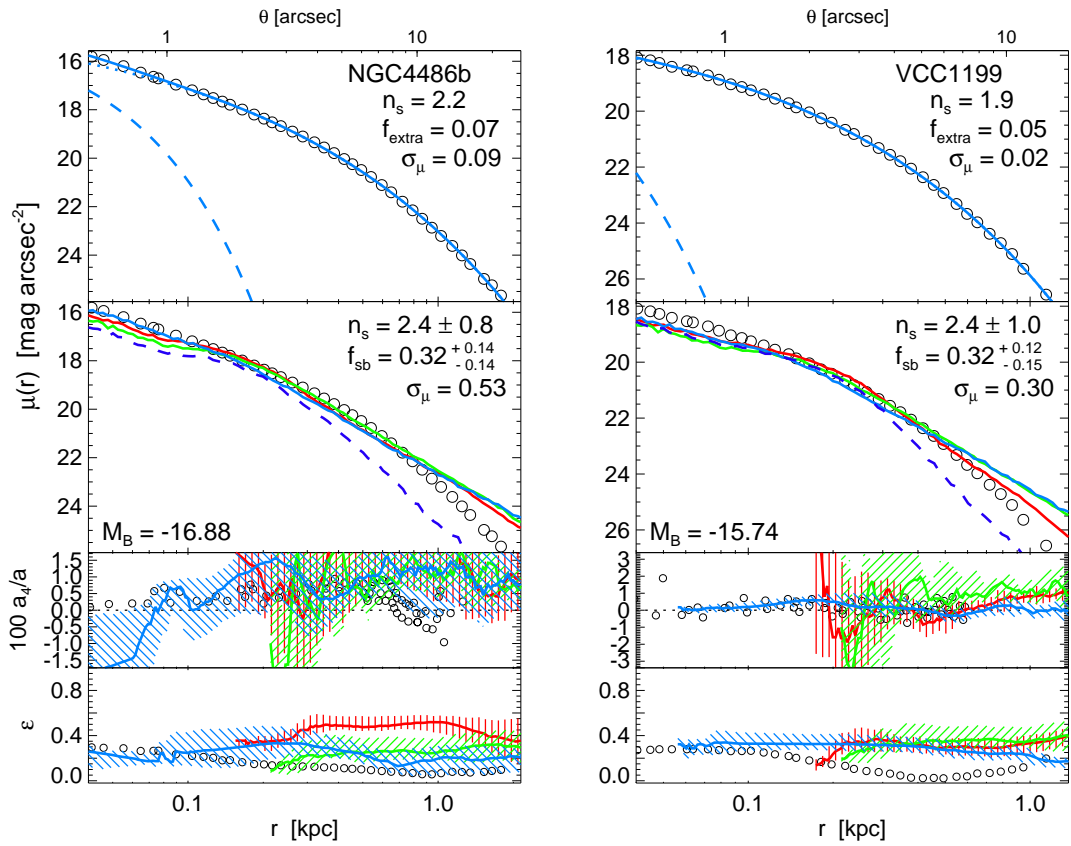}
    \caption{``Compact ellipticals.'' None of our simulations are 
    as compact as these objects (effective radii $\sim200\,$pc). 
    (As Figure~\ref{fig:jk10}, but in log-log space.)
    \label{fig:jk10.log}}
\end{figure*}

In Figures~\ref{fig:jk1.log.b}-\ref{fig:jk10.log} we reproduce Figures~\ref{fig:jk1}-\ref{fig:jk10}, 
but with profiles shown in log-log projection as opposed to $r^{1/4}$ projection. 


\clearpage 
\appendixcolumns

\section{Are Gas-Rich Merger Remnants ``Cusps''? Resolution Tests as $r\rightarrow0$}
\label{sec:appendix:resolution}

In \paperone, we conduct resolution tests and demonstrate that the primary 
quantities of interest here, namely the extra light fraction and 
outer \Sersic\ index, are reasonably well converged for the mass 
range of interest given our typical $<100$\,pc spatial resolution. 
In that paper, however, the observations had comparable (or poorer) spatial 
resolution to our simulations; the HST observations of the nuclear regions of 
ellipticals, on the other hand, resolve extremely small scales 
$\sim 1-10\,$pc, well below our typical simulations. It is therefore of some 
interest to examine the behavior at small radii in our simulations. 

\begin{figure}
    \centering
    \scaleup
    \plotter{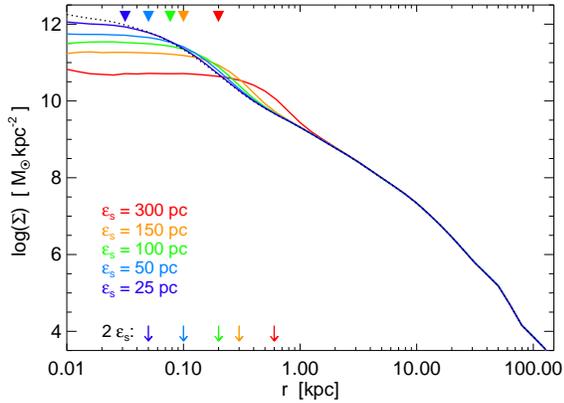}
    \caption{Effects of resolution on a remnant from a simulation of a gas-rich merger. 
    We show profiles of otherwise identical simulations with a large dissipational 
    starburst, but different gravitational softening lengths $\epsilon_{s}$. 
    Dotted line is the profile (extrapolated to $r\rightarrow0$) of a typical 
    ``cusp'' elliptical (logarithmic slope ${\rm d}\ln{I}/{\rm d}\ln(r)\rightarrow0.5$ 
    as $r\rightarrow0$). Finite resolution limits mean our profiles will always 
    artificially flatten within some radius: triangles at top show where each simulation profile 
    artificially flattens below this slope (${\rm d}\ln{I}/{\rm d}\ln(r)=0.5$); 
    this tends to happen at our softening length $\approx \epsilon_{s}$. 
    Arrows are shown for each simulation at $2\,\epsilon_{s}$: at 
    $\sim2-3\,\epsilon_{s}$ the profiles are converged within $\sim 0.1\,$mag 
    (good for our purposes in this paper). At $\gtrsim5\,\epsilon_{s}$ the 
    profiles and kinematics are fully converged. 
    \label{fig:res.test}}
\end{figure}

Figure~\ref{fig:res.test} shows an example of the (sightline-averaged) 
surface brightness profile of a simulated major merger remnant 
as a function of numerical resolution, reflected in the gravitational 
softening length $\epsilon_{s}$. 
Clearly, at some point around our resolution limits, the profiles 
artificially flatten (according to the softening) and 
become flat (essentially creating an artificial ``core'' at the 
center).\footnote{Note that this is different from the behavior 
seen in earlier generations of numerical simulations such as those in 
\citet{mihos:cusps}.
In those simulations, time integration inaccuracies at the
highest densities led the stellar cusps to artificially 
contract to the spatial resolution set by the gravitational
softening length, yielding sharp, compact 
``spikes'' in the surface brightness profile.
This did not affect any other aspects of the evolution and,
indeed, if these spikes are smoothed on the spatial scale 
set by the condition that the starburst component be 
self-gravitating, the results of \citet{mihos:cusps}
agree in detail with those presented here
and in 
\citet{hopkins:cusps.mergers}.} 
It is clear that, as we increase our resolution, 
the profiles continue to rise towards smaller and smaller radii. 
In detail, 
outside of $\sim2-3\,\epsilon_{s}$, we find that the profiles are 
sufficiently well-converged for our purposes in this paper. 
However, more subtle, detailed features in the galaxies such as 
the boxy or disky-ness of the isophotal shapes $a_{4}/a$ 
(typically a $\sim1\%$ effect in the deviation of the shapes 
from ellipses) and central kinematics, converge less rapidly. Nevertheless 
we find that outside of $\sim 5\,\epsilon_{s}$, there are 
no measurable resolution effects in any of these parameters. 

Where do our resolution limits flatten the profiles into 
false cores? We check this by simply adopting the common observational 
definition of a ``core'' \citep{faber:ell.centers,lauer:bimodal.profiles}, 
namely where the logarithmic derivative 
of the surface brightness profile flattens 
below a threshold $I\propto r^{-1/2}$ ($-{\rm d}\ln{I(r)}/{\rm d}\ln{r}$ < 0.5). 
This is shown in Figure~\ref{fig:res.test}. On average, our systems 
only significantly flatten into false cores at $r\lesssim 1\,\epsilon_{s}$. 
In other words, nearly all of our gas-rich merger 
simulations, including those pushing our spatial 
resolution to $\lesssim 20\,$pc, have ``cuspy'' nuclear profiles 
(by the observational definition) all the way down to the 
gravitational softening length. 

\begin{figure}
    \centering
    \scaleup
    \plotter{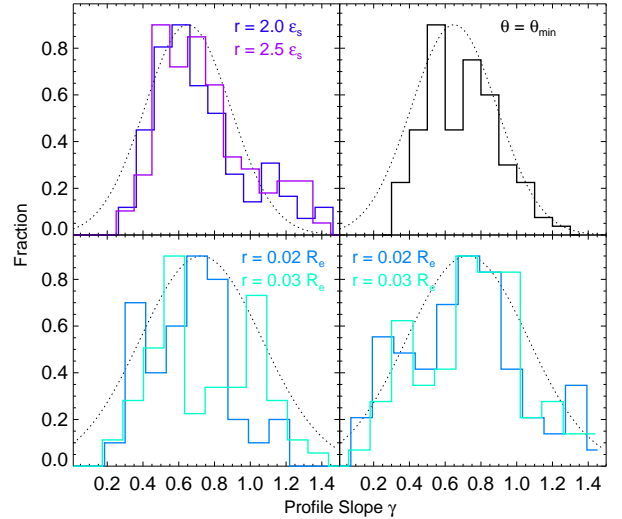}
    \caption{``Cuspy-ness'' of our simulations at reasonably resolved radii. 
    {\em Top Left:} Distribution of brightness profile 
    slopes ($\gamma \equiv -{\rm d}\ln{I}/{\rm d}\ln(r)$)
    in the remnants of 
    gas-rich simulations, at the smallest radii where 
    resolution limits do not completely flatten the profiles 
    (measured at $\sim2.0\,\epsilon_{s}$ and $2.5\,\epsilon_{s}$). 
    {\em Top Right:} Same, but slopes are 
    measured from the observed profiles at the HST resolution limits for 
    each object. 
    Dotted line in both panels is a fit to the observed distribution 
    \citep[see][]{lauer:bimodal.profiles}. 
    {\em Bottom Left:} Profile slopes from the simulations, measured at fixed 
    (fractional) radius ($0.02\,R_{e}$ and $0.03\,R_{e}$; we include only 
    simulations where these radii are $>3\,\epsilon_{s}$). 
    {\em Bottom Right:} Observed profile slopes measured at the 
    same fractional radii ($0.02\,R_{e}$ and $0.03\,R_{e}$). 
    Dotted line in the lower panels is a fit to the observed 
    distribution at $0.02\,R_{e}$. 
    Down to our best resolution limits, the simulated gas-rich merger 
    remnants show typical steep ``cuspy'' slopes similar to 
    those in the observed cusp population.     
    \label{fig:cusp.slopes}}
\end{figure}

This approaches (and in the best cases overlaps with) 
the radii where the observed slopes are classified as ``cusps'' or ``cores.'' 
We therefore directly check whether the slopes of our simulated systems are similar 
to those observed. We already know that many of our simulations match observed 
profiles down to $\sim50$\,pc, from Figures~\ref{fig:jk1}-\ref{fig:lauerpp1}; here 
we use our highest-resolution simulations and test whether the slopes at the 
smallest radii we can reliably say anything about are reasonable or not. 
We consider the distribution in logarithmic slopes of our simulations 
at $2\,\epsilon_{s}$, the smallest radii where they are not strongly 
flattened by resolution effects, and compare with the observed nuclear 
slope distribution from \citet{lauer:bimodal.profiles} at the observational resolution limits. 
In a more physically motivated manner, we compare the slopes measured 
at fixed radii relative to the effective radius 
$\sim0.02-0.03\,R_{e}$ in both our simulations and observations 
(comparison at fixed absolute radius yields similar results), only including 
simulations (and observations) where these radii are well-resolved ($>3$ times 
the resolution limit). In either case, the agreement is good. 
This strongly suggests that, down to 
$\sim10\,$pc where the physics of e.g.\ individual star-forming sites becomes 
important, our gas-rich 
merger remnants genuinely have ``cuspy'' or ``power-law'' central profiles, and 
that the agreement seen in Figures~\ref{fig:jk1}-\ref{fig:lauerpp1} 
between simulated and observed profiles would continue 
down to such radii if we only had the numerical resolution. 
Similar conclusions were obtained in \citet{cox:feedback}, who 
also demonstrated that these conclusions (given a fixed amount of 
gas at the time of the final merger) are insensitive to the details of the 
numerical algorithm and feedback prescriptions in the 
simulations.

\begin{figure}
    \centering
    \scaleup
    \plotter{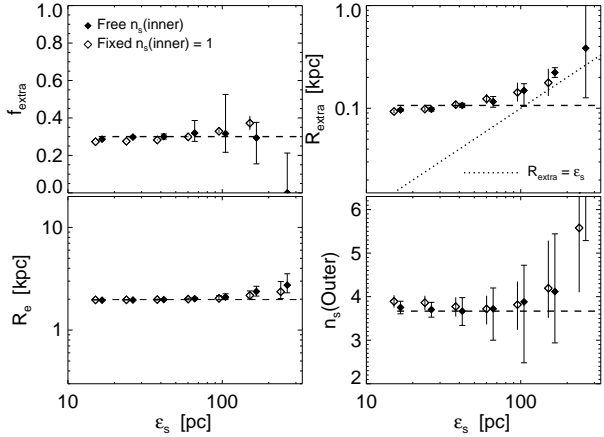}
    \caption{Convergence properties of the quantities of interest in this 
    paper, for an otherwise identical simulation 
    (shown in Figure~\ref{fig:res.test}) 
    as a function of gravitational softening length $\epsilon_{s}$. 
    Filled diamonds show results from fitting the projected profile 
    to our two-component decomposition, with a free inner \Sersic\ 
    index/shape parameter $n_{s}^{\prime}$; open circles 
    adopt a fixed $n_{s}^{\prime}=1$. Error bars show the $\pm1\,\sigma$ 
    range owing to shot noise, statistical fit degeneracies, 
    and variation between different realizations. The quantities of interest 
    here are well converged for our typical resolution limits 
    $\sim30-50\,$pc. For all resolutions ($\lesssim100\,$pc) where 
    the quantities are converged, we obtain the same average answers 
    using a free $n_{s}^{\prime}$ or fixed $n_{s}^{\prime}=1$; but when 
    resolution or seeing 
    is relatively poor ($\gtrsim30\,$pc), a fixed $n_{s}^{\prime}$ is useful 
    to minimize the noise (simply owing to the limited dynamic range in the fits). 
    \label{fig:convergence}}
\end{figure}

Because the inner shape of the extra light clearly changes (continuing to rise towards 
small $r$) in Figure~\ref{fig:res.test}, we examine the convergence properties in 
a similar resolution test in Figure~\ref{fig:convergence}. 
We plot the fitted extra light fraction $f_{\rm extra}$, effective radius of 
the extra light component $R_{\rm extra}$, 
effective radius of the whole elliptical inferred from the fit $R_{e}$, and outer \Sersic\ index $n_{s}$, 
in otherwise identical simulations as a function of $\epsilon_{s}$. We show the 
median values across a large number of sightlines, but the sightline-to-sightline 
distribution behaves as a whole in the same manner (see \paperone\ for resolution 
tests demonstrating the convergence of the distribution of $f_{\rm extra}$ 
and $n_{s}$ across sightlines). For each simulation, we generate alternative realizations 
by randomly re-scattering the stars according to the smoothing kernel, and 
show the error bars corresponding to the $\pm1\,\sigma$ range in fits to these realizations
(unlike e.g.\ the variation from sightline-to-sightline, much of which is real in that it 
reflects actual asymmetries in the galaxy, this variation is purely a resolution 
effect, and should vanish as $\epsilon_{s}\rightarrow0$). The error bars 
therefore effectively include the formal statistical errors and fit degeneracies 
as well. We show results both 
for a fixed inner $n_{s}^{\prime}=1$ (the \Sersic\ index of the extra light component itself, 
as described in \S~\ref{sec:fits}), and free inner $n_{s}^{\prime}$ (i.e.\ both assuming a fixed 
extra light component shape or fitting for the shape). 

Two broad rules of thumb consistently emerge from our resolution tests. 
First, regardless of the exact details of our fit methodology, the quantities of 
interest here, especially $f_{\rm extra}$, $n_{s}$, and $R_{e}$ are well-converged
for spatial resolution below $\sim100\,$pc ($\sim1''$ at the distance to Virgo, 
a factor $\sim10$ larger than the HST diffraction limit). We demonstrate in \paperone\ that 
the distribution of these quantities across sightlines is also reasonably well-converged 
below this threshold. In terms of general criteria, global quantities (such as $R_{e}$) 
converge quickly (at a resolution 
$\sim$ a couple hundred pc), followed by integral quantities such as 
$f_{\rm extra}$ and quantities related to the ``outer'' (large-scale) profile ($n_{s}$). 
The effective radius of the extra light 
($R_{\rm extra}$) is more demanding, because of course the fitted size will 
not be smaller than the resolution limits (generally we find this is 
converged up to a smoothing length $\lesssim1/2$ of the converged or ``true'' 
extra light size). For this reason, we do not reproduce observed systems with 
$R_{\rm extra}\ll100\,$pc (see Figure~\ref{fig:sizes} 
and Figures~\ref{fig:jk7.log}-\ref{fig:jk10.log}), but this is only relevant for a few of the 
very lowest-mass $\sim 0.01\,L_{\ast}$ ellipticals in our sample. 
Most demanding, of course, is the detailed shape of the extra light component itself 
(as $r\rightarrow0$), 
which Figures~\ref{fig:res.test} and \ref{fig:cusp.slopes} demonstrate we are only marginally 
resolving in our highest-resolution simulations. We reach almost identical conclusions 
regarding convergence if 
we repeat this study by artificially degrading the seeing in the observed profiles. 
Note that the absolute values here are for $\sim 0.1-1\,L_{\ast}$ ellipticals -- in much larger, more 
massive systems (especially those with flat central 
cores extending to $\sim100-300\,$pc), the resolution limits can be much less restrictive. 
In any case our resolution studies and experiments with observed profiles suggests 
we are not significantly biased in our estimates of the most important quantity here, 
$f_{\rm extra}$ (or $f_{\rm sb}$). 

Second, Figure~\ref{fig:convergence} explicitly demonstrates that our results 
are not changed whether we (for convenience) adopt a fixed inner extra light 
component shape $n_{s}^{\prime}$ when we fit to our 
simulations, or leave it as a free parameter. In the mean, the two recover the 
same answer (so long as our resolution is below the $\sim100\,$pc threshold 
needed to resolve the structures of interest in the first place). 
Unsurprisingly, when the resolution is extremely good, the results are most 
robust when we allow a free $n_{s}$ -- this allows the fit the freedom to 
deal with small scale features and e.g.\ the broad range in inner profile 
``cusp'' slopes (Figure~\ref{fig:cusp.slopes}). Fixing $n_{s}^{\prime}$ in 
such cases (or in e.g.\ observations with $\sim1-5\,$pc resolution) 
can produce a higher rate of catastrophic failure owing to
the presence of small-scale features that are unimportant for the
overall profile. For example, a strict inner exponential ($n_{s}^{\prime}=1$) 
implies that the logarithmic slope of the 
surface brightness profile goes to zero as $r\rightarrow0$, 
whereas the small-scale ($\ll 30$\,pc) profiles of cusp galaxies have a wide 
range of non-zero logarithmic slopes. 

When the resolution is sufficient then, 
one can (and should) free $n_{s}^{\prime}$ and fit for the shape of 
the inner component as well as its radius and mass fraction (although, 
again, fixing it introduces no bias, just a higher failure/confusion rate). 
However, Figure~\ref{fig:convergence} also shows the expected 
behavior as the resolution is downgraded: up to the limits where 
properties are converged, fits with a free inner component shape 
recover the same answer on average, but the uncertainties (realization-to-realization 
noise and fit degeneracies) grow rapidly (owing largely to the poor resolution 
of the converged extra light shape, which gives the fit too much freedom to 
trade off extra light and outer components). The result is that, as noted in 
\S~\ref{sec:fits}, our results for this paper are entirely unchanged if we 
re-fit all our simulations with a free inner $n_{s}^{\prime}$, but 
the scatter in predicted quantities increases significantly. When the resolution 
is poor, then, especially if the data sets of interest are limited (i.e.\ one does 
not have so large a number of objects that very large scatter is not a problem), 
fixing the inner shape is a convenient assumption that greatly reduces fitting 
degeneracies and, for the choice of inner shape $n_{s}^{\prime}\sim1$, 
introduces no significant systematic bias. 
Fortunately, the behavior in Figure~\ref{fig:convergence} and our other 
resolution tests is reasonably simple, and yields a useful rule of thumb: 
it is appropriate to free the inner component $n_{s}^{\prime}$ and fit for the shape 
of the extra light with resolution better than $\sim20-30\,$pc (ideally $\lesssim10\,$pc), 
or equivalently with resolution such that the extra light size $R_{\rm extra}$ is 
resolved with at least $\sim5-10$ resolution elements. Again, we find 
the same is true for observations (the observations used in this paper -- at least 
for the intermediate and massive galaxies -- 
easily meet this criterion, with $\sim10-20\,$pc corresponding to $\sim0.1-0.3''$ 
at Virgo, a factor of a couple larger than the HST diffraction limit). 

Our dynamic range at large radii is, of course, not significantly limited in the simulations 
(rather we restrict to a comparable range to that observed). For a 
detailed study of the dynamic range requirements at large radii for fitting 
e.g.\ the outer $n_{s}$, we refer to \citet{jk:profiles}. We note though their conclusions 
(which we also find in limited experiments changing our sampled dynamic range) 
that for the observations and simulations here, the dynamic range is sufficient so 
as not to introduce bias or significantly larger uncertainties.

\section{The Distinction Between Stellar Nuclei and Extra Light}
\label{sec:appendix:nuclei}

\begin{figure*}
    \centering
    \scaleup
    \plotterr{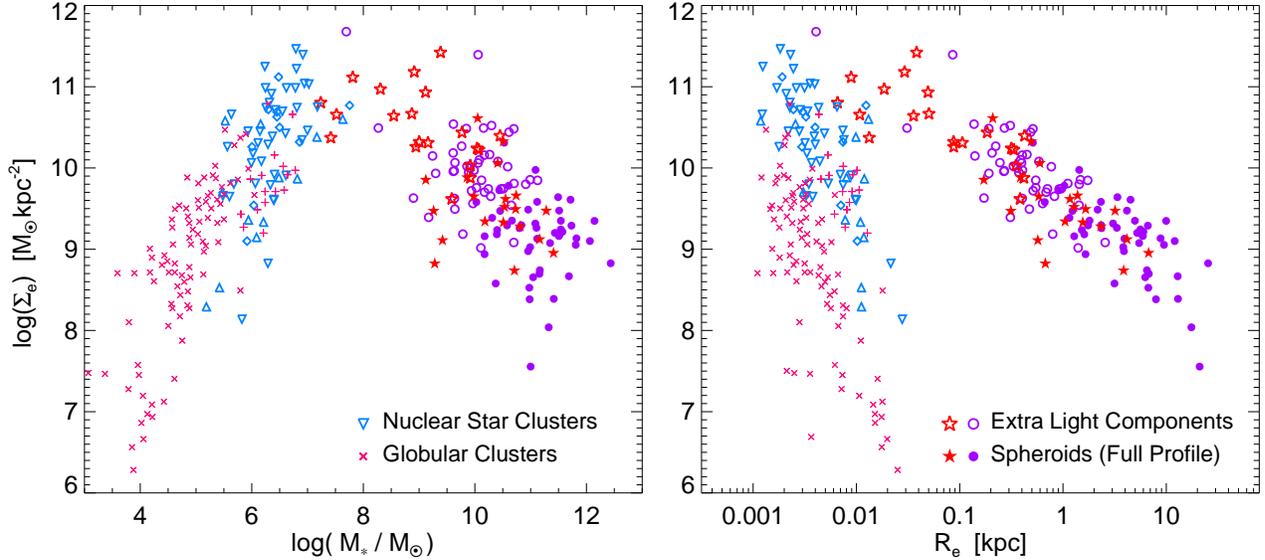}
    \caption{Fundamental parameter correlations of 
    extra light components (open red stars and violet circles; style 
    as in Figure~\ref{fig:ns.mass}), compared to 
    those of stellar nuclei (nuclear stellar clusters or 
    ``central massive objects''; blue symbols). 
    {\em Left:} Effective surface mass density versus stellar 
    mass (a nearly identical plot is obtained if we consider 
    surface brightness versus luminosity). 
    Observed stellar nuclei parameters are from 
    \citet[][triangles]{geha02:dE.nuclei}, 
    \citet[][inverted triangles]{boker04:nuclei.scalings}, and 
    \citet[][diamonds]{walcher05:nuclei.mdyn}. 
    The two classes of systems not only separate strongly in 
    mass, but trace nearly perpendicular correlations with 
    {\em opposite} physical senses. 
    For comparison, we plot the corresponding points for 
    the entire elliptical profiles (filled symbols) and 
    for globular clusters (pink symbols; $\times$'s and crosses  
    are Milky Way and NGC 5128 (Cen A) globulars from \citet{harris96:mw.gcs} and 
    \citet{harris02:cenA.gcs,martiniho04:cenA.gcs}, respectively). 
    Extra light components 
    form a continuous extension of the spheroid/elliptical 
    population (``classical'' stellar bulges also lie along this correlation 
    when plotted); unsurprising given that we argue they drive the 
    effective radii and central surface brightness of the ellipticals. 
    Stellar nuclei, on the other hand, appear to follow (at least roughly) 
    an extension of  the correlations for globular clusters. 
    {\em Right:} Same, but showing surface density versus effective 
    radius of each component/population. 
    Although the sequences may approach each other 
    at the very lowest elliptical masses (where classical bulges 
    and ellipticals are rare, and where we may have 
    misidentified some dwarf spheroidals as ellipticals), 
    there is a sharp division between their structural properties, 
    masses, radii, and parameter correlations (also, stellar population 
    ages and metallicities; see text). The two are in general easy to separate 
    and should not be confused. 
    \label{fig:xl.nuc.compare}}
\end{figure*}

Many spiral galaxies and dwarf spheroidals show 
a central excess in their light profiles associated with dense stellar 
nuclei (nuclear star clusters) \citep{phillips96:nuclei.imaging,
carollo97:nuclei.morphology,carollo98:nuclei.morphology,carollo99:nuclei.scalings,
matthews99:nuclei.statistics,boker02:nuclei.ids,boker04:nuclei.scalings,seth:nuclear.star.clusters}. 
In their 
analysis of the ACS Virgo data, this is what \citet{cote:virgo} 
and \citet{ferrarese:profiles} 
identify as the central ``excess'' component (largely in their 
dwarf spheroidal sample). 
In rough terms, 
this aspect of these nuclei is superficially similar to what we identify as 
extra light. However, as we discuss in \S~\ref{sec:fits} 
and has been demonstrated with detailed HST observations of these 
objects \citep[see e.g.][]{carollo99:nuclei.scalings,matthews99:nuclei.statistics,
boker04:nuclei.scalings,walcher06:nuclei.ssp}, 
on closer examination it is immediately clear 
that these nuclei are very different and 
physically distinct from the extra light or starburst 
components we identify in the observations and model in our 
simulations. 

Figure~\ref{fig:xl.nuc.compare} compares the parameter correlations 
of extra light components and stellar nuclei; specifically their 
stellar masses, effective radii, and effective surface brightness 
or surface mass density $\sim M / (2\pi\,R_{e}^{2})$. The sequence of 
stellar nuclei is clearly distinct from that of extra light. 
The typical extra light component (effective radius $\sim100-500$\,pc)
is $\sim100$ times 
larger in spatial extent than a stellar nucleus (effective radius $\sim1-5\,$pc)
at the same surface brightness. The slopes of the sequences in either 
projection shown in Figure~\ref{fig:xl.nuc.compare} are nearly 
perpendicular: for starburst/extra light components, we find 
$I_{e}\propto M_{\ast}^{-0.8}$ (less massive systems are 
{\em more} dense and compact); for 
the stellar nuclei, $I_{e}\propto M_{\ast}^{+1.3}$ (less massive 
systems are {\em less} dense).
If we include the limited subsample of stellar nuclei with velocity 
dispersions measured in \citet{geha02:dE.nuclei} 
and \citet{walcher05:nuclei.mdyn}, we find similar results 
in fundamental projections involving $\sigma$. 

In fact, it is well established that the parameter correlations 
of stellar nuclei are similar to those of globular clusters, not ellipticals 
\citep[e.g.][and references therein]{carollo99:nuclei.scalings,geha02:dE.nuclei,
boker04:nuclei.scalings,walcher05:nuclei.mdyn}, 
and we show this in Figure~\ref{fig:xl.nuc.compare}. 
The extra light components, on the other hand, form a 
relatively smooth extension of the sequence obeyed by ellipticals 
\citep[in terms of the elliptical total half-mass radius and mass; see e.g.][]{kormendy:spheroidal1}
towards smaller radii and higher surface densities. This is 
a natural prediction of our models: in \S~\ref{sec:structural.fx} and 
\citet{hopkins:cusps.fp} we argue that ellipticals and bulges 
are (when analyzed as a single entity) driven along these correlations 
by the properties of their starburst/extra light, beginning 
from a location in parameter space (occupied by the least dissipational 
ellipticals with large radii and low surface brightness) 
similar to their progenitor disks. 
In short, dissipational/starburst components represent a smoothly 
rising excess at $\sim0.5-1$\,kpc from an outer dissipationless 
component extending to $\sim10-100$\,kpc in classical bulges and ellipticals, 
with smooth associated gradients; stellar nuclei represent 
a sharp excess \citep[being a distinct {\em object} with a fairly 
steep internal density profile; see e.g.][]{walcher05:nuclei.mdyn} at $\sim1-5$\,pc, 
with very distinctive properties. To the extent that 
stellar nuclei exist in some ellipticals, they would sit ``on top'' of the 
extra light that dominates the profile within $\sim100-500\,$pc, 
and exist entirely below the radius regime we model or fit. 

As discussed in \S~\ref{sec:fits}, this is borne out by a detailed comparison 
between our fits and those in e.g.\ \citet{cote:virgo} and \citet{ferrarese:profiles}, 
who fit multi-component profiles to identify stellar nuclei in 
the very small-scale nuclear regions of Virgo galaxies. 
The ``outer'' profile in their images is 
based on the HST ACS profiles, which extent to outer 
radii $\sim1$\,kpc (almost entirely dominated by the 
starburst component of the galaxy), and their identified stellar clusters typically dominate 
the light profile at very small radii $\lesssim0.01\,R_{e}$. 
This is akin to separating our 
``inner'' component itself into multiple sub-components -- i.e.\ a starburst 
stellar component that blends relatively smoothly 
onto the outer, violently relaxed stars and an innermost nuclear 
star cluster. 

\begin{figure}
    \centering
    \plotter{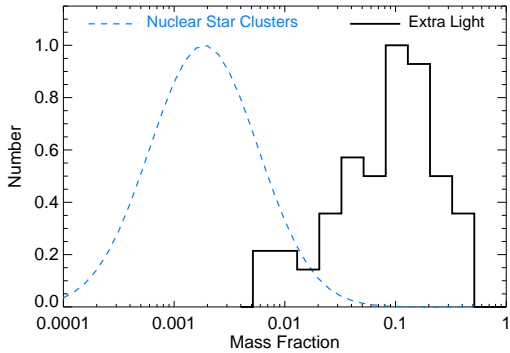}
    \caption{Distribution of 
    the mass/light fraction of the galaxy bulge 
    in stellar nuclei \citep[dashed; from the fit to the distribution observed 
    in][]{cote:virgo} versus our 
    fitted extra light components (solid). 
    Stellar nuclei have typical mass fractions $\sim2-3\times10^{-3}\,M_{\rm bul}$, 
    whereas extra light is typically $\sim0.1\,M_{\rm bul}$, a two order-of-magnitude 
    difference. The gap between the populations is even larger 
    if we consider e.g.\ absolute masses, effective radii, 
    masses relative to the whole galaxy (as opposed to the bulge), 
    or our best-fit simulation $f_{\rm sb}$ (instead of the fitted $f_{\rm extra}$, 
    which has somewhat more scatter). The systems are more than just distinct; it 
    is not possible to assemble typical extra light components from any amount of 
    hierarchical merging/aggregation of stellar nuclei. 
    \label{fig:xl.nuc.mf.compare}}
\end{figure}

Moreover, the mass fractions of these components are very 
different. Figure~\ref{fig:xl.nuc.mf.compare} compares the distribution 
of stellar nucleus mass fractions (relative to their host bulges) 
versus the mass fractions of extra light/starbursts. 
The characteristic stellar nucleus has a mass fraction 
$\sim 0.001-0.003\,M_{\rm gal}$ 
\citep[we plot the distribution fitted in \citet{cote:virgo}, but 
other studies find very similar distributions; see e.g.][]{carollo97:nuclei.morphology,
matthews99:nuclei.statistics}, compared to $\sim0.1\,M_{\rm gal}$ 
in extra light. Because the star clusters are typically identified 
in much less massive galaxies, the discrepancy is even larger 
($\sim4-5$ orders of magnitude) if we consider absolute masses. 
No amount of hierarchical merging, then (which would roughly 
conserve mass fractions in these components), would 
move a stellar cluster into the regime of extra light 
components. 

A number of other properties reveal the sharp division between 
these populations. The stellar populations in the extra 
light (see \S~\ref{sec:ssp.fx}) tend to be of similar (albeit slightly younger) 
stellar age to their hosts ($\sim3-10\,$Gyr), and are highly metal enriched 
(typical central metallicities $\sim1.5-3$ times solar) and 
moderately $\alpha$-enhanced 
(${\rm [\alpha /Fe]}\sim0.2-0.3$). 
Stellar nuclei tend to be extremely young 
(ages $\lesssim100\,$Myr) and have somewhat sub-solar
metallicities \citep{walcher06:nuclei.ssp}. 
There are also kinematic differences, with characteristically 
less rotation in nuclear clusters \citep{geha02:dE.nuclei}. 

Altogether, there should be no risk of confusing 
stellar nuclei and dissipational components with detailed 
observations. It is worth noting that at the very lowest elliptical masses 
($M_{\ast}<10^{9}\,M_{\sun}$, $M_{B}\gtrsim-17$)
and very highest stellar nuclei masses, 
the parameter sequences in Figure~\ref{fig:xl.nuc.mf.compare} 
approach one another, albeit still with opposite slopes. The most likely 
explanation is that this is just a coincidental overlap of their structural 
scalings, and in any case, true ellipticals and classical bulges 
become extremely rare at such low masses (likewise, stellar 
nuclei at and above these masses also become rare). 
It is also possible that, at these masses, some of our smallest ellipticals 
are really misclassified dwarf spheroidals (which have significant 
stellar nuclei; our fitting 
procedures might then mistakenly call that nucleus the ``extra light'') 
or that we are in these couple of cases finding a real stellar nucleus 
in an elliptical
and accidentally calling it the extra light (as opposed to 
recognizing the larger, less dense starburst component, on top of which 
such a stellar cluster would sit at the center of the 
surface brightness profile). Figures~\ref{fig:jk4} \&\ \ref{fig:jk4.log} 
show examples where our parameterized fitting may fall victim to 
this misclassification. 
In general however, even at the lowest classical bulge masses and highest 
stellar nuclei masses, the other striking differences in the 
systems (their stellar populations and kinematics) remain distinct: there 
is no continuity or ``intermediate'' class between the two populations in 
our data or the literature. 

It is in principle possible that there would be correlations between 
extra light and stellar nuclei, in that both might be formed by 
dissipational processes \citep[see e.g.][]{milosavljevic:diss.nuclei.formation,
seth:nuclear.star.clusters}. Moreover, at some level similar physics 
may be involved in determining e.g.\ the competition between star formation 
and gravitational collapse that determines their size-mass relations. 
However, given 
their separation in stellar populations, and given the prevalence 
of stellar nuclei in systems such as bulgeless disks and 
dwarf spheroidals that have manifestly {\em not} experienced 
major dissipational angular momentum loss, they probably are not 
directly coupled. 

\end{appendix}

\clearpage
\lscapeopen
\pagestyle{empty}
\begin{\tableset}{lccccccccccccccccccc}
\rotator
\tablecolumns{18}
\sizer
\tablecaption{Extended Fit Results\label{tbl:cusp.fits.extended}}
\tablewidth{0pt}
\tablehead{
\colhead{Name} &
\colhead{Morph.} &
\colhead{Ref.} &
\colhead{$\mu_{e}$ (extra)} &
\colhead{$R_{e}$ (extra)} &
\colhead{$n_{s}^{\prime}$ (extra)} &
\colhead{$\mu_{e}$ (out)} &
\colhead{$R_{e}$ (out)} &
\colhead{$n_{s}$ (out)} &
\colhead{$f_{e}$ (fit)} &
\colhead{$n_{s}$ (sim)} &
\colhead{$f_{sb}$ (sim)} \\
\colhead{(1)} &
\colhead{(2)} &
\colhead{(3)} &
\colhead{(4)} &
\colhead{(5)} &
\colhead{(6)} &
\colhead{(7)} &
\colhead{(8)} &
\colhead{(9)} &
\colhead{(10)} &
\colhead{(11)} &
\colhead{(12)} 
}
\startdata
NGC 4621 & E4/E/E5 & 1,2,3 & 
$17.38^{+4.82}_{-1.02}$ & $0.24^{+2.25}_{-0.21}$ & $3.91^{+4.50}_{-2.91}$ & 
$21.42^{+0.62}_{-2.20}$ & $ 4.33^{+2.35}_{-1.48}$ & $3.23^{+1.51}_{-0.75}$ & 
$0.053^{+0.005}_{-0.044}$ & $3.06^{+1.70}_{-0.18}$ & $0.140^{+0.168}_{-0.050}$ \\
\enddata
\tablenotetext{ \, }{{\footnotesize As Table~\ref{tbl:cusp.fits}, but with a complete 
list of fit parameters (included as supplemental material in the on-line edition of the 
journal; here, we show an illustrative example of one table entry). 
Parameters are the same as in Table~\ref{tbl:cusp.fits}, with additional 
fitted parameters listed. 
Added columns include: (2) Morphology (taken from the sources given in column (3), 
in the same order). 
(3) Source for surface brightness 
profiles (as in Table~\ref{tbl:cusp.fits}), 
where $1=$\citet{jk:profiles}, $2=$\citet{lauer:bimodal.profiles}, $3=$\citet{bender:data}, 
$4=$\citet{rj:profiles}. 
(4)-(10) Parameters of our two-component fits: 
(4) Effective surface brightness (in ${\rm mag\,arcsec^{-1}}$), i.e.\ $\mu(R_{\rm extra})$, 
for the fitted inner (starburst) component. (5) Effective radius of the extra 
light component $R_{\rm extra}$ [kpc]. (6) Sersic index of the extra light component 
$n_{s}^{\prime}$ (cases with $n_{s}^{\prime}=1$ and errors $=0$ are where 
$n_{s}^{\prime}$ is held fixed as described in the text). (7) Effective surface brightness 
of the outer (violently relaxed envelope) component. (8) Effective radius of this 
component. (9) Sersic index of this component ($n_{s}$(fit) in Table~\ref{tbl:cusp.fits}). 
(10) Integrated mass/light fraction in the fitted ``extra'' (inner) component. 
(11) Range of outer \Sersic\ indices (equivalent of $n_{s}$(out)) 
fit in the same manner to the best-fit simulations, 
at $t\approx1-3$\,Gyr after the merger when the system has relaxed.
(12) Fraction of light from stars produced in the central, merger-induced starburst 
in the best-fit simulations ($\pm$ the approximate interquartile range allowed). 
These values are medians for the fits to the available photometric profiles 
for each galaxy; they 
do not, together, represent the best fit to any particular individual 
measurements.}}
\end{\tableset}
\clearpage
\lscapeclose

\end{document}